%% file: push4_arxiv.tex
\documentclass[twocolumn]{aastex62}
\pdfoutput=1 
\usepackage{amsmath,amstext}
\usepackage[T1]{fontenc}
\usepackage{apjfonts} 
\usepackage[figure,figure*]{hypcap}
\usepackage{comment}

\usepackage{graphicx}
\usepackage{natbib}
\usepackage{epsfig}
\usepackage{epstopdf}
\usepackage{enumerate}
\usepackage{longtable} 
\usepackage{multirow, bigdelim}

\usepackage{tabularx}
\usepackage{multirow}
\usepackage{booktabs}

\newcommand{\push}{\mathrm{push}}
\newcommand{\msun}{$M_{\odot}$}
\newcommand{\kpush}{$k_{\rm push}$}
\newcommand{\trise}{$t_{\rm rise}$}

\graphicspath{{./}{figures/}{figures_metallicity/}}

\received{\today}
\revised{}
\accepted{}
\submitjournal{ApJ}

\shorttitle{Pushing 1D CCSNe to explosions: Low and zero metallicity}
\shortauthors{Ebinger et al.}

\begin{document}

\title{PUSHing core-collapse supernovae to explosions in spherical symmetry IV:\\ 
Explodability, remnant properties and nucleosynthesis yields of low metallicity stars\footnote{Dedicated to Margaret Burbidge on the occasion of her 100th birthday.}}

\author[0000-0002-0023-0864]{Kevin Ebinger}
\affiliation{GSI Helmholtzzentrum f\"ur Schwerionenforschung, D-64291 Darmstadt, Germany}
\author[0000-0002-3211-303X]{Sanjana Curtis}
\affiliation{Department of Physics, North Carolina State University, Raleigh NC 27695}
\author[0000-0001-7868-6944]{Somdutta Ghosh}
\affiliation{Department of Physics, North Carolina State University, Raleigh NC 27695}
\author[0000-0003-0191-2477]{Carla Fr\"ohlich}
\affiliation{Department of Physics, North Carolina State University, Raleigh NC 27695}
\author{Matthias Hempel}
\affiliation{Department f\"ur Physik, Universit\"at Basel, CH-4056 Basel, Switzerland}
\author[0000-0002-0936-8237]{Albino Perego} 
\affiliation{Dipartimento di Fisica, Universit\`a degli Studi di Trento, via Sommarive
14, 38123 Trento, Italy}
\affiliation{INFN, Sezione di Milano-Bicocca, Piazza della Scienza 3, 20126 Milano, Italy}
\author{Matthias Liebend\"orfer}
\affiliation{Department f\"ur Physik, Universit\"at Basel, CH-4056 Basel, Switzerland}
\author[0000-0002-7256-9330]{Friedrich-Karl Thielemann}
\affiliation{Department f\"ur Physik, Universit\"at Basel, CH-4056 Basel, Switzerland}
\affiliation{GSI Helmholtzzentrum f\"ur Schwerionenforschung, D-64291 Darmstadt, Germany}
\email{k.ebinger@gsi.de}
\email{ssanjan@ncsu.edu}

\begin{abstract}
In this fourth paper of the series, we use the parametrized, spherically symmetric explosion method PUSH to perform a systematic study of two sets of non-rotating stellar progenitor models. Our study includes pre-explosion models with metallicities Z=0 and Z=Z$_{\odot}\times 10^{-4}$ and covers a progenitor mass range from 11 up to 75 M$_\odot$. We present and discuss the explosion properties of all models and predict remnant (neutron star or black hole) mass distributions within this approach. We also perform systematic nucleosynthesis studies and predict detailed isotopic yields as function of the progenitor mass and metallicity. We present a comparison of our nucleosynthesis results with observationally derived $^{56}$Ni ejecta from normal core-collapse supernovae and with iron-group abundances for metal-poor star HD~84937. Overall, our results for explosion energies, remnant mass distribution, $^{56}$Ni mass, and iron group yields are consistent with observations of normal CCSNe. We find that stellar progenitors at low and zero metallicity are more prone to BH formation than those at solar metallicity, which allows for the formation of BHs in the mass range observed by LIGO/VIRGO.
\end{abstract}


\section{Introduction} \label{sec:intro}

At the end of their lives, stars more massive than 8~$M_{\odot}$ undergo gravitational collapse of their core, either of an ONeMg core for stars of $\sim 8$--10~M$_{\odot}$ \citep{Doherty2017} or of an iron core for stars $>10$~$M_{\odot}$. This collapse marks the onset of a core-collapse supernova (CCSN), a violent event resulting in an explosion that disrupts the star, leaves behind a compact object as a remnant, and synthesizes and disperses many chemical elements in the host galaxy of the progenitor star, contribution to its chemical enrichment. However, exactly which stars successfully explode and form neutron stars (NSs) after core collapse and which stars fail to explode and eventually form black holes (BHs) remains an open question.

Observationally, estimates of explosion energies and pre-explosion masses are available for many CCSNe. For the well-studied case of SN1987A, additional information on the ejected masses of $^{56,57,58}$Ni and $^{44}$Ti is also available \citep{Fransson.Kozma:2002,Seitenzahl2014,Boggs2015}. Failed SNe provide complimentary data relevant to the question of which massive stars successfully explode and which stars fail to explode. The LIGO/VIRGO collaboration has observed gravitational wave signals from 10 BH-BH mergers \citep{ligo-GWTC1} during the first two observing runs (and several more during the ongoing third observing run). The individual BH masses span a range from 7.6 to 50.6~$M_{\odot}$, with most of the masses being above $\sim 23$~$M_{\odot}$. A possible formation pathway of these black holes are failed SNe of low-metallicity massive stars \citep{ligo_GW150914}. In addition, \citet{adams16} have optically confirmed the disappearance of a 25~$M_{\odot}$ zero age main sequence (ZAMS) mass red supergiant star. 

There exist several studies that have investigated the connection between the final stellar pre-collapse structure of a massive star and the outcome of neutrino-driven explosions in effective models \citep{OConnor.Ott:2011,ugliano12,pejcha2015,push1,ertl16,sukhbold16,mueller2016,push2,push3,couch19,murphy19}. These studies use different approaches and have different scopes. However, overall they all agree that there is no single stellar mass that divides the massive stars into exploding and imploding stars. Instead, a more complex picture with ``islands of explodability'' has emerged.

Another open question related to CCSNe is what are the detailed yields from CCSNe for the entire range of initial masses and initial metallicities. To date only a few nucleosynthesis predictions from multi-dimensional CCSN simulations are available \citep{Harris.Hix.ea:2017,wanajo2d,Eichler.Nakamura.ea:2017,yoshida2017}. Currently, multi-dimensional simulations are computationally still too expensive to be performed over the full range of initial masses and metallicities, as required for Galactic chemical evolution (GCE) simulations.  Nucleosynthesis predictions from piston and thermal/kinetic bomb models in spherical symmetry are very abundant in the literature \citep{ww95,thielemann96,rauscher02,limongi06,nomoto06,hw07,umeda08,heger10,limongi12,chieffi13,nomoto13,chieffi17,nomoto17,limongi2018}, but these calculations do not include the physics of the collapse and of the explosion phase. In particular, the neutrino interactions that set the electron fraction in the innermost ejecta are omitted. This uncertainty most strongly affects the iron group yields \citep{cf06a}. In addition, the explosion energy and the mass cut are treated as two independent free parameters. Moreover, at the present time, only a few sets of yield predictions that include the entire mass range of CCSNe at more than one initial metallicity exist in the literature \citep{ww95,nomoto13,limongi2018}. Such yields are required to interpret the observed abundances in metal-poor stars and also as input to GCE simulations.

The present work is part of a series of investigations using the PUSH method, first introduced in \citet{push1} (hereafter Paper~I). The PUSH method is an effective method that relies on the neutrino-driven mechanism for the central engine of CCSNe. It mimics, in spherically symmetric simulations, the enhanced neutrino energy deposition of multidimensional models. Particularly important for the nucleosynthesis predictions is the fact that the PUSH method allows us to trigger explosions without modifying the electron-flavor neutrino/anti-neutrino luminosities. In addition, the bifurcation between ejecta and PNS matter (the mass cut) emerges naturally from the PUSH simulations. Both of these aspects are crucial for accurate predictions of the conditions in the innermost ejecta, where in particular the iron group elements are synthesized.

In Paper~I, we calibrated the PUSH method using SN~1987A and pre-explosion models with ZAMS masses in the range 18--21~$M_{\odot}$. We showed in a proof-of-principle study that explosion energies and yields of $^{56,57,58}$Ni and $^{44}$Ti consistent with observationally derived values can be obtained. 
In \citet{push2} --- hereafter Paper~II --- we extended and refined the PUSH method and applied it to two large sets of pre-explosion models of massive stars at solar metallicity. We predicted explosion properties and remnant properties in Paper~II. In \citet{push3} --- hereafter Paper~III --- we made use of the simulations from Paper~II to predict detailed nucleosynthesis yields. 
In this work, we apply the PUSH method, using the calibration obtained in Paper~II, to predict explosion outcomes and nucleosynthesis yields for pre-explosion models of massive stars at low and zero initial metallicity.

This article is organized as follows: Section \ref{sec:method} summarizes the input models, the CCSN simulations, and the nuclear reaction network used. We present the explosion outcomes in Section \ref{sec:exposions} and the detailed nucleosynthesis yields in Section \ref{sec:scan_nucsyn}. We also discuss trends in the explosion properties and yields as a function of progenitor properties. The explosion properties and nucleosynthesis yields are compared to observations in Section~\ref{sec:observations}. In Section \ref{sec:remnants}, we present and discuss the properties of the compact remnants (NSs and BHs) from our simulations. We conclude with a discussion and a summary of our results in Section \ref{sec:discuss}.

\section{Method and Input} \label{sec:method}

\subsection{Initial Models} \label{subsec:progenitors}

In this study, we explore pre-explosion models of two different metallicities: low-metallicity ($Z=10^{-4} Z_{\odot}$, ``u-series'') and zero metallicity ($Z=0$, ``z-series'') from \cite{Woosley.Heger:2002}. These models are based on the stellar evolution code KEPLER and represent non-rotating single stars. We investigate all models of the z-series from 11 to 40 $M_\odot$ ZAMS mass, increasing in steps of 1~$M_{\odot}$. From the u-series, we investigate the corresponding pre-explosion models of the same mass and six additional models between $45 M_{\odot}$ and $70 M_{\odot}$. A list of all pre-explosion models used in this study is given in Table~\ref{tab:progenitors}.
We use the same naming convention as in our previous papers: Each model is labelled by its ZAMS mass and the letter `u' or `z' indicating which series it belongs to (see also Table \ref{tab:progenitors}).
We compare our results for the u- and z-series to those of the two previously investigated series of pre-explosion models with solar metallicity: the ``s-series'' \citep{Woosley.Heger:2002} and the ``w-series'' \citep{hw07}, as presented in Paper~II and Paper~III.

\begin{figure}[]  
	\includegraphics[width=0.48\textwidth]{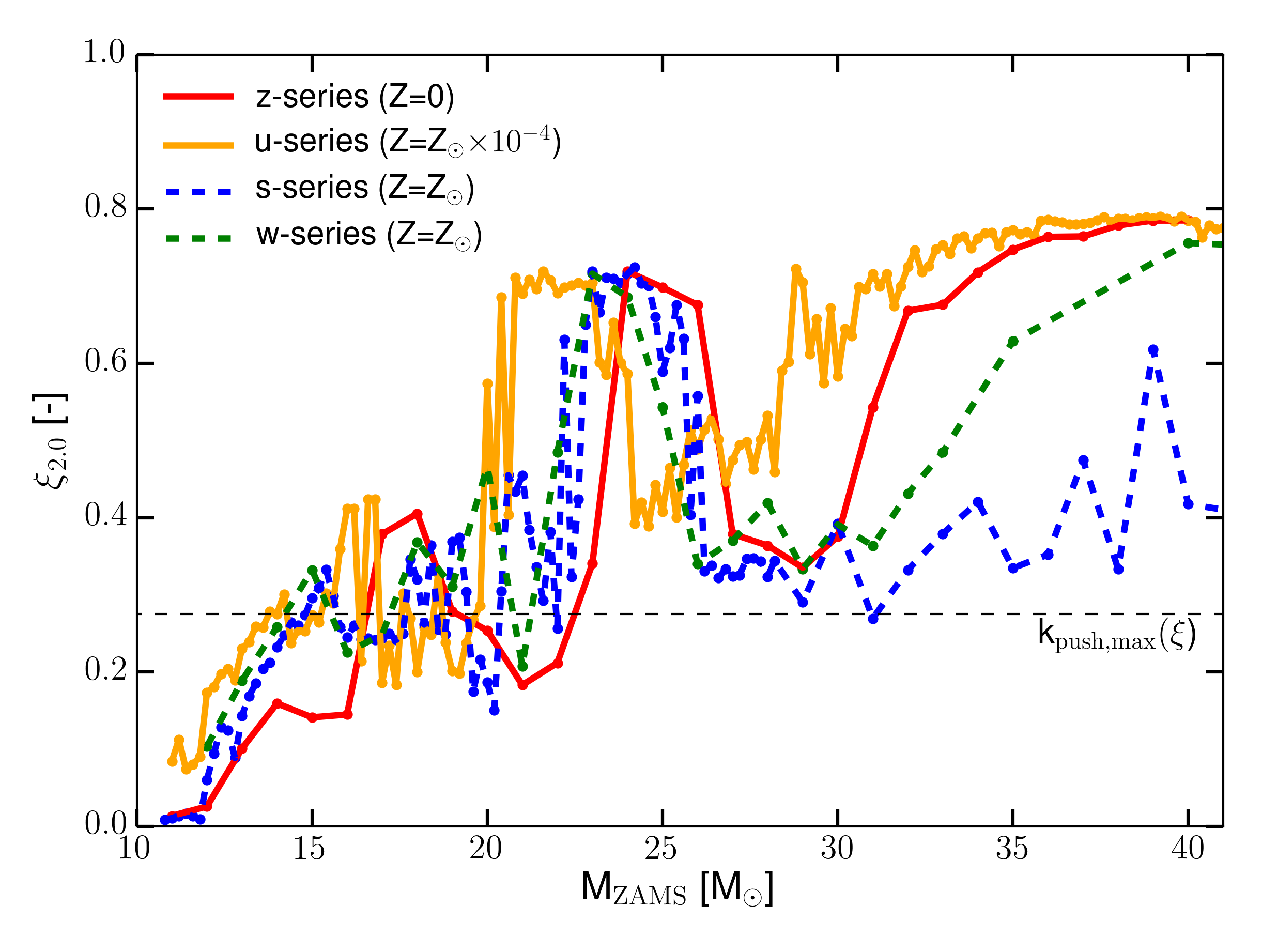}
	\caption{Compactness  $\xi_{2.0}$ (evaluated at $2.0$~$M_{\odot}$) as function of the ZAMS mass of the pre-explosion models. Four sets of initial models are shown: z-series (red), u-series (yellow), and s-series (blue) from \cite{Woosley.Heger:2002}, and also the solar metallicity models (green) from \cite{hw07}.
	The horizontal dashed black line denotes the compactness for which the standard calibration of \kpush reaches the maximum value.
		\label{fig:prog_compactness}
        }
\end{figure}

\begin{figure}[]  
	\includegraphics[width=0.48\textwidth]{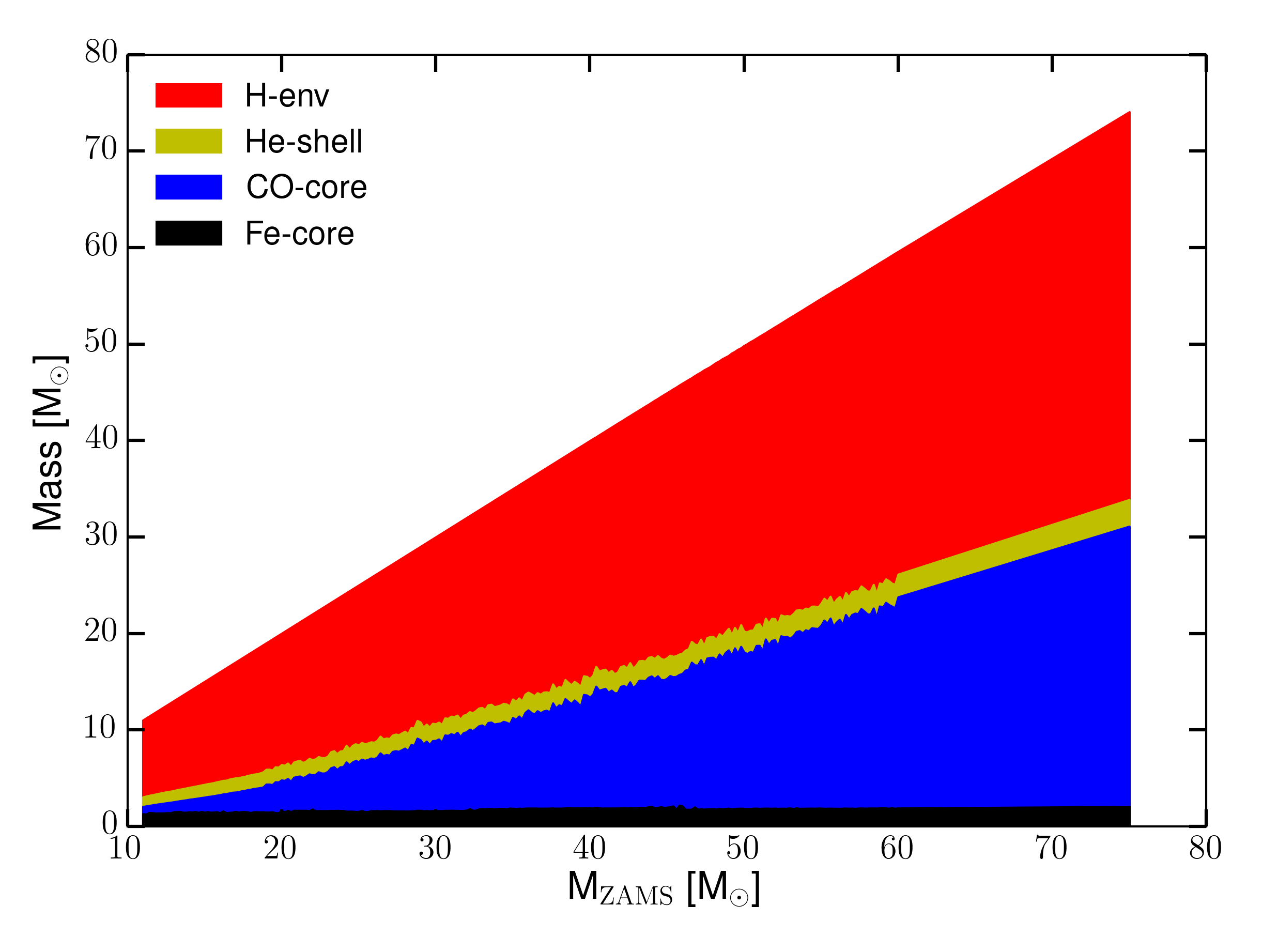}
	\includegraphics[width=0.48\textwidth]{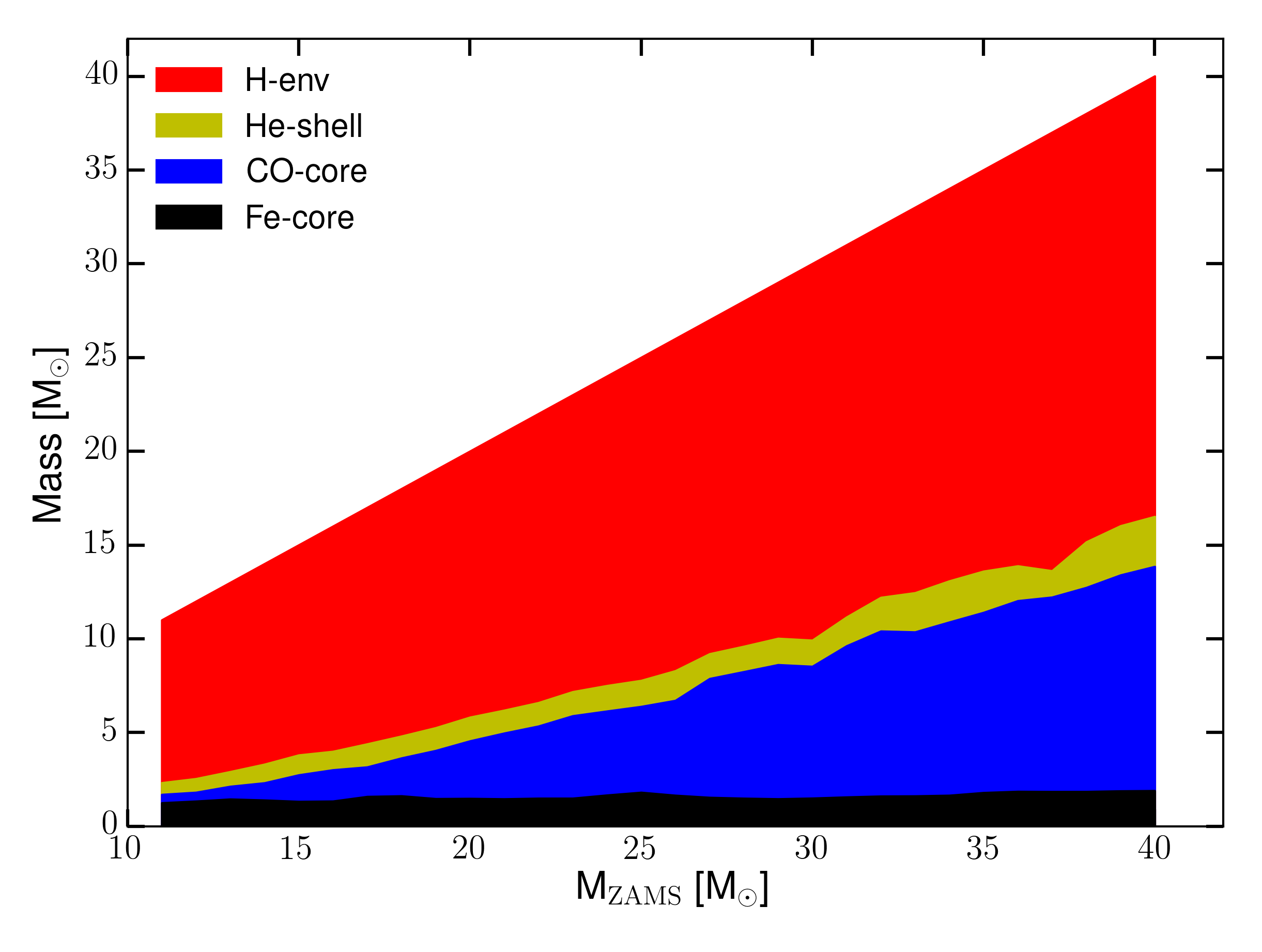}
	\caption{Fe-core mass (black), CO-core mass (blue), He-shell mass (yellow), and total mass including the H-envelope (red) as function of ZAMS mass at the onset of collapse for the pre-explosion models from the u-series (top) and z-series (bottom). 
		\label{fig:prog_u02_z02_cores}
        }
\end{figure}

As in our previous studies, we make use of the compactness, introduced in \cite{OConnor.Ott:2011} and given by
\begin{equation}
\xi_M = \frac{M/M_{\odot}}{R(M)/100\mathrm{km}},
\end{equation}
 where we use $M=2.0$~$M_{\odot}$ for the mass enclosed by the radius $R(M)$ in our investigations. In Figure ~\ref{fig:prog_compactness} we show the compactness for the low and zero metallicity pre-explosion models used in this study and the solar metallicity models used in our previous studies, which illustrates the non-monotonic relationship between compactness and ZAMS mass. This is due to a complex interplay of different burning shells and the efficiency of semiconvection and overshooting \citep{sukhbold.woosley:2014,sukhbold18}. Furthermore, at zero metallicity hydrogen burning proceeds initially solely via the pp chains which do not produce enough energy to prevent the star from contracting during H burning. Once the temperature is high enough, the triple alpha reaction switches on and H burning continues as in low metallicity stars but at higher temperature. The total mass at collapse, however, follows a monotonic relationship with the ZAMS mass for these low/zero metallicity pre-explosions models. Stellar winds, which cause mass loss from the stellar surface, are metallicity dependent and can significantly influence the stellar evolution \citep{chiosi86}. The mass loss rate scales with the initial metallicity of the stellar model, such that zero metallicity stars experience virtually no mass loss during their evolution. In Figure~\ref{fig:prog_u02_z02_cores}, we show the internal structure of the pre-explosion models used in this work. We define the Fe-core as the layers with $Y_e < 0.495$), the carbon-oxygen (CO) core mass as the enclosed mass with $X_{\mathrm{He}} \leq 0.2$, i.e. up to the beginning of the He-shell, and the He-core mass as the mass regions with $X_{\rm H}\leq 0.2$, i.e.\ up to the beginning of the H-shell. 
For ZAMS masses up to $\sim 27$~$M_{\odot}$, the CO-core masses are very similar for all four series of pre-explosion models and increase monotonically with ZAMS mass. Above 27~$M_{\odot}$, the CO-core mass continues to increase monotonically for the u- and z-series. For the s- and w-series (at solar metallicity), mass loss can be strong enough that even the outermost layers of the CO-core can be stripped (see Figures 3 and 4 in Paper~II).

\begin{table}    
\begin{center}
	\caption{Pre-explosion models used in this study.
    	\label{tab:progenitors}
	}
	\begin{tabular}{lllllc}
	\tableline \tableline 
Series & Label & Min Mass & Max Mass & $\Delta m$ & Ref. \\
 &  & ($M_{\odot}$) & ($M_{\odot}$) & ($<_{\odot}$) &  \\
	\tableline
u-series & u & $11.0$ & $40.0$ & $1.0$ & 1 \\
         &   & $45.0$ & $70.0$ & $5.0$ & 1 \\
z-series & z & $11.0$ & $40.0$ & $1.0$ & 1 \\
	\tableline
	\end{tabular}
\end{center}
\tablecomments{The u-series has sub-solar metallicity ($Z=10^{-4}Z_{\odot}$). The z-series has zero metallicity ($Z=0$). 
}

\tablerefs{(1)~\citet{Woosley.Heger:2002}}
\end{table}

\subsection{Hydrodynamic Simulations}
\label{subsec:hydro}

We use the same setup as in Paper~II and Paper~III to simulate and investigate the collapse, bounce, and post-bounce phases of all models in Table~\ref{tab:progenitors}. The hydrodynamic simulations are performed with the general-relativistic, adaptive-grid code Agile which is coupled to neutrino transport. We use the IDSA scheme \citep{Liebendoerfer.IDSA:2009} for the electron-flavor neutrino transport and the advanced spectral leakage scheme (ASL) for heavy-flavor neutrino transport \citep{perego16}. For matter in nuclear statistical equilibrium (NSE) conditions we use the finite temperature, nuclear equation of state (EOS) HS(DD2)  \citep{Hempel.SchaffnerBielich:2010}. Matter in the non-NSE regime is described by an ideal gas EOS coupled with an approximate alpha-network. We trigger explosions in spherical symmetry using the PUSH method \citep{push1}. PUSH is a physically motivated, effective method that mimics in one-dimensional simulations the additional neutrino heating caused by accretion and convection present in multi-dimensional simulations. This is achieved via the parametrized heating term $Q^+_{\push} (t,r)$ which deposits a fraction of the heavy-flavor neutrino energy behind the shock (see Equation (4) in Paper~I). This heating term depends on the spectral energy flux for a single heavy lepton neutrino flavor and includes both a spatial term (which ensures that heating only takes place where electron neutrinos also heat) and a temporal term $\mathcal{G}(t)$ which contains the two free parameters of the method, \kpush and \trise. 

We follow Paper~II and use the there presented standard calibration of the PUSH method which is in good agreement with observations. This calibration uses $t_{\mathrm{rise}}=400$~ms and a parabolic dependence of \kpush on the compactness $\xi$:  
$k_{\rm{push}}(\xi)=a\xi^2+b\xi+c$ where the calibrated parameters are $a=-23.99$, $b=13.22$\, $c=2.5$, and $\xi$ denotes the compactness at bounce. For the standard calibration we use $\xi_{2.0}$, i.e.\ the compactness evaluated at $M=2.0$~$M_{\odot}$. To investigate the dependence of the explodability and remnant properties on the PUSH calibration we also study the `second calibration' from Paper~II\footnote{The parameters for the second calibration parabola are:$a=-25.05$, $b=13.96$, $c=2.5$}, which is based on the compactness value $\xi_{1.75}$ instead of $\xi_{2.0}$. Hence, the `second calibration' depends more strongly on the iron core mass resulting in a setup more prone to BH formation.

We include matter of the pre-explosion model from the center to the He-layer, corresponding to a radial coordinate of $\sim10^{10}$cm. For the pre-explosion models above $\sim 30$~$M_{\odot}$ which collapse to BHs we include $\sim 10$~$M_{\odot}$.
The hydrodynamic simulations are run for a total time of 5~s. The outcome of a run for a given model corresponds to one of the following three options: (i) a successful explosion if the explosion energy at the end of the simulation is saturated and positive, (ii) the formation of a black hole if the central density exceeds $\sim 10^{15}$~g~cm$^{-3}$ during the 5~s simulations time, or (iii) a failed explosion if the explosion energy is negative at the final simulation time for which PUSH no longer is active. We calculate the explosion energy as in Paper~I, where we defined it as the sum of thermal, kinetic, and gravitational energy integrated from the mass cut to the stellar surface (see also Equations (14)--(16) in Paper~I) .

\subsection{Nucleosynthesis Postprocessing}
\label{subsec:nucpost}

The detailed nucleosynthesis of our CCSN simulations is calculated in a post-processing approach using the nuclear reaction network {\sc CFNET} \citep{cf06a}, as in  Paper III. We follow 2902 isotopes, including free neutrons, protons, and isotopes on both sides of the valley of $\beta$-stability, up to $^{211}$Eu. For the reaction rates we use the reaction rate library REACLIB \citep{reaclib}, which is based on experimentally known rates wherever available, and theoretical predictions for n-, p-, and $\alpha$-capture reactions from \citet{Rauscher.FKT:2000}. For the weak interactions, we include electron and positron capture rates from \cite{lmp}, $\beta^{\pm}$ decays from the nuclear database \textit{NuDat2}\footnote{http://www.nndc.bnl.gov/nudat2/} and from \cite{Moller.ea:1995}, and also neutrino/anti-neutrino captures on free nucleons.

We divide the ejecta into mass elements (called `tracers') of $10^{-3} M_{\odot}$ each. For every tracer, the thermodynamic history is known from the beginning of the hydrodynamical simulation until the final simulation time of 5.0~s. Following Papers~II and III, only tracers which reach a peak temperature $\geqslant 1.75$~GK are processed with the nuclear reaction network. 
For tracers which are heated to $\geqslant 10$~GK, we start the nucleosynthesis calculations when the temperature starts dropping below that value. We assume a NSE abundance distribution for the initial abundances at 10~GK. 
For the other tracers which never reach 10~GK, we start the nucleosynthesis calculations at the beginning of the hydrodynamic simulation and follow the full thermodynamic history of the tracer. In both cases, we use the electron fraction from the hydrodynamic simulation as the initial value and evolve the electron fraction in the nuclear reaction network consistent with the nuclear reactions occurring. The extrapolation of the trajectories of the tracers beyond the end of the hydrodynamic simulations is given by:
\begin{eqnarray}
r(t) &=& r_{\rm final}  + t {\rm v}_{\rm final} \label{eq:extrapol_rad}, \\
\rho(t) &=& \rho_{\rm final} \left( \frac{t}{t_{\rm final}} \right)^{-3}, \\
T(t) &=& T[s_{\rm final},\rho(t),Y_e(t)], 
\label{eq:extrapol_t9}
\end{eqnarray}
where $r$ is the radial position, ${\rm v}$ the radial velocity, $\rho$ the density, $T$ the temperature, $s$ the entropy per baryon, and $Y_e$ the electron fraction of the tracer.  The subscript ``final'' indicates the end time of the hydrodynamical simulation. To calculate the extrapolated temperature, we use the equation of state of \citet{Timmes.Swesty:2000}. The end point of the nucleosynthesis calculations is set when the temperature falls below 0.05~GK.

\section{Systematic Explosion Properties}
\label{sec:exposions}

\subsection{Explosion Properties} 
\label{subsec:expl-properties}

\begin{figure*}  
\begin{center}
	\begin{tabular}{cc}
	\includegraphics[width=0.48\textwidth]{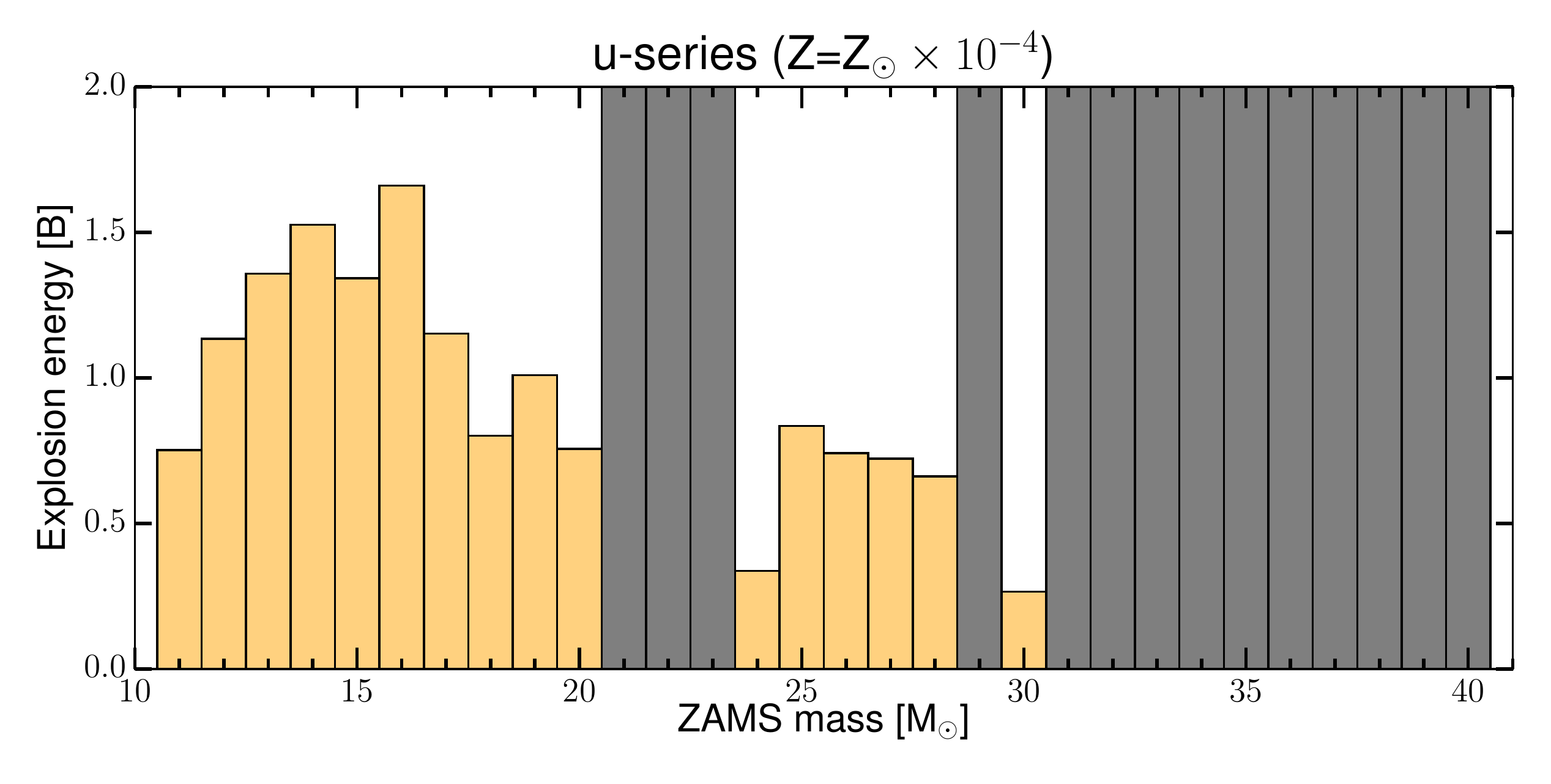} & 
    \includegraphics[width=0.48\textwidth]{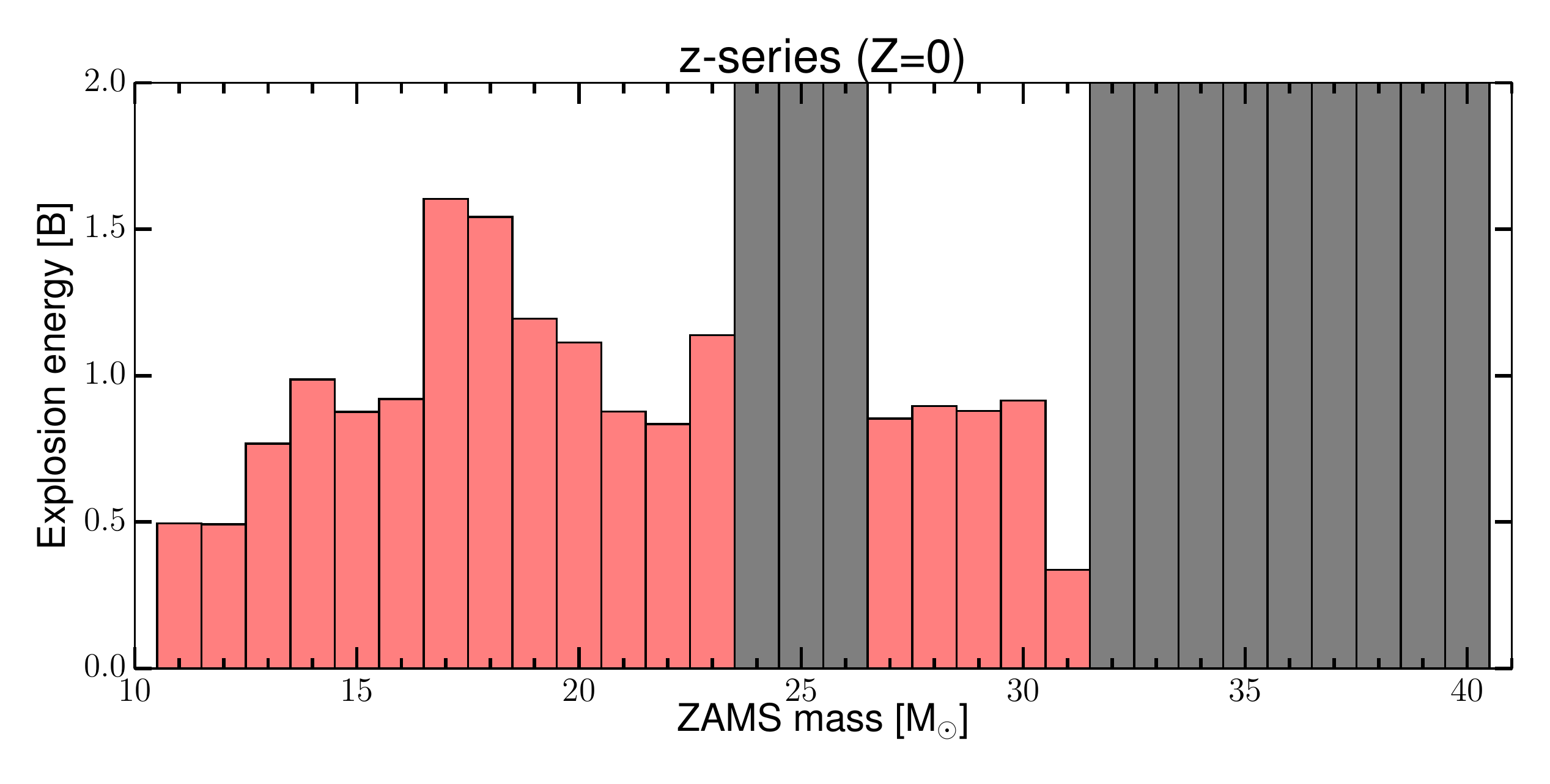} \\ 	
	\includegraphics[width=0.48\textwidth]{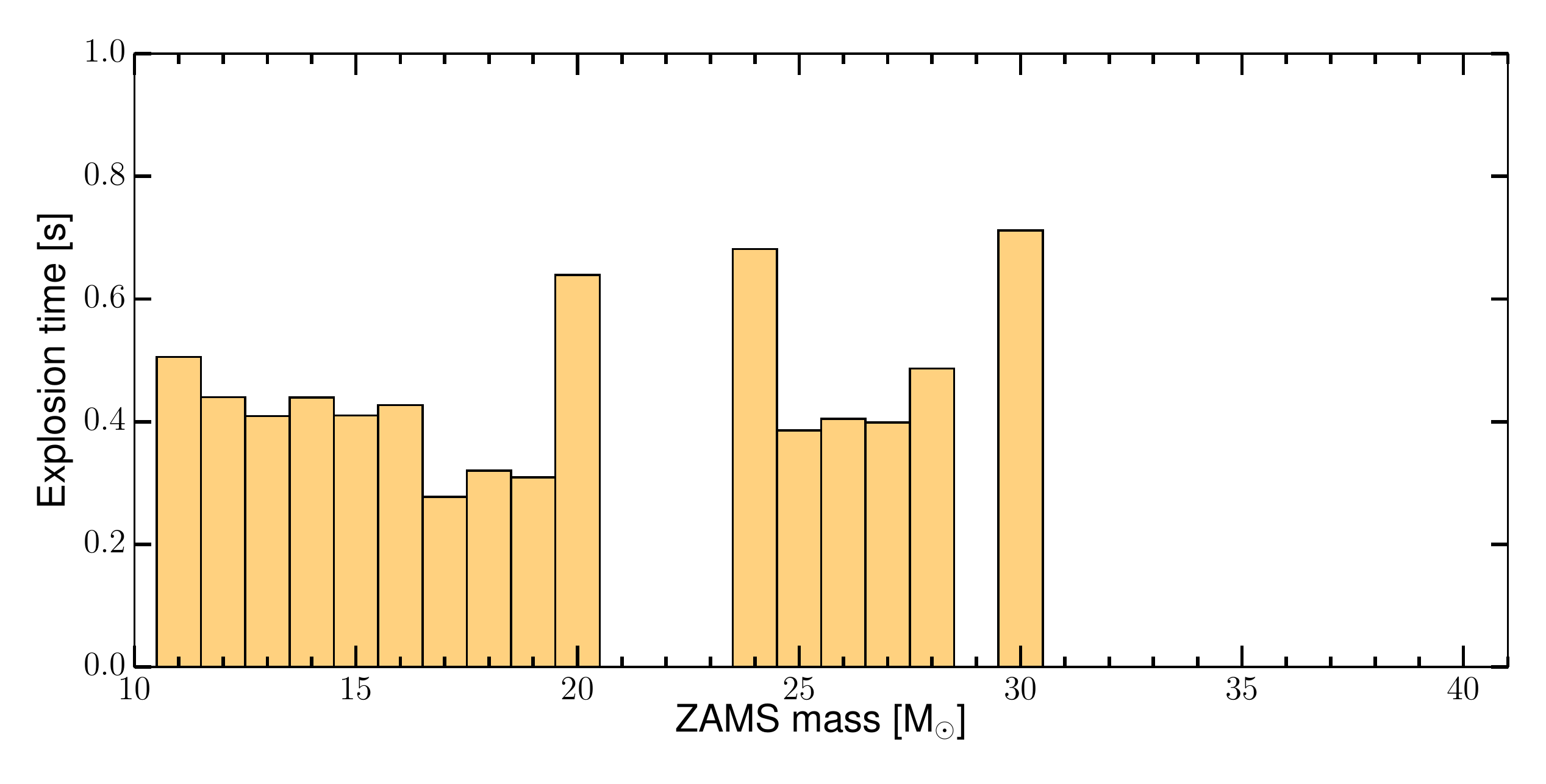} & 
    \includegraphics[width=0.48\textwidth]{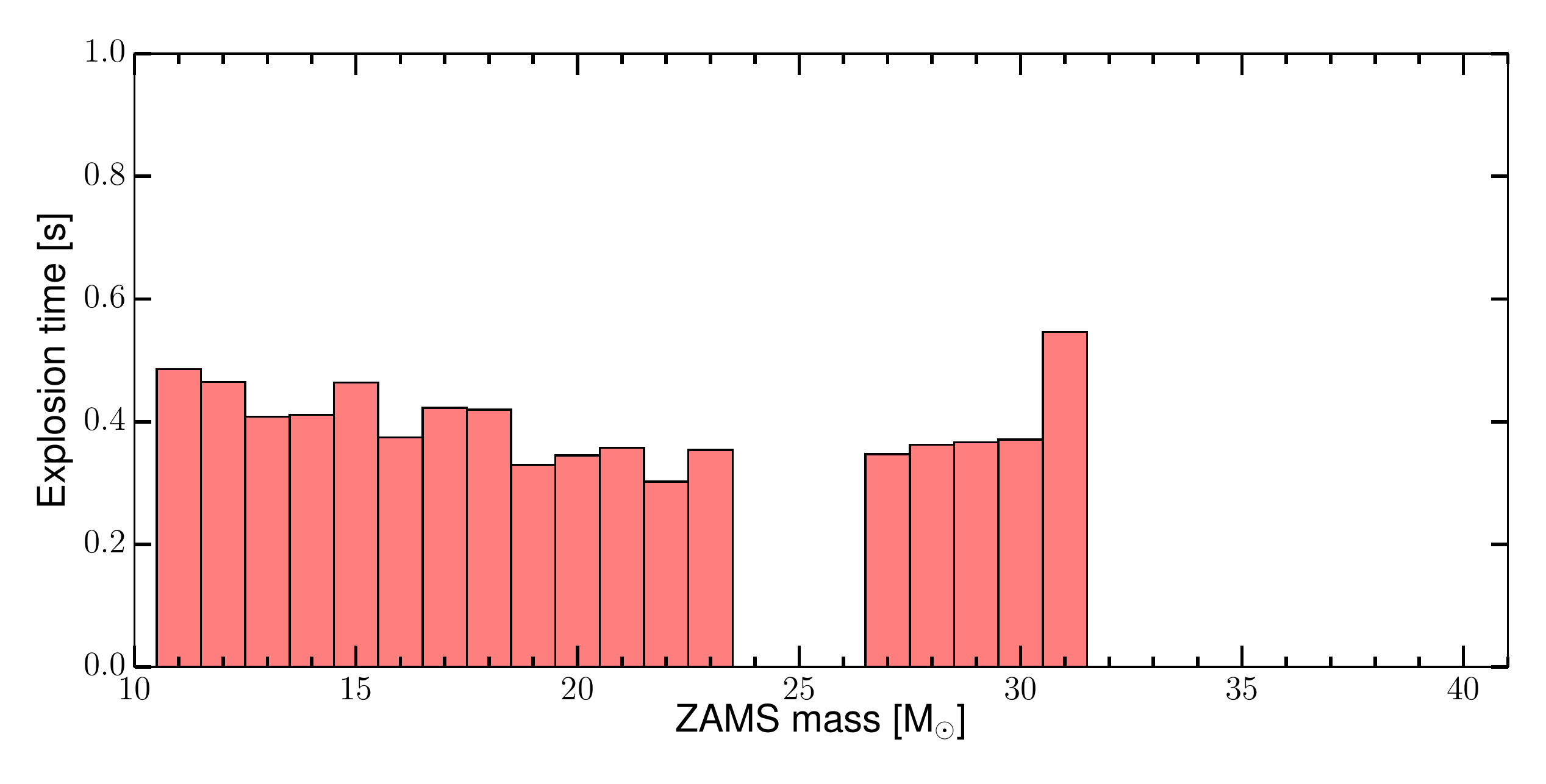} \\
	\includegraphics[width=0.48\textwidth]{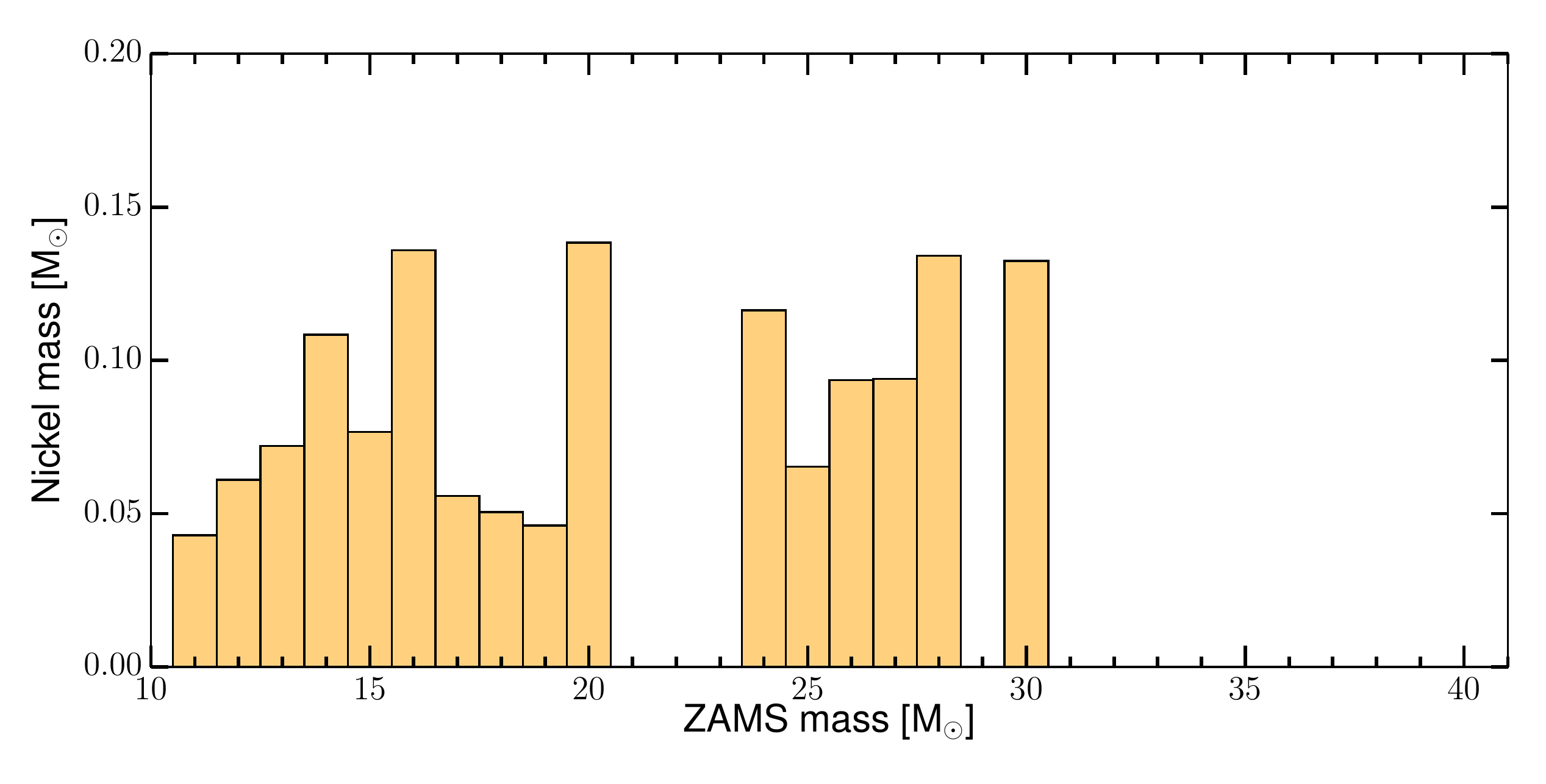} & 
    \includegraphics[width=0.48\textwidth]{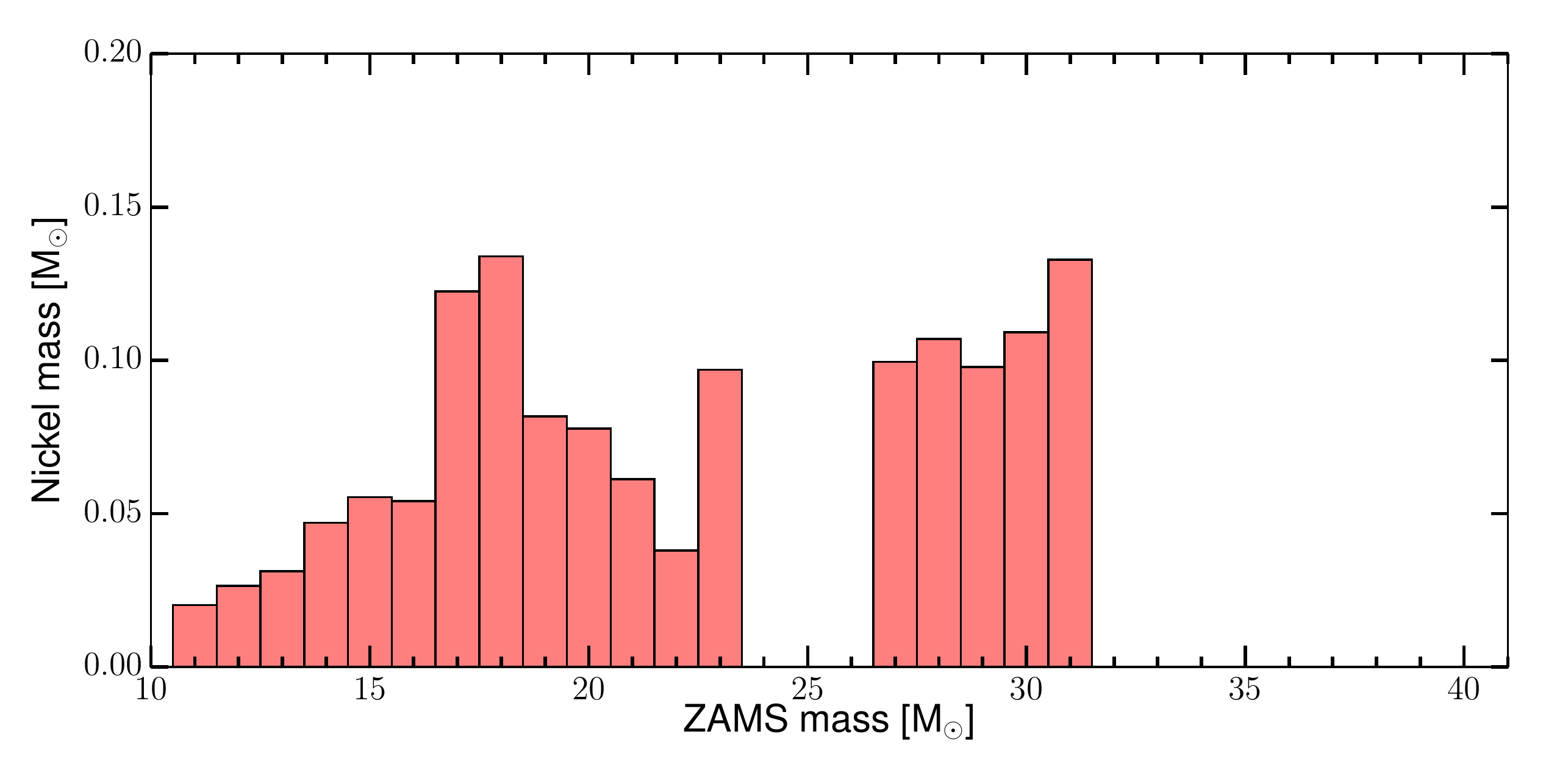} \\
	\includegraphics[width=0.48\textwidth]{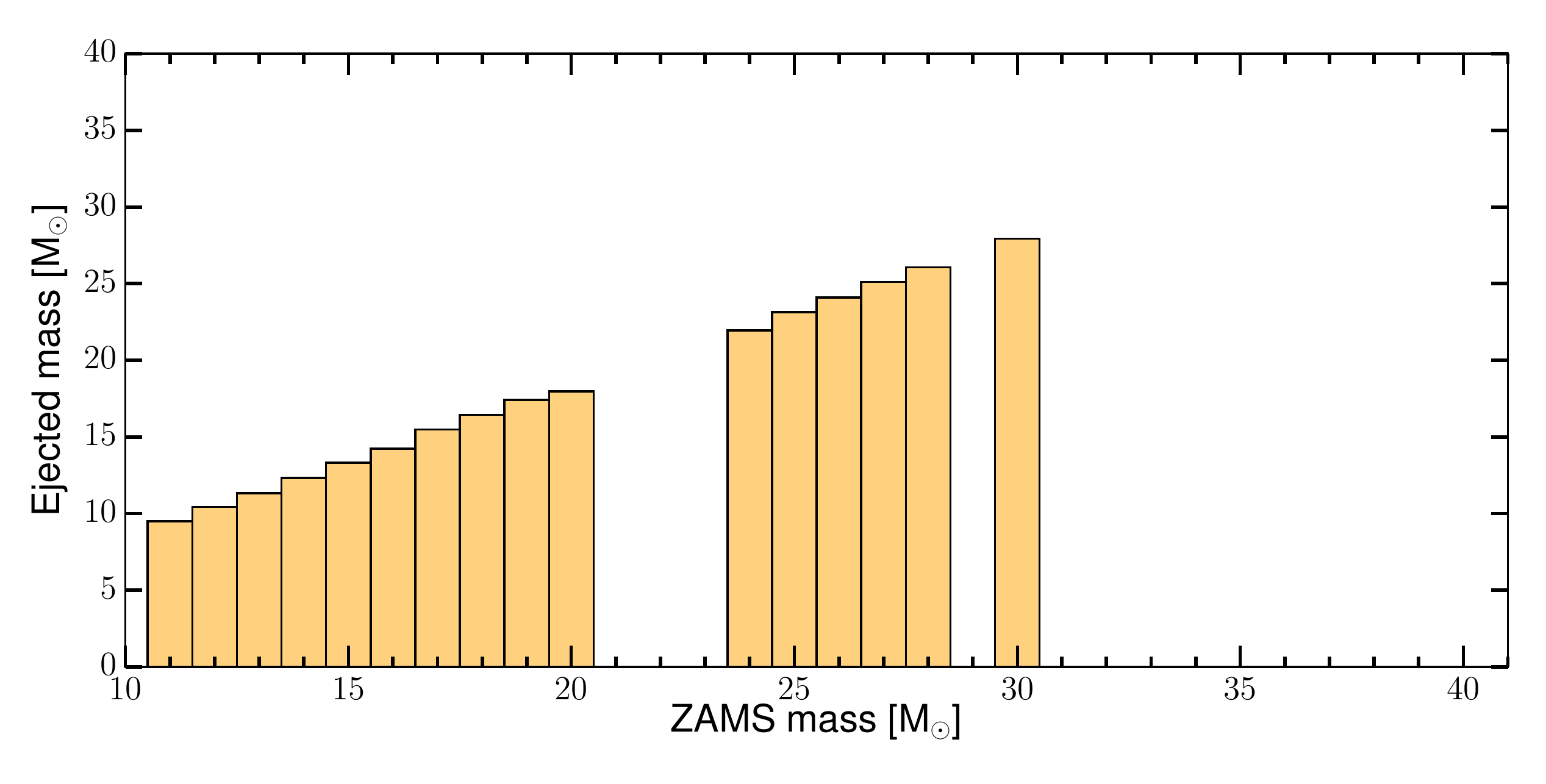} & 
    \includegraphics[width=0.48\textwidth]{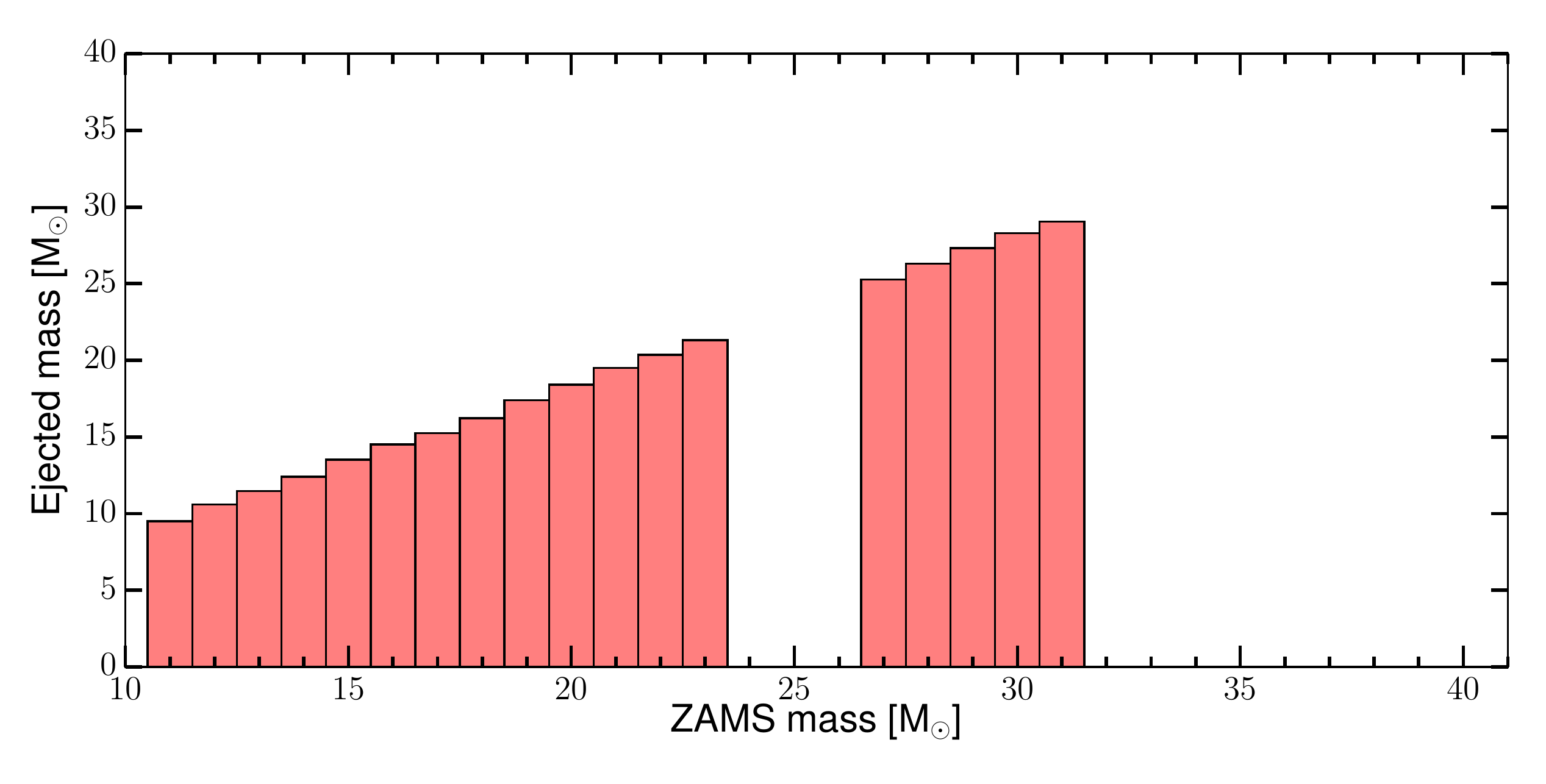} \\
    \includegraphics[width=0.48\textwidth]{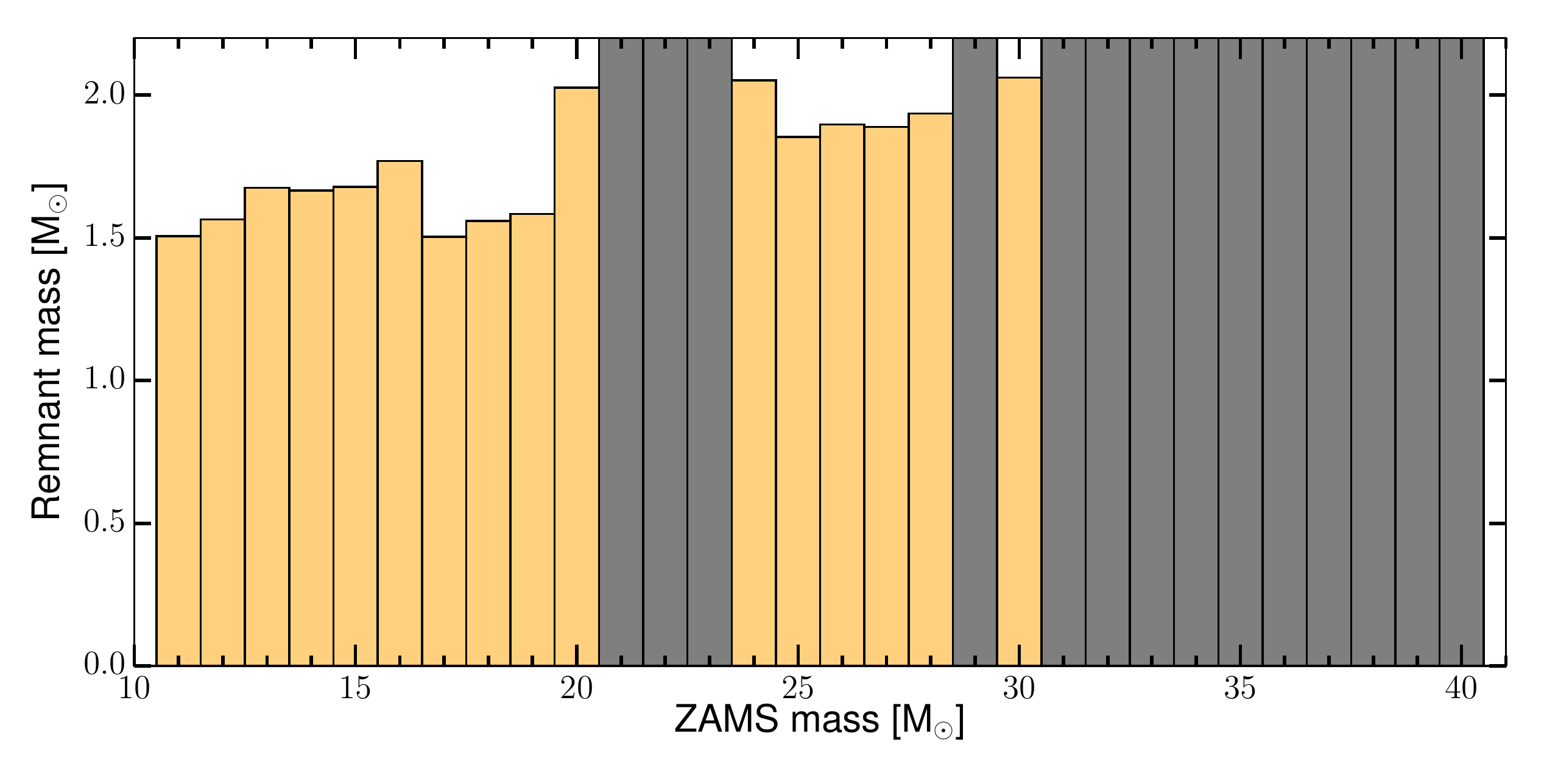} & 
    \includegraphics[width=0.48\textwidth]{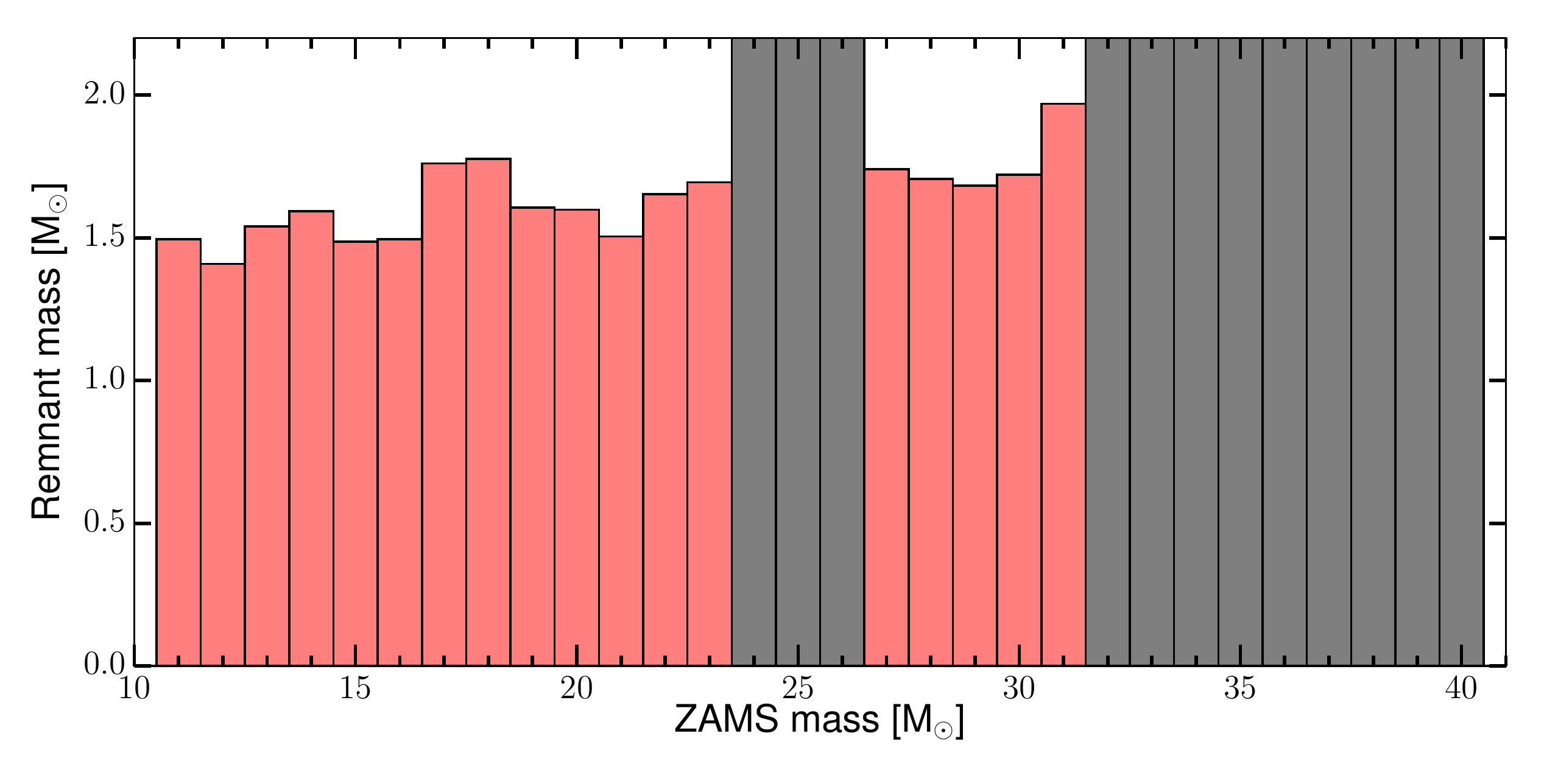} \\
	\end{tabular}
 	\caption{From top to bottom: explosion energy, explosion time, ejected Ni mass, total ejecta mass, and remnant mass (baryonic mass) for the u-series (left column) and z-series (right column) as function of the ZAMS mass using the standard calibration. Dark bars in the explosion energy and remnant mass panels indicate models that did not explode, i.e.\ ultimately formed black holes. The presented data is available as machine-readable Table. Sample Tables can be found in Appendix~\ref{appendix:hydro}.
		\label{fig:properties-zams}
    }
\end{center}
\end{figure*}

In this Section, we present and discuss the explosion properties of our simulations for the u-series and z-series. 
An overview of the explosion properties and predicted remnant properties for all pre-explosion models of this study is given in Figure \ref{fig:properties-zams}. We show the explosion energy, the explosion time, the ejected $^{56}$Ni mass, the total ejecta, and the baryonic remnant mass (top to bottom) for the u-series (left column) and the z-series (right column) for the standard calibration. 

We obtain explosion energies from $\sim$0.2 to 1.6~Bethe. The lowest explosion energies are obtained for the lowest ZAMS mass progenitors ($\leq 12$~\msun),  as well as for the heaviest ZAMS mass progenitors that still explode for both series. In addition, we find very low explosion energies for a few models with ZAMS masses that are located directly next to a region of BH formation (almost failing supernovae, see Section~\ref{subsec:highY-lowE}). The highest explosion energies are obtained for pre-explosion models around $\sim 15$~\msun ZAMS mass for the u-series and around $\sim 17$~\msun for the z-series. This corresponds to the slight shift of the peak compactness values to higher ZAMS masses for the u-series when compared to the z-series.

The obtained ejected $^{56}$Ni masses range from $\sim 0.025$ to 0.14~\msun. 
Below 20~\msun ZAMS mass, the highest and lowest values of the ejected $^{56}$Ni coincide with the highest and lowest explosion energies, respectively. Above 20~\msun, the almost failing models do not follow this trend. Instead they eject the largest amount of $^{56}$Ni at the lowest explosion energies in delayed explosions. This aspect is discussed in more detail in Section~\ref{subsec:highY-lowE}. 

The total ejecta mass increases with ZAMS mass below $\sim 20$~$M_{\odot}$ for all exploding models. Above $\sim 20$~$M_{\odot}$ ZAMS mass, the total ejecta mass continues to increase with increasing ZAMS mass for models at low/zero initial metallicity while at solar metallicity the total ejecta mass decreases with increasing ZAMS mass above $\sim 20$~$M_{\odot}$. This is because at low metallicity massive stars experience almost no mass loss during their evolution and as a result they collapse with their initial mass, while at solar metallicity mass loss is quite strong which reduces the total stellar mass significantly.

For the exploding models, we obtain gravitational NS masses from $\sim$1.3 to 1.8~\msun. Overall, the u-series is more prone to BH formation than the z-series, consistent with the results of \cite{OConnor.Ott:2011}. Also in agreement with their results, we find that low metallicity stars above $\sim 30$~$M_{\odot}$ robustly form BHs. These low metallicity sets were studied by \cite{pejcha2015} as well. While they also find successful explosions interwoven with BH formation, their detailed predictions differ from ours. In particular, one of their parametrizations predicts a larger fraction of BHs compared to our standard calibration, including BHs below 20~\msun ZAMS mass, for both progenitor sets. Their second parametrization is more optimistic and predicts explosions below 20~\msun, however, it also finds many exploding models above 30~$M_{\odot}$, in contrast with our results. In the u-series, BH formation starts to occur at lower ZAMS masses than in the z-series. Exploding models beyond 20~\msun ZAMS mass also have lower explosion energies in the u-series than their counterparts in the z-series. As we have speculated in Paper~II --- based on the higher compactness and larger stellar masses at collapse --- the low metallicity models indeed do not lead to successful explosion above $\sim 30$~$M_{\odot}$, and therefore constitute a potential origin of the BHs seen by LIGO/VIRGO (see also Section~\ref{sec:remnants}).

\begin{figure}[]  
	\includegraphics[width=0.48\textwidth]{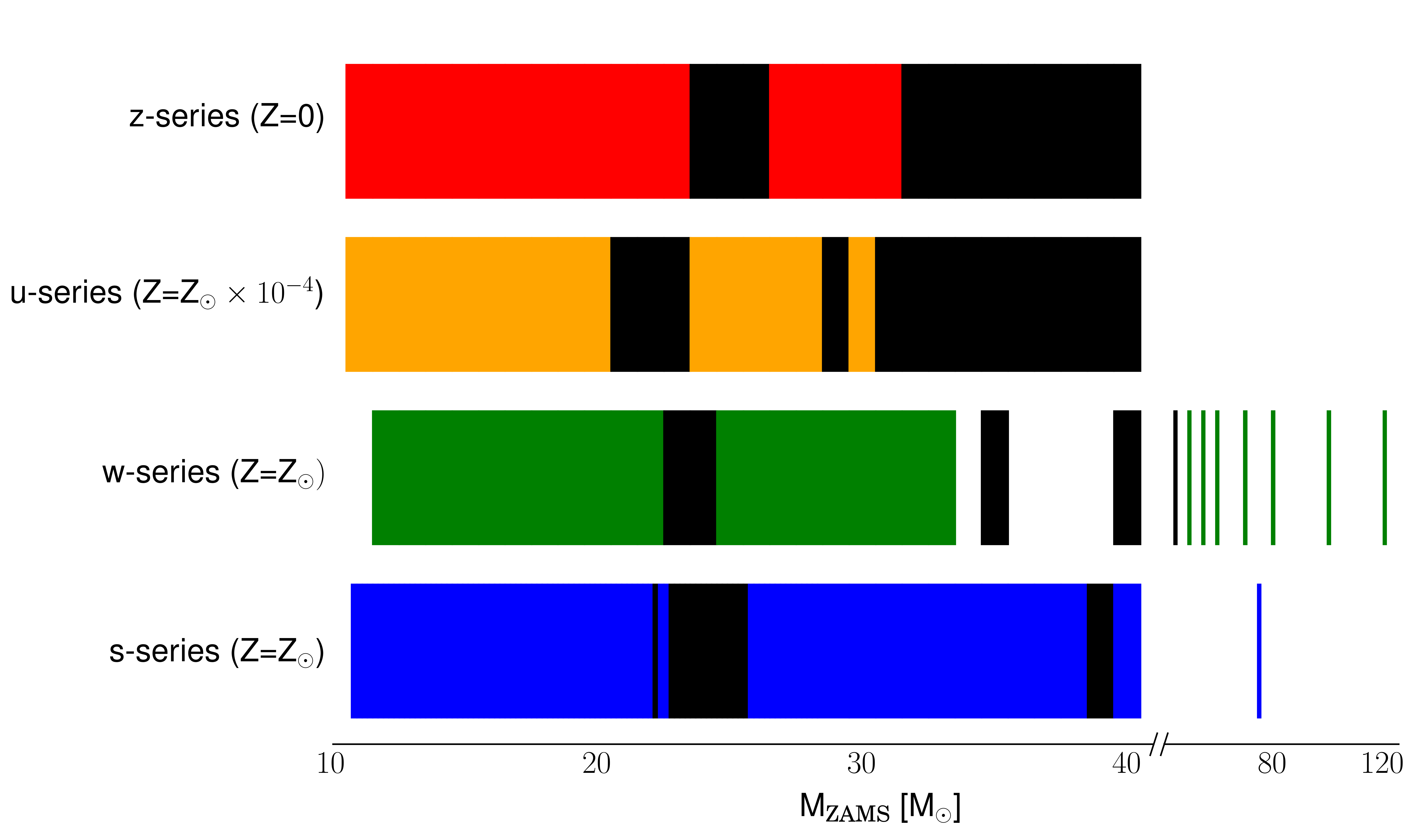}
	\includegraphics[width=0.48\textwidth]{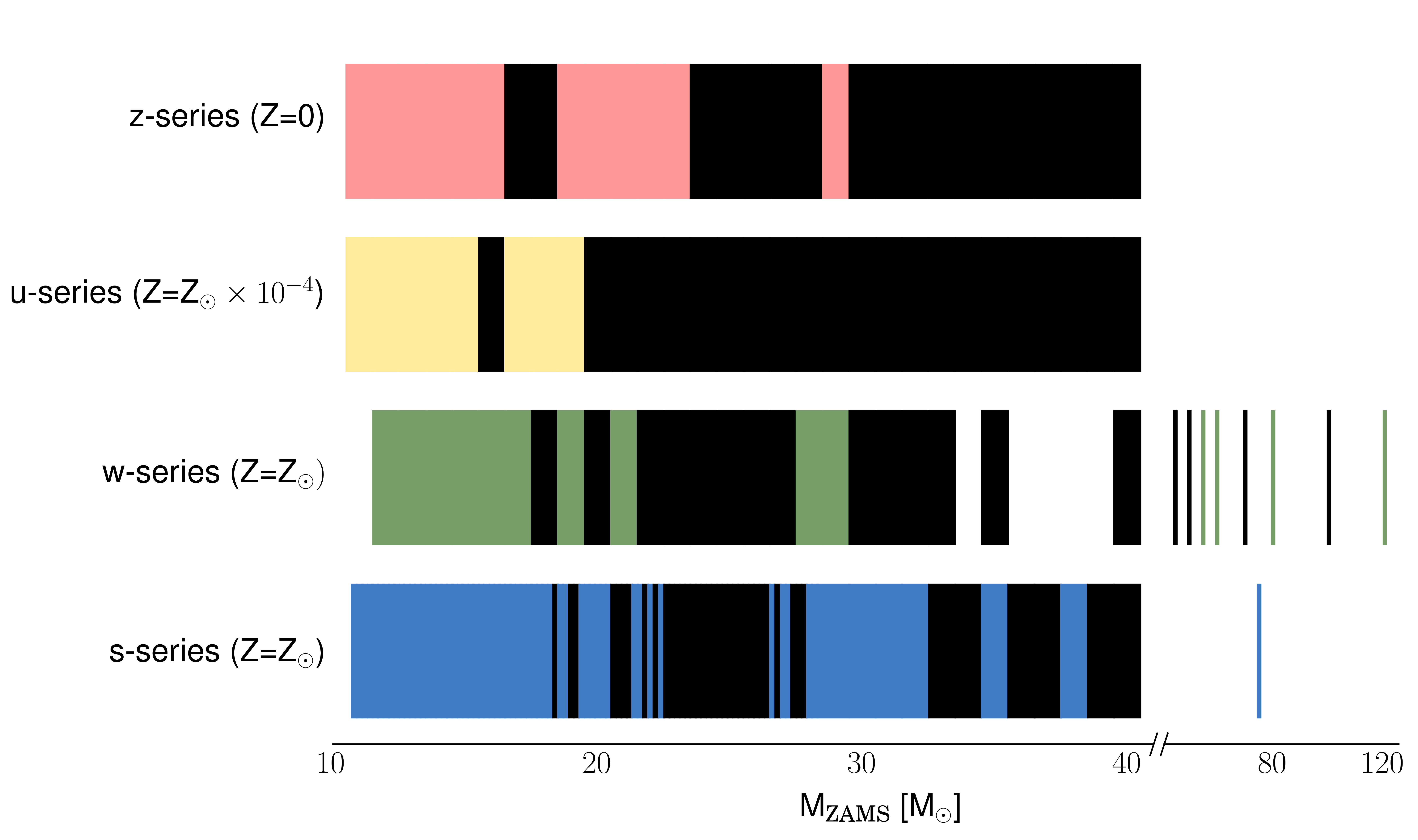}
	\caption{Explosion outcomes for the four sets of pre-explosion models: z-series (red), u-series (yellow), s-series (blue), and w-series (green). The colored areas indicate exploding models which leave behind a neutron star as a remnant and black areas indicate failed explosions which lead to BH formation. 
	Dark colors are used for the standard calibration (top panel), lighter colors are used for the second calibration (bottom panel).
		\label{fig:rectangles-engine}
        }
\end{figure}

In Figure~\ref{fig:rectangles-engine}, we give a summary of the final outcomes of the `standard calibration' and of the `second calibration' for both samples of pre-explosion models presented in this paper as well as the two samples at solar metallicity discussed in Paper~II. In agreement with our previous work, we see that the second calibration results in a lower explodability and leads to a larger fraction of BHs, which is of particular interest for the resulting birth mass distributions of NSs and BHs discussed in Section~\ref{sec:remnants}. Additionally, the low and zero metallicity series have a continuous region of BH formation above certain values of ZAMS mass for both calibrations. The one outlier to this in the z-series is a model with very low explosion energy and almost fails to explode (see also Section~\ref{subsec:highY-lowE}.

\subsection{Trends with Compactness}
\label{subsec:scan_expls}

\begin{figure}  
\begin{center}
   \includegraphics[width=0.48\textwidth]{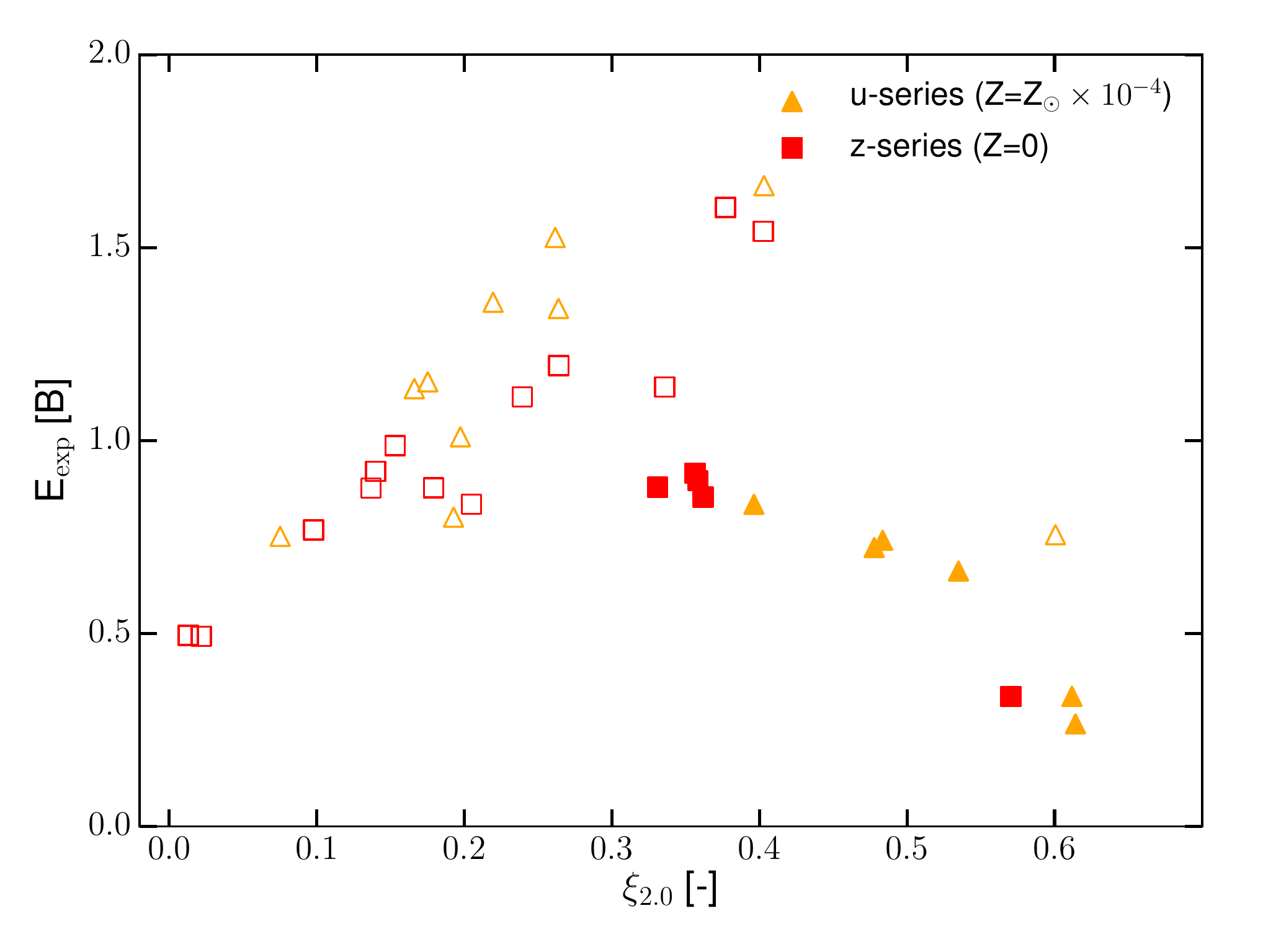} 
   \includegraphics[width=0.48\textwidth]{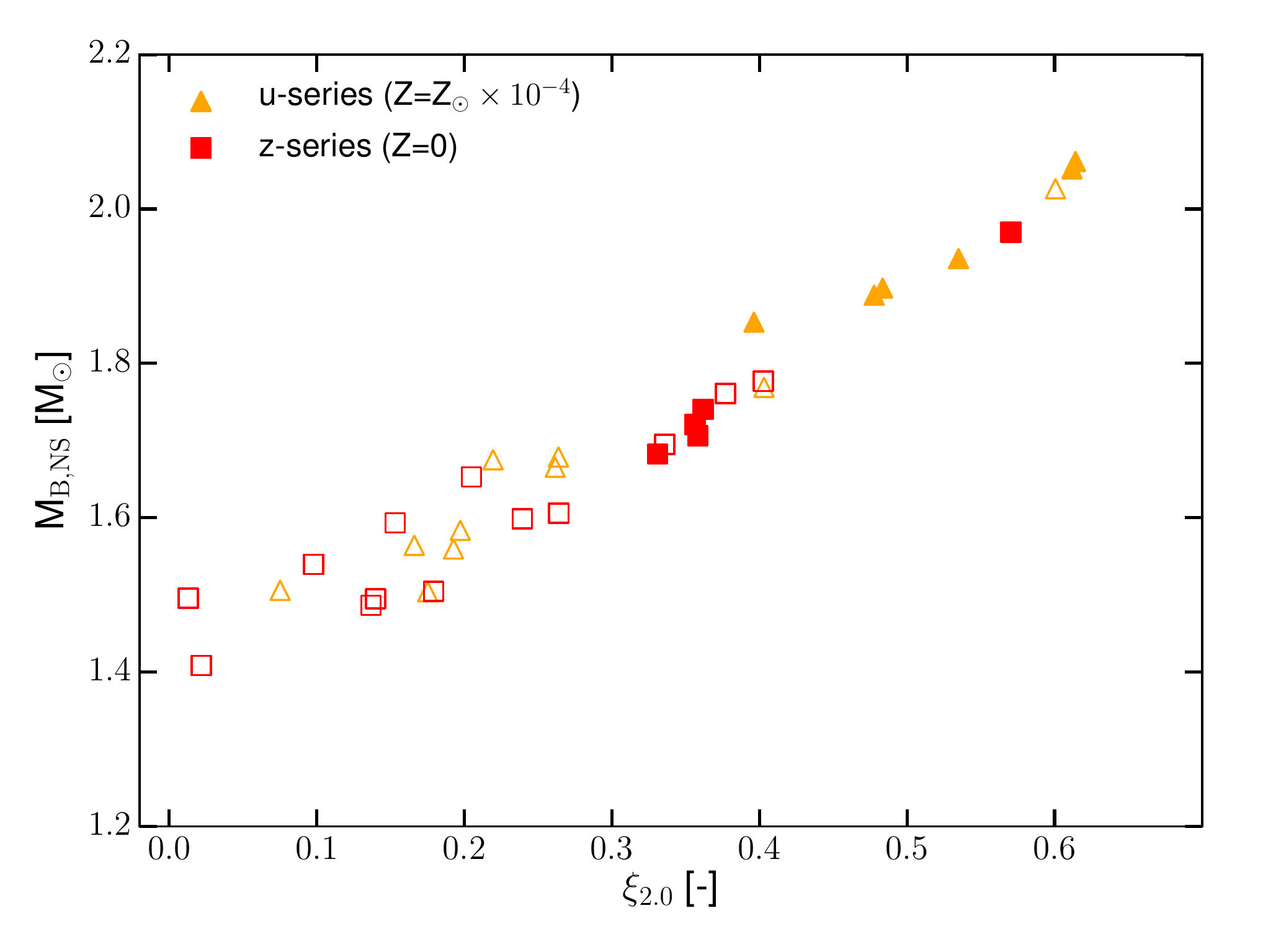} 
   \includegraphics[width=0.48\textwidth]{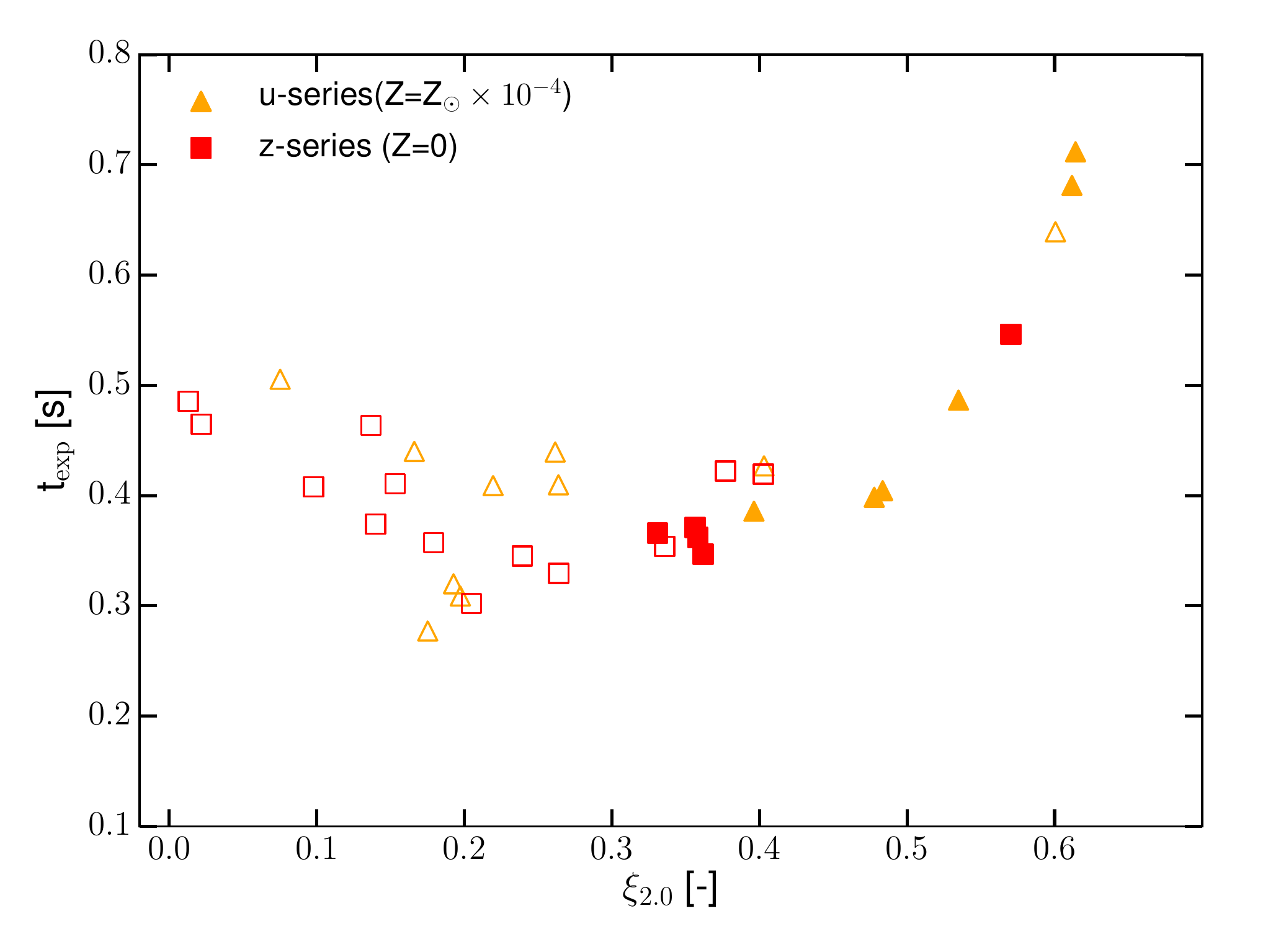} 
 	\caption{From top to bottom: explosion energy, remnant mass (baryonic mass), and explosion time for the u-series (orange triangles) and z-series (red squares) as function of the compactness $\xi_{2.0}$. Only exploding models are shown. Open symbols correspond to models to the left of the compactness peak; filled symbols indicate models to the right of the compactness peak.
		\label{fig:scan_properties-compactness}
    }
\end{center}
\end{figure}

In this Section, we examine the outcomes of our simulations in terms of the compactness $\xi_{2.0}$ at bounce and discuss emerging trends. The explosion energy, the baryonic remnant mass, and the explosion time for both series of pre-explosion models are shown in Figure \ref{fig:scan_properties-compactness}. For all three quantities we find similar trends as for the samples at solar metallicity presented already in Paper~II.

For the explosion energy as a function of the compactness we find two trends within our models. Below compactness values of $\xi_{2.0} \sim 0.3$ the explosion energy increases roughly linearly with compactness (\kpush increases with compactness up to $\xi_{2.0} \sim 0.3$ , see Figure~8 in Paper~II). Above compactness $\xi_{2.0} \gtrsim 0.3$, we observe a bifurcation into two branches of explosion energies. One branch (consisting mostly of filled symbols), the explosion energies decrease with compactness, slightly following the decrease in \kpush for $\xi_{2.0} \gtrsim 0.3$ of our calibration. In the other branch (consisting of open symbols) continues the increasing relation between compactness and explosion energy from $\xi_{2.0}<0.3$. The two branches of explosion energy correspond to the pre-explosion models with similar compactness but distinct ZAMS masses. In Figure~\ref{fig:prog_compactness} we see that there are two regions of ZAMS masses, separated by a peak in compactness around $\sim 25$~$M_{\odot}$, which have similar values of compactness. These models, however, differ in their total mass at collapse and also in the mass of the CO-core (see Figure~\ref{fig:prog_u02_z02_cores}). These differences lift the degeneracy in compactness and explain the two branches of explosion energies. In Figure~\ref{fig:scan_properties-compactness} we use open markers for models to the left of the peak in compactness and filled markers for models to the right of the peak to emphasize this point. Furthermore, the models in the lower explosion energy branch have higher pre-explosion model mass and an overall flatter density profile with in average higher densities above an enclosed mass of $\sim$1.25~\msun. Therefore, in these models more mass needs to be unbound in the course of the explosion and the shock front needs to pass through denser in falling matter.

We find a strong correlation between the neutron star mass and the compactness. Models with a higher compactness experience higher mass accretion rates and therefore accrete more matter onto the PNS before they ultimately explode. In our framework, the models with the highest compactness values do not explode and instead collapse to BHs. This sets an upper mass limit for the NSs that are formed in CCSNe.
The explosion times somewhat reflect our choice of calibration and as such are not a true prediction from our models. The models with the lowest and the highest compactness values have lower values of \kpush and hence take longer to explode than the models with intermediate compactness values where \kpush reaches the largest values. There the explosion times become comparable to the set value of \trise.

\subsection{Almost Failing Supernovae} \label{subsec:highY-lowE}

\begin{figure}[]  
	\includegraphics[width=0.47\textwidth]{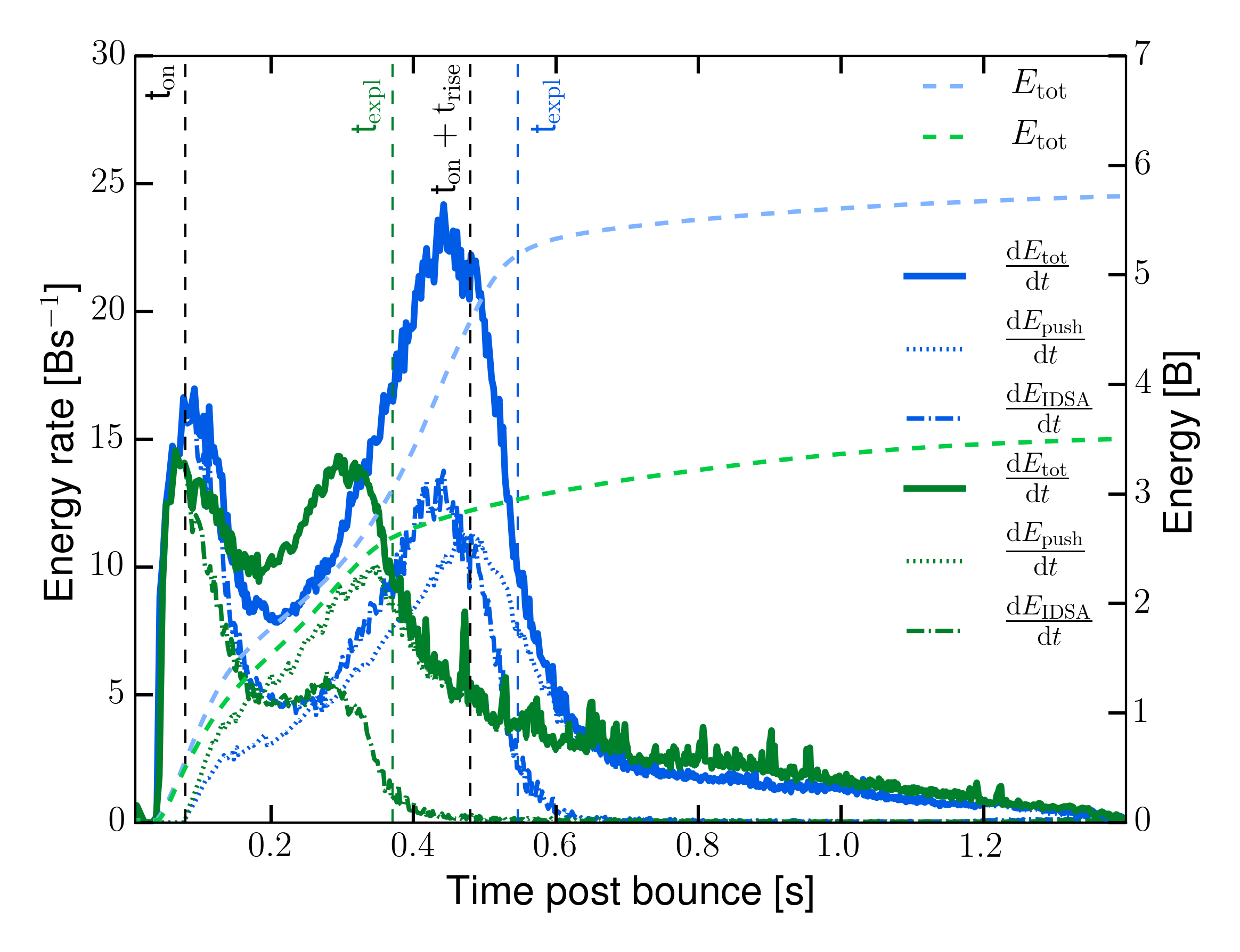}
	\caption{Heating rates from electron neutrinos (dash dotted line), from PUSH (dotted line), and total heating rates (solid line) for models z30 (green) and z31 (blue) as function of the post-bounce time. The light colored dashed lines denote the total energy deposited in the gain region. Vertical lines indicate different times during the simulations. The explosion time is set as the time when the shock goes beyond a radius of 500~km. 
		\label{fig:energyrate}
        }
\end{figure}

As we have seen in Section~\ref{subsec:expl-properties} there are some models with low explosion energies around 0.3~Bethe and relatively high masses of ejected $^{56}$Ni around 0.1~\msun. From Figure~\ref{fig:properties-zams} we immediately find that these are the u24, u30 and z31 models, which have very low explosion energies and large values of ejected $^{56}$Ni mass, coupled with very delayed explosion times and among the highest remnant masses in the investigated samples. 
All three models have high compactness values ($\xi_{2.0} \sim 0.6$) and are right next to regions of BH formation.  In our calibration of PUSH, a high compactness value implies a small \kpush value and hence little extra heating is provided which results in delayed explosions (or no explosion). A reduction of \kpush by a small amount leads to the failing of the explosion of these models and an increase of \kpush would result in earlier explosions with increased explosion energies. For delayed explosions, the mass accretion onto the freshly formed PNS extends to later times. This leads to higher luminosities of the electron neutrinos and antineutrinos during this extended accretion. As a consequence the neutrino heating from electron neutrinos and antineutrinos is enhanced at later times. This is illustrated with Figure~\ref{fig:energyrate} which shows the contributions from electron neutrinos/antineutrinos ($dE_{\mathrm{IDSA}}/dt$) and from PUSH ($dE_{\mathrm{push}}/dt$) as well as the total heating rate $dE_{\mathrm{tot}}/dt$ for z31 (blue) and z30 (green). In the almost-failing models such as u24, u30, and z31 this extended heating leads to very marginal and delayed explosions ($t_{\mathrm{expl}} \gtrsim 0.5$~s). A delayed explosion time also means that some of the energy is deposited in layers that will be accreted later onto the PNS, which further reduces the final explosion energy. 
As long as the mass accretion and hence the high neutrino luminosities persist, the material above the PNS is being heated to temperatures $\gtrsim 6$~GK, sufficient for the synthesis of $^{56}$Ni. In models with delayed explosions, this is $\sim 0.1$~$M_{\odot}$ of material (a similar amount of $^{56}$Ni is synthesized in models with more canonical explosion energies around 1~B).  In spherically symmetric models the accretion onto the PNS effectively shuts off when the explosion sets in irrevocably. In models with short explosion times, the accretion onto the PNS turns off early, allowing less material to be neutrino heated to temperatures high enough for the synthesis of $^{56}$Ni. In models that ultimately fail to explode, similar (or larger) amounts of mass are neutrino-heated above 6~GK, however all of this material eventually accretes on the central compact object.

Based on the low explosion energies, the delayed explosion times, and the fact that these models are directly next to BH forming regions we argue that they could easily experience sufficient fallback to transition into the failed SN branch of neutrino-driven SNe.  We will discuss the effect the assumed collapse of the critical models has on the resulting distribution of the NS birth mass and black hole birth mass in Section~\ref{sec:remnants}.
Due to the large amount of synthesized $^{56}$Ni the weakly exploding CCSNe could result in comparably bright events.

\section{Nucleosynthesis yields} \label{sec:scan_nucsyn}

In the previous Section, we applied the `standard calibration' from Paper~II to the u- and z-series allowing us to predict explosion outcomes. In this Section, we present and discuss the nucleosynthesis yields computed for all exploding models. 
These yields are  available electronically as machine-readable table (see Appendix~\ref{appendix:yields}, Tables~\ref{tab:finab-u} and \ref{tab:finab-z}).

CCSNe make an important contribution to the iron-group elements. The iron group elements are synthesized in the inner layers of the ejecta which undergo either complete or incomplete silicon burning. The detailed nucleosynthesis pathways for all iron group isotopes is discussed already in Paper~III. The yields of these elements are quite sensitive to the conditions in these layers. The interaction of neutrinos with matter in these layers sets the local electron fraction $Y_e$ which determines whether the nucleosynthesis takes place under proton-rich or neutrino-rich conditions. In our models, we consistently find a proton-rich ($Y_e > 0.5$) environment (a few $10^{-3}$~$M_{\odot}$ close to the mass cut, above the late neutron-rich wind. In these layers, some isotopes beyond iron can be formed through the $\nu p$-process \citep{cf06b}. The late time neutron-rich neutrino-driven wind ejecta is quite uncertain in our models. We use a mass resolution of $10^{-3}$~$M_{\odot}$ for the nucleosynthesis post-processing. This is quite coarse for the neutrino-driven wind. In our models, the wind includes a couple of hundreths of a solar mass of material. The electron fraction can be quite low, and, in a few models, we see some production of elements up to mass number $A \sim 140$. However, the conditions are not sufficient for a full r-process \citep{farouqi2010,kratz14}, as discussed already in Paper~I. A more detailed analysis of the conditions and nucleosynthesis in the late neutrino-driven wind is beyond the scope of this paper.

CCSNe also make significant contributions to the alpha elements ($^{16}$O, $^{20}$Ne, $^{24}$Mg, $^{28}$Si, $^{32}$S, $^{36}$Ar, $^{40}$Ca, $^{48}$Cr, $^{52}$Fe). These elements have contributions from hydrostatic and explosive burning. Intermediate mass elements are primarily synthesized in explosive oxygen burning, which only reduces the overall oxygen yield by a small fraction. The majority of the oxygen-neon rich layer does not reach high enough temperatures in the explosion to significantly alter the composition from its pre-explosion state. 

The light elements (such as H, He, and C) depend on the details of stellar evolution (e.g.\ mixing) and the mass loss during the pre-explosion evolution. The abundances of these elements are essentially unaltered by the explosion.

\begin{figure}[]  
    \includegraphics[width=0.48\textwidth]{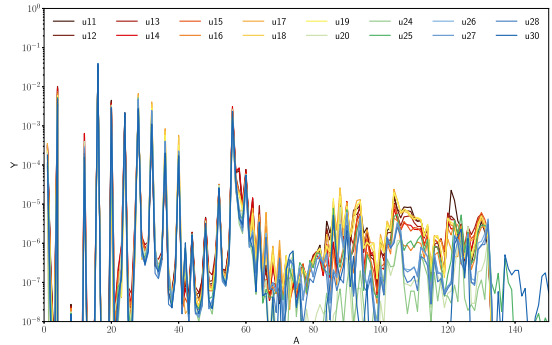}
\caption{
    Final abundances after decay for all exploding models of the u-series ($Z=10^{-4}Z_{\odot}$) as a function of the the mass number $A$.
	\label{fig:YvsA_u}
    }
\end{figure}

\begin{figure}[]  
    \includegraphics[width=0.48\textwidth]{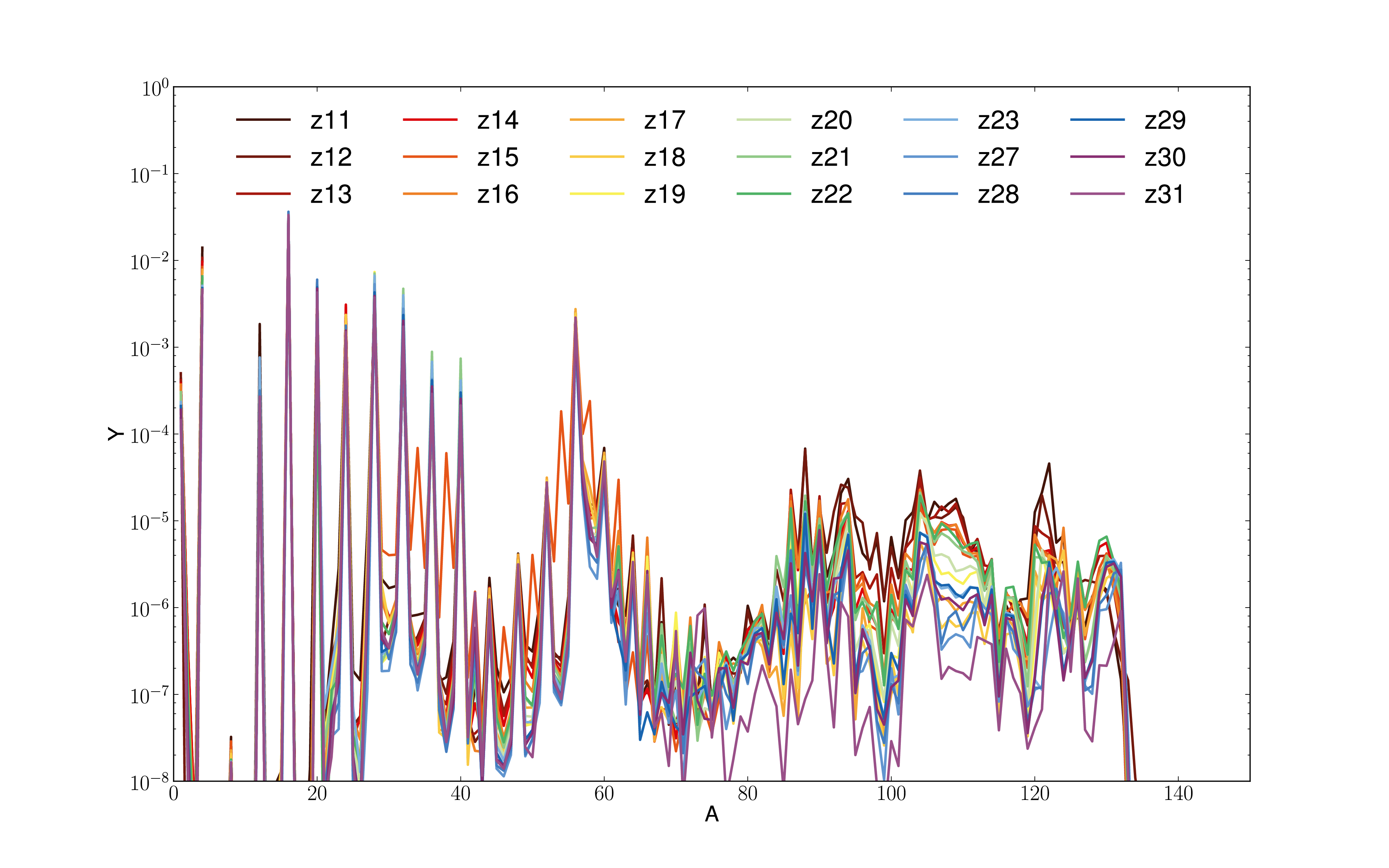}
\caption{
    Same as Figure~\ref{fig:YvsA_u} but for the z-series ($Z=0$).
	\label{fig:YvsA_z}
    }
\end{figure}

Figures~\ref{fig:YvsA_u} (u-series) and \ref{fig:YvsA_z} (z-series) show the final abundances after decay to stability for all exploding models as function of the mass number $A$. In both series, we find a pronounced iron peak around $A=56$ and high abundances of alpha elements. There are some variations in the synthesis of elements beyond the iron peak between different models, depending on the exact $Y_e$ value in the innermost ejected layers.

\subsection{Trends with Progenitor and Explosion Properties} \label{subsec:yield_trends}

\begin{figure*}[]  
    \includegraphics[width=0.48\textwidth]{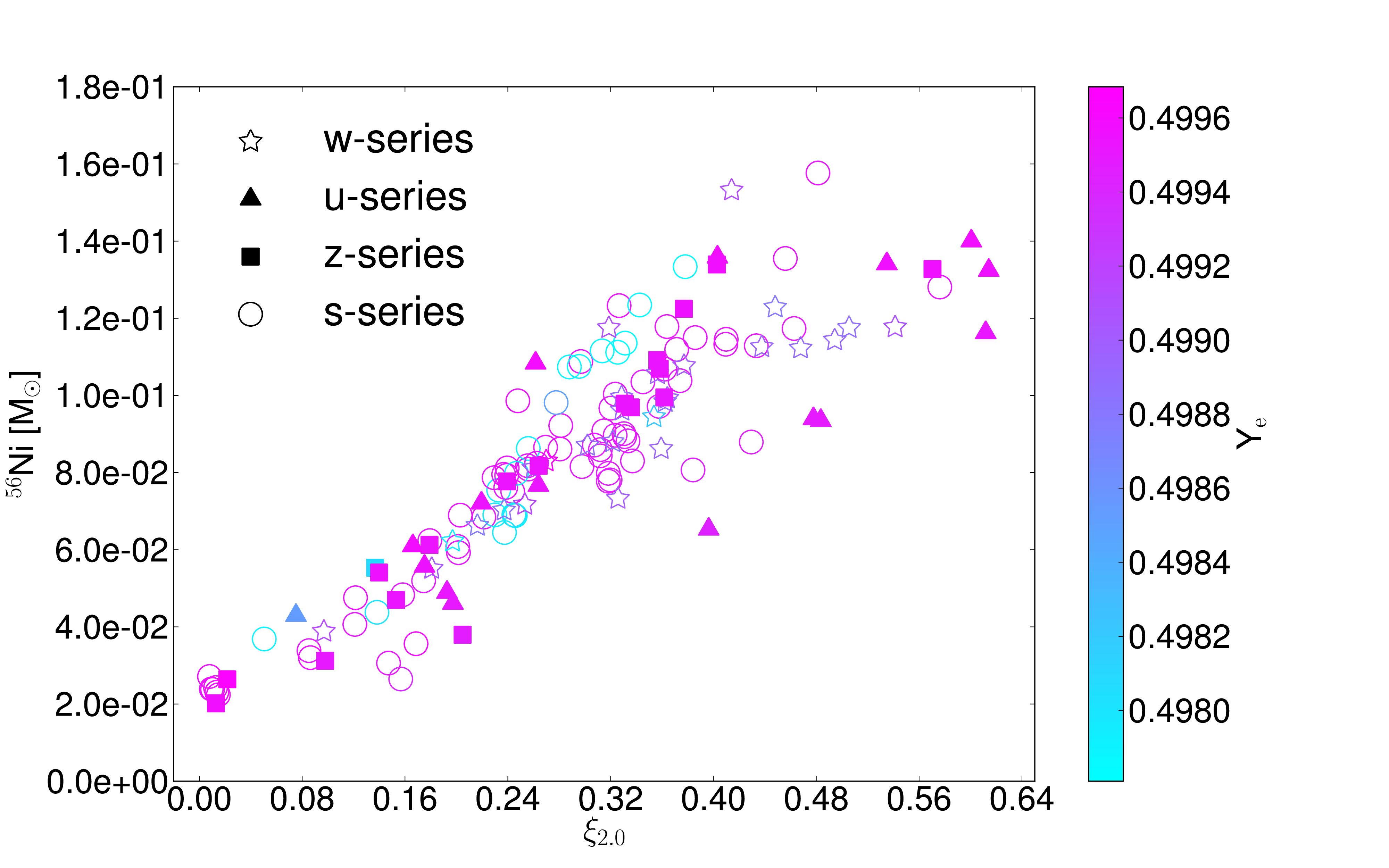}
    \includegraphics[width=0.48\textwidth]{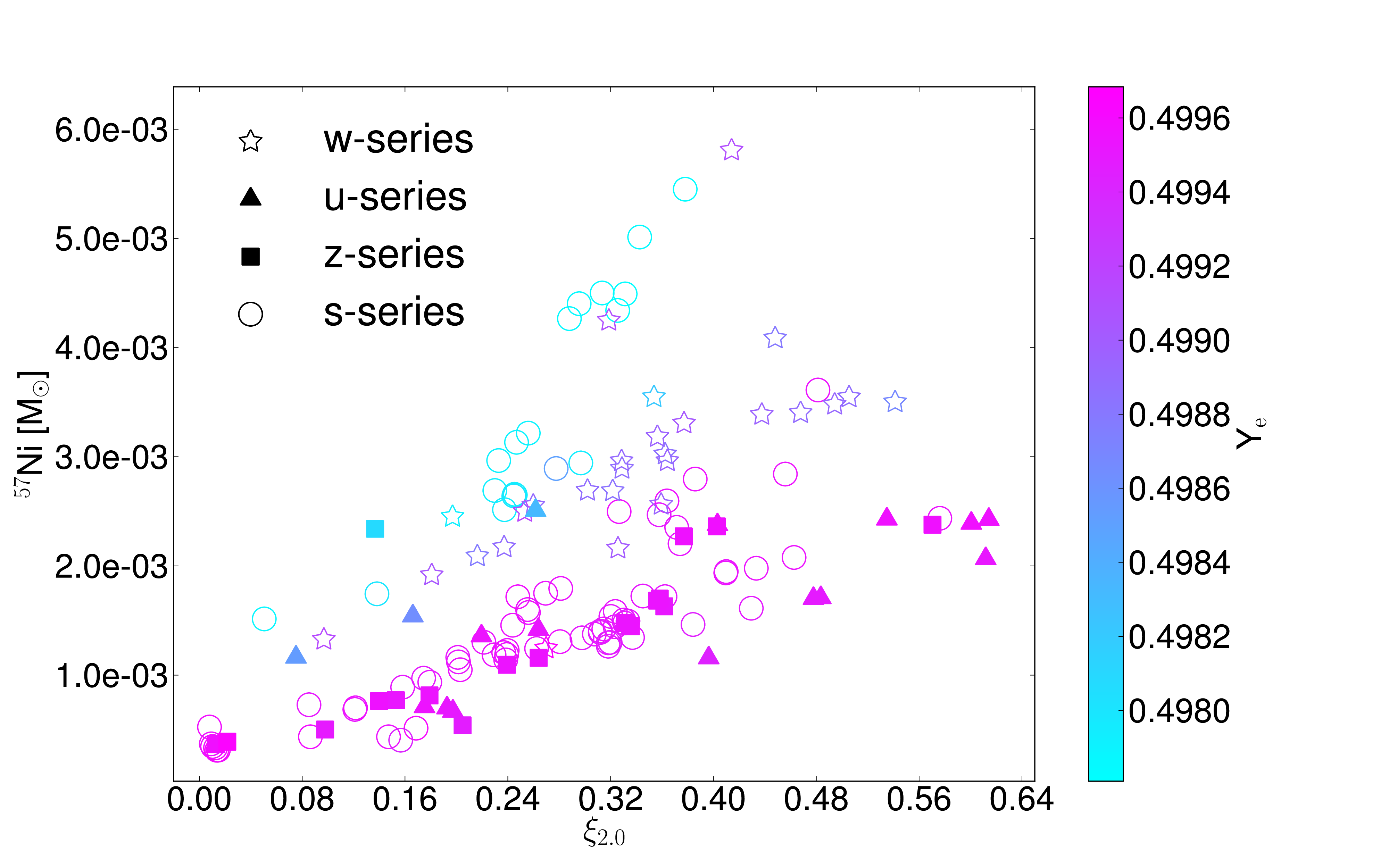} \\
    \includegraphics[width=0.48\textwidth]{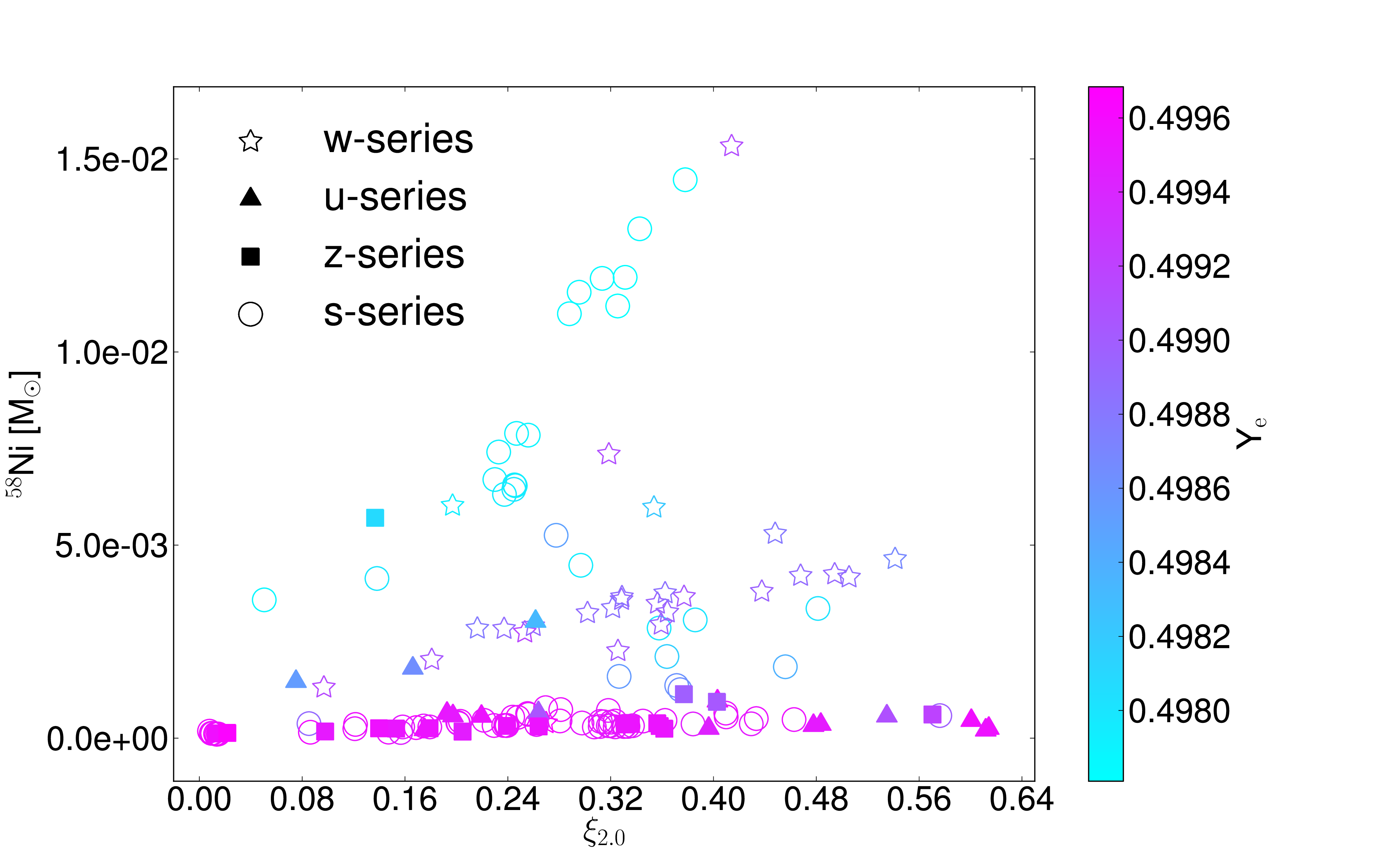}
    \includegraphics[width=0.48\textwidth]{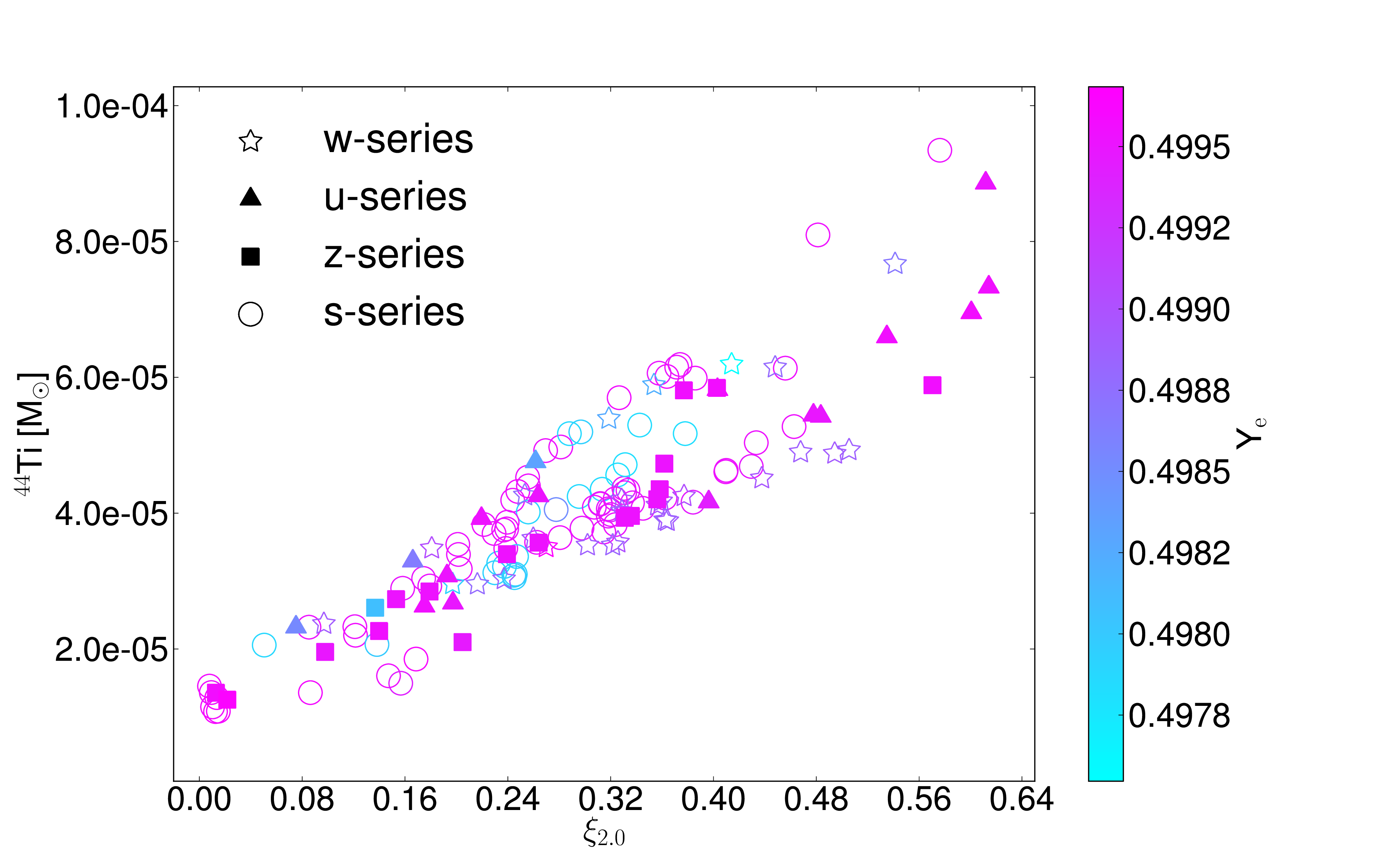} \\
    \includegraphics[width=0.48\textwidth]{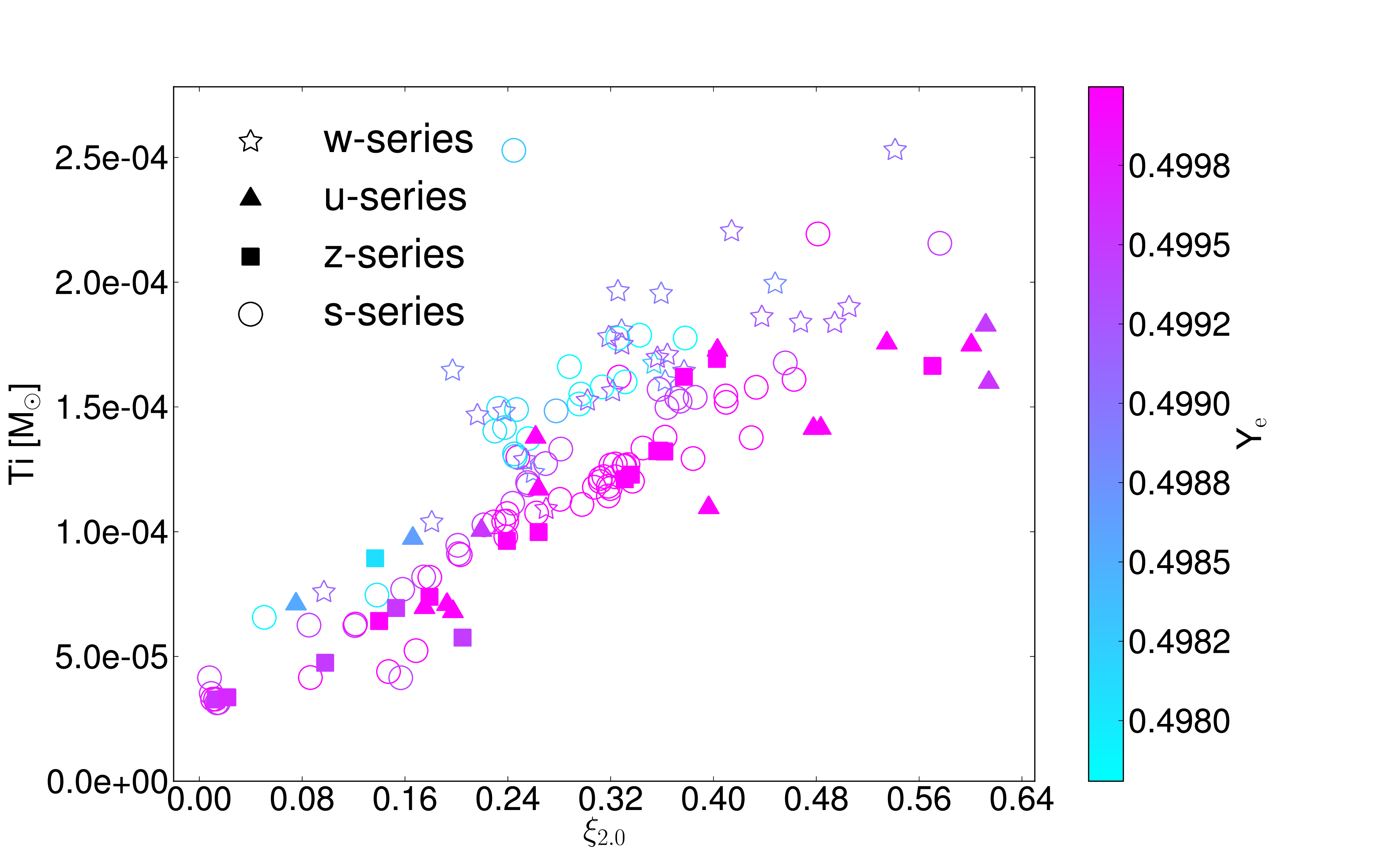}
    \includegraphics[width=0.48\textwidth]{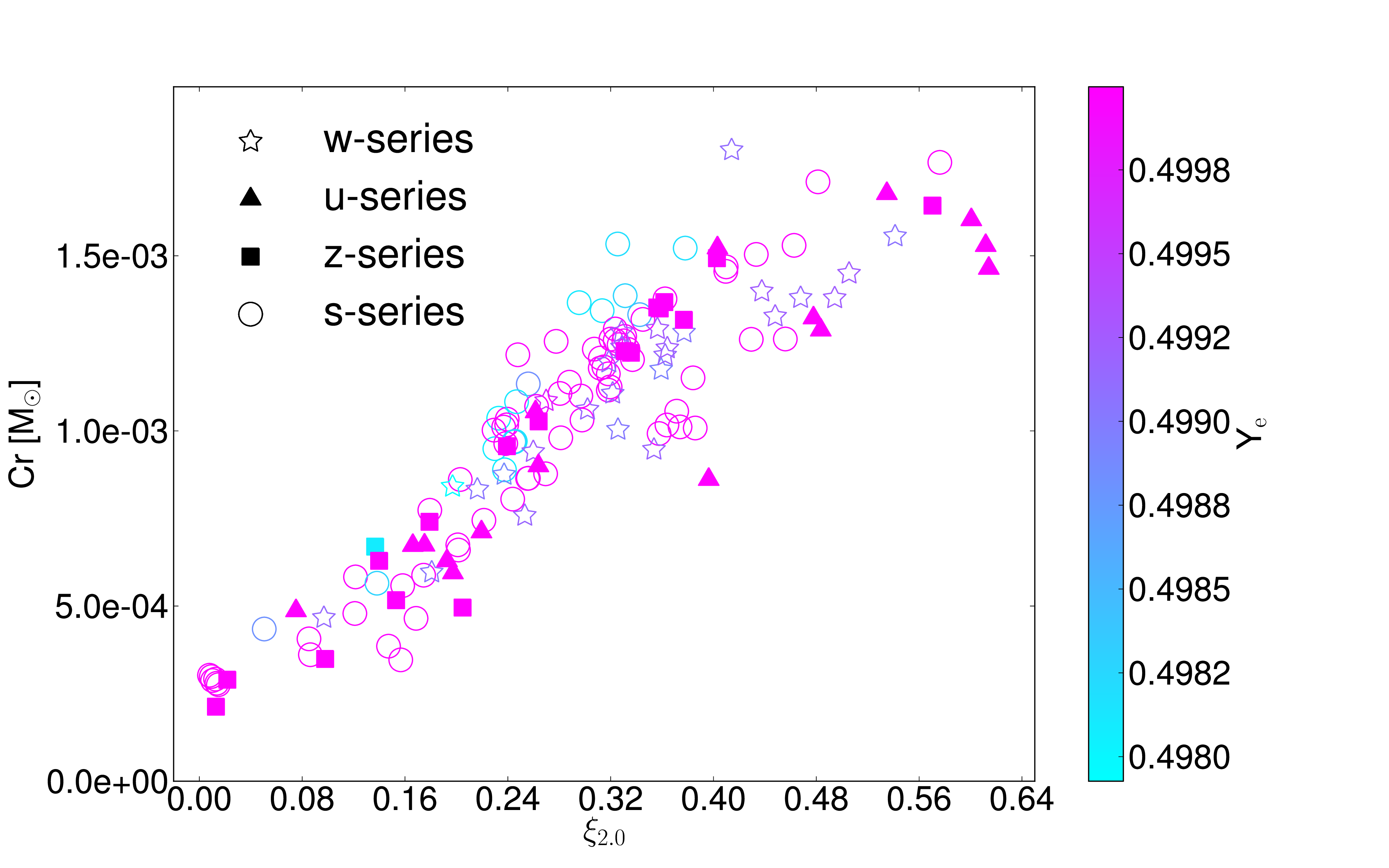} \\
    \includegraphics[width=0.48\textwidth]{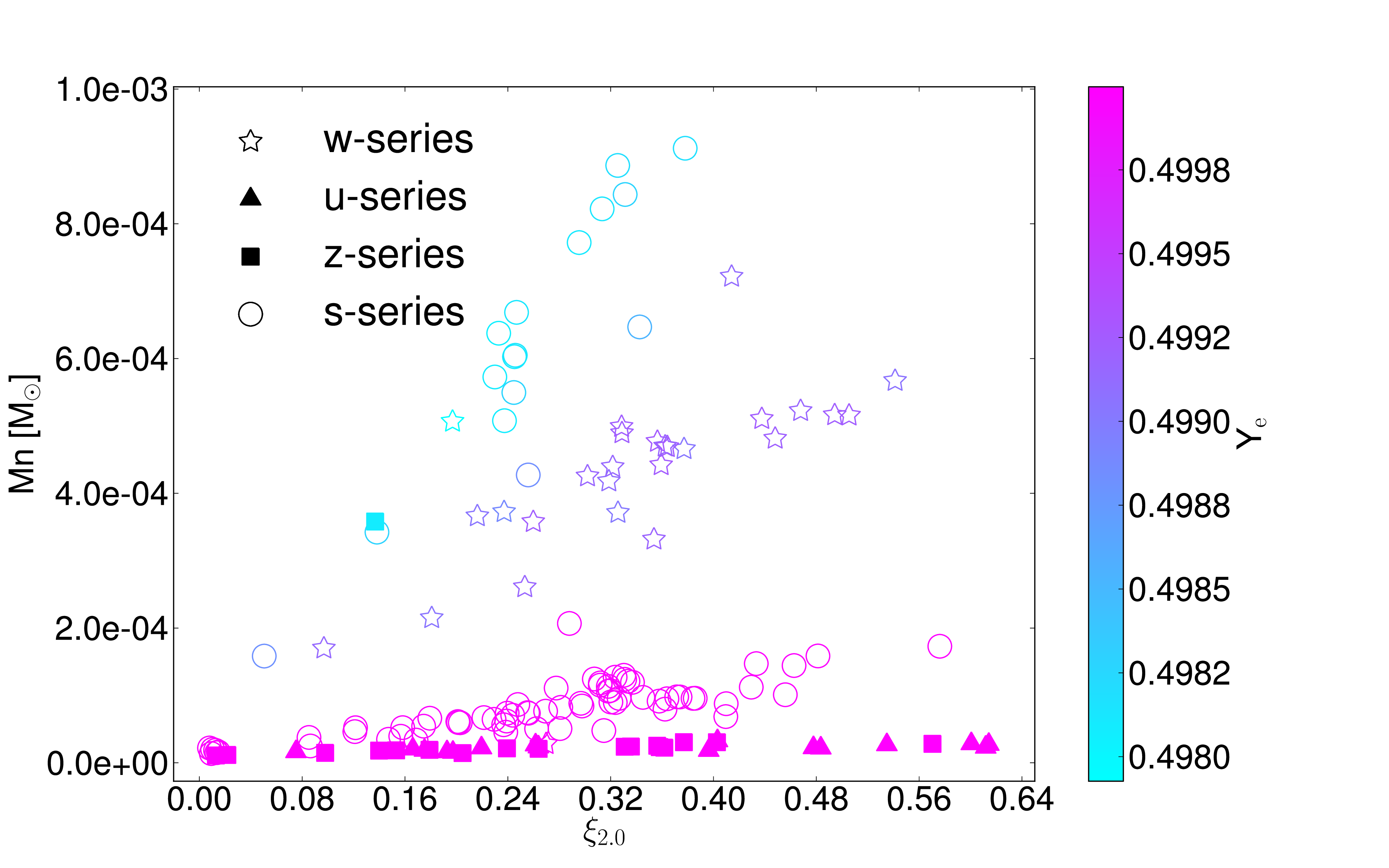}
    \includegraphics[width=0.48\textwidth]{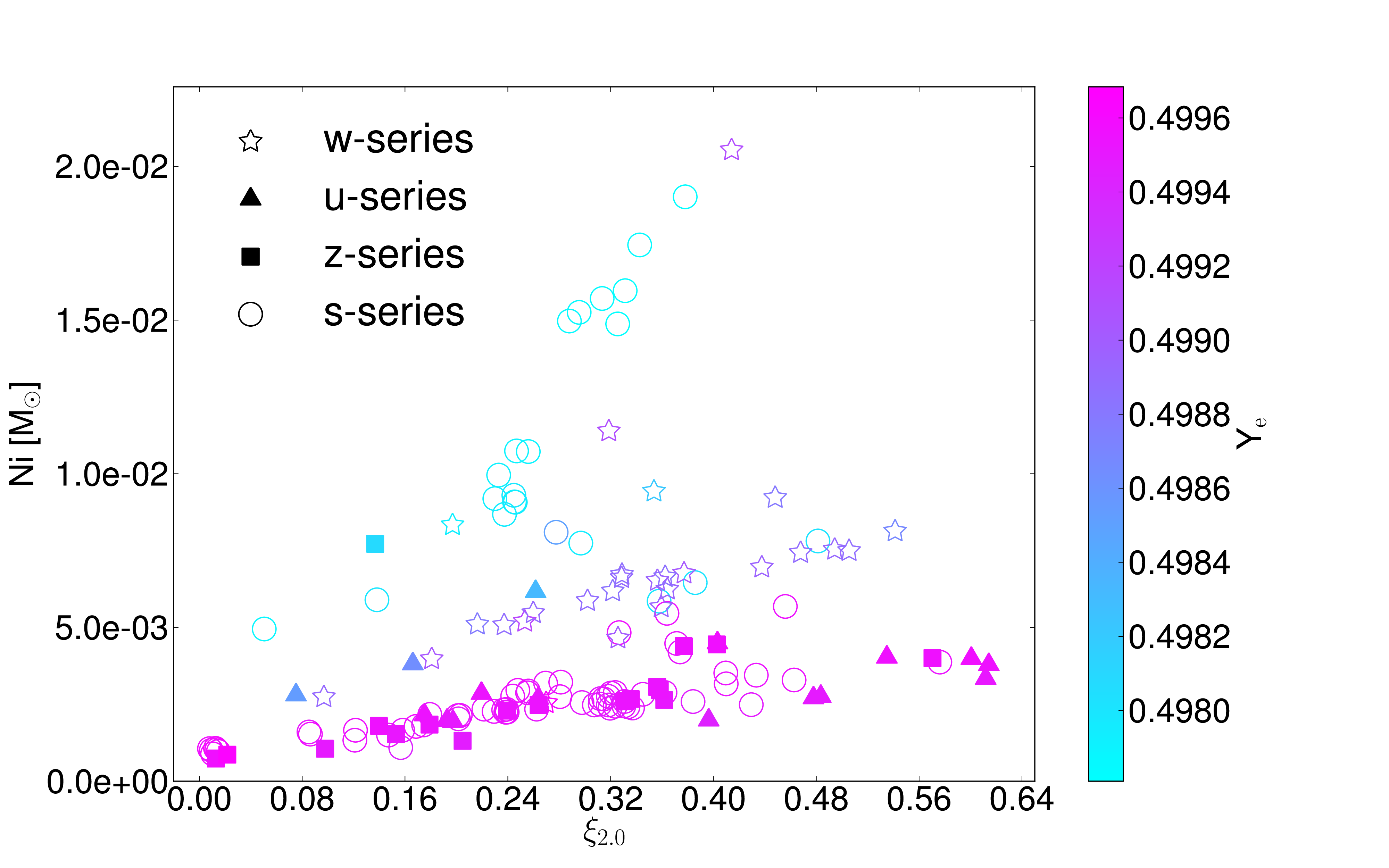}
\caption{
Top two rows: Isotopic yields of $^{56}$Ni(top left), $^{57}$Ni (top right), $^{58}$Ni (2nd row left), and $^{44}$Ti (2nd right) after explosive processing as function of compactness for all four sets of pre-explosion models: squares for the z-series ($Z=0$), triangles for the u-series ($Z=10^{-4}Z_{\odot}$), circles for s-series ($Z=Z_{\odot}$), and stars for w-series ($Z=Z_{\odot}$). 
Bottom two rows: elemental yields of titanium (3rd left), chromium (3rd right), manganese (bottom left), and nickel (bottom right).
The color-coding represents the average $Y_e$ value in the layers that made the highest contribution to the yield of each isotope shown. 
		\label{fig:Y_compactness_yecoded}
    }
\end{figure*}

Here, we want to examine the yields of all models for trends with ZAMS mass and compactness of the pre-explosion models. 
As seen in Figure \ref{fig:properties-zams}, we do not find a monotonic behavior of ejected $^{56}$Ni as function of the ZAMS mass. However, we find a correlation of the symmetric ($N=Z$) isotopes $^{56}$Ni and $^{44}$Ti with compactness $\xi_{2.0}$, as seen in the first and fourth panel of Figure \ref{fig:Y_compactness_yecoded}. The correlation is similar for the u-series and z-series presented in this paper as for the s- and w-series series presented in Paper~III. When combining the results from all four sets of pre-explosion models it becomes visible that the correlation has some width to it, which indicates that an exact one-to-one connection between compactness and $^{56}$Ni ejecta may not exist. Some of our models with the highest yields of ejected $^{44}$Ti are the almost failing models with low explosion energy and high $^{56}$Ni yields discussed in Section~\ref{subsec:highY-lowE}. The asymmetric nickel isotopes $^{57}$Ni and $^{58}$Ni behave differently. The yields of these asymmetric isotopes strongly depend on small changes in the local $Y_e$, with lower values of $Y_e$ favoring higher production of $^{57,58}$Ni (see second and third panel of Figure~\ref{fig:Y_compactness_yecoded}). The color-coding in Figure~\ref{fig:Y_compactness_yecoded} denotes the electron fraction of the layers where each isotope is made. In Paper~III, we found two (or more) branches of yields that independently correlate with compactness for the pre-explosion model sets at solar metallicity. The u-series and the z-series predominantly populate the branch with lower yields of the asymmetric $^{57,58}$Ni isotopes. For almost all models of the z-series, the final mass cut lies in layers with $Y_e\sim 0.5$, either in the Si-layer but outside of the region with $Y_e < 0.5$ or in the O-rich layer. The only exception is z15.0 where the mass cut is located in the inner parts of the Si-rich layer where the pre-explosion $Y_e$ is lower ($Y_e \sim 0.498$). This allows for the ejection of slightly neutron-rich material where asymmetric isotopes such as $^{57,58}$Ni can be synthesized. In the u-series, only u11.0 and u12.0 are on the branch with higher yields of $^{57}$Ni. In these two models, the final mass cut is in the Si-rich layer where $Y_e \sim 0.498$, allowing for the synthesis of asymmetric nickel isotopes.

The elemental yields of the iron group elements show similar trends (see bottom four panels of Figure \ref{fig:Y_compactness_yecoded}). To understand these trends we need to look at the isotopes from which the elemental yields originate. For example, elemental titanium is mainly synthesized as $^{48}$Cr (a symmetric isotope) and chromium is synthesized as $^{52}$Fe (also a symmetric isotope). Thus, the elemental yields of Ti and Cr, similar to $^{44}$Ti and $^{56}$Ni, show linear correlations with compactness, having experienced the highest temperatures and an alpha-rich freeze-out. 
Manganese is made as (asymmetric) $^{55}$Co and stable nickel is dominated by $^{58}$Ni and $^{60}$Ni which are made as $^{58}$Ni and $^{60}$Cu. Hence, the yields of elemental Mn and Ni strongly depend on the local $Y_e$ value. With the exception of z15.0, all models of the u-series and z-series result in low elemental Mn yields. Only u11.0, u12.0, u14.0, and z15.0 have relatively high elemental Ni yields (synthesized at relatively low $Y_e$ values). 

\begin{figure}[]  
    \includegraphics[width=0.48\textwidth]{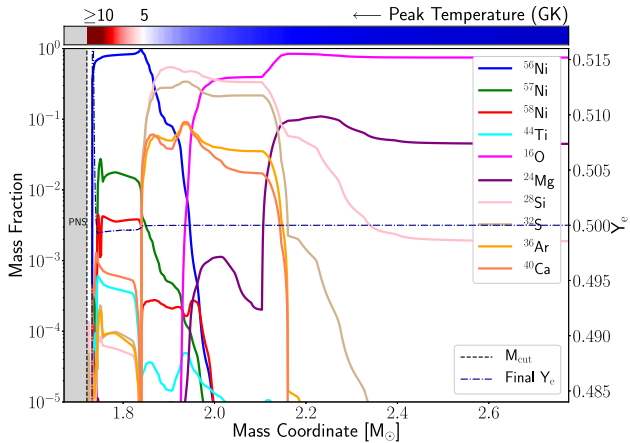}
\caption{
Post-explosion composition profile for model z30.0 as a function of the mass coordinate. The dot-dashed lines indicate the pre-explosion and the final electron fraction. The vertical dashed line indicates the mass cut. 
		\label{fig:outlier-spaghetti-z30}
    }
\end{figure}

\begin{figure}[]  
    \includegraphics[width=0.48\textwidth]{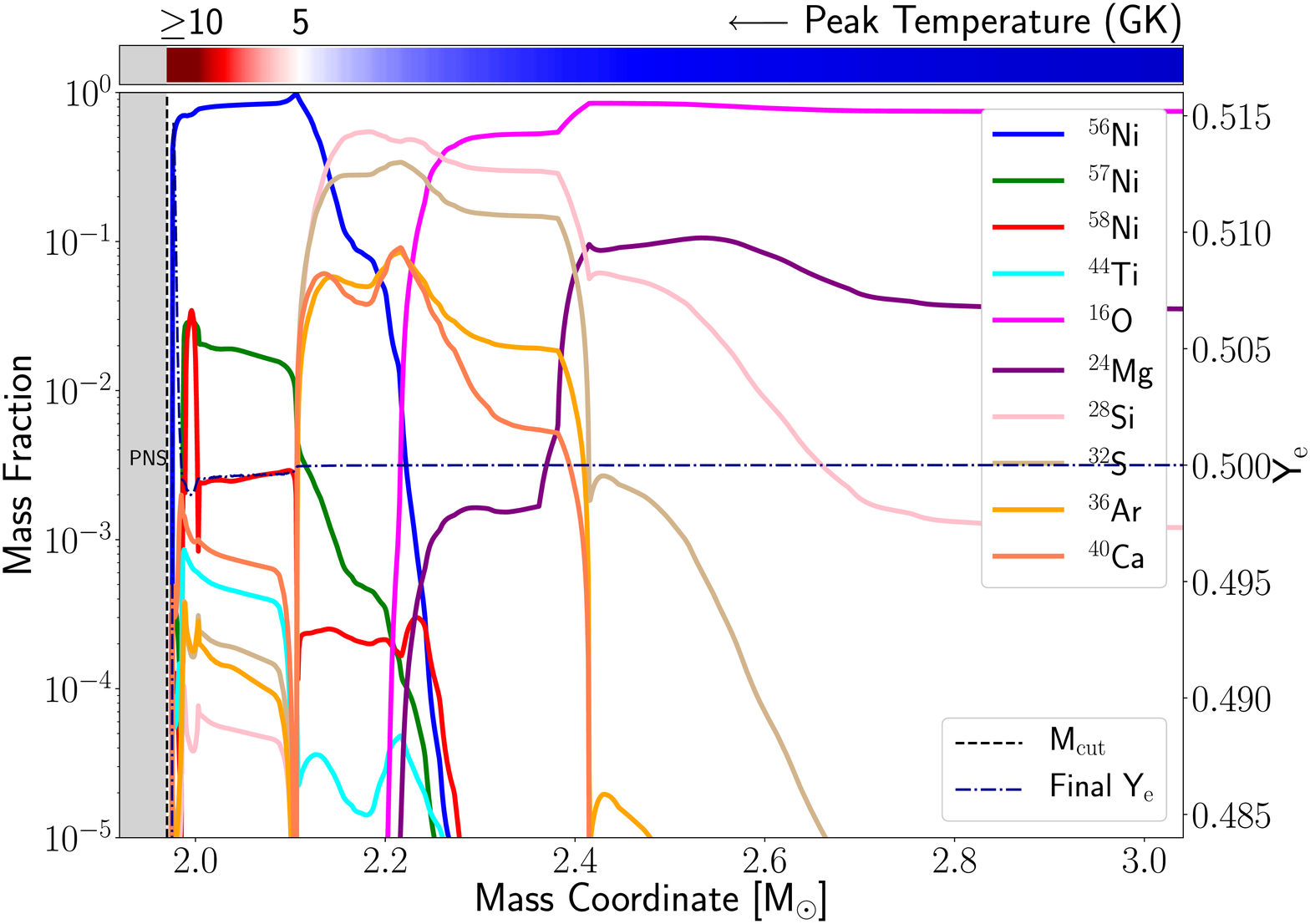}
\caption{
Same as Figure \ref{fig:outlier-spaghetti-z30} but for z31.0
		\label{fig:outlier-spaghetti-z31}
    }
\end{figure}

In Section~\ref{subsec:highY-lowE}, we discussed explosion details of the models which do not follow the general correlation between explosion energy and $^{56}$Ni yields. Here, we analyze the detailed nucleosynthesis yields of these models using the z31.0 model, which we compare to model z30.0 to illustrate the differences. Figures~\ref{fig:outlier-spaghetti-z30} and \ref{fig:outlier-spaghetti-z31} show the post-explosion profiles of the innermost $\sim 1$~$M_{\odot}$ of ejecta above the mass cut. For z30.0 (a model which follows the general $E_{\mathrm{expl}}$-$^{56}$Ni trend) the mass cut resides inside the Si layer. The ejected $^{56}$Ni is explosively synthesized from $^{28}$Si. For z31.0 (an almost failing model with low explosion energy and high Ni yields), the final mass cut is located further from the center in the O-rich layer. Here, the ejected $^{56}$Ni originates from explosively processed $^{16}$O instead. However, the total amount of ejecta heated to temperatures above 6~GK (and hence resulting in $^{56}$Ni) is similar in z30.0 and z31.0.
While z30.0 and z31.0 both have ejecta with similar peak electron fractions of $Y_e \sim 0.515$, the z31.0 model has some ejecta just above the mass cut with low enough $Y_e$ values that the $^{58}$Ni production is enhanced to approximately the same level as $^{57}$Ni. 
The yields of alpha-elements and iron-group elements are very similar between z30.0 and z31.0. Only beyond mass number $A \approx 80$ there are some (small) differences in individual isotopes, but the overall abundance pattern is the same.

\subsection{Metallicity Dependence} \label{sec:gce}

\begin{figure*}[]
	\includegraphics[width=0.48\textwidth]{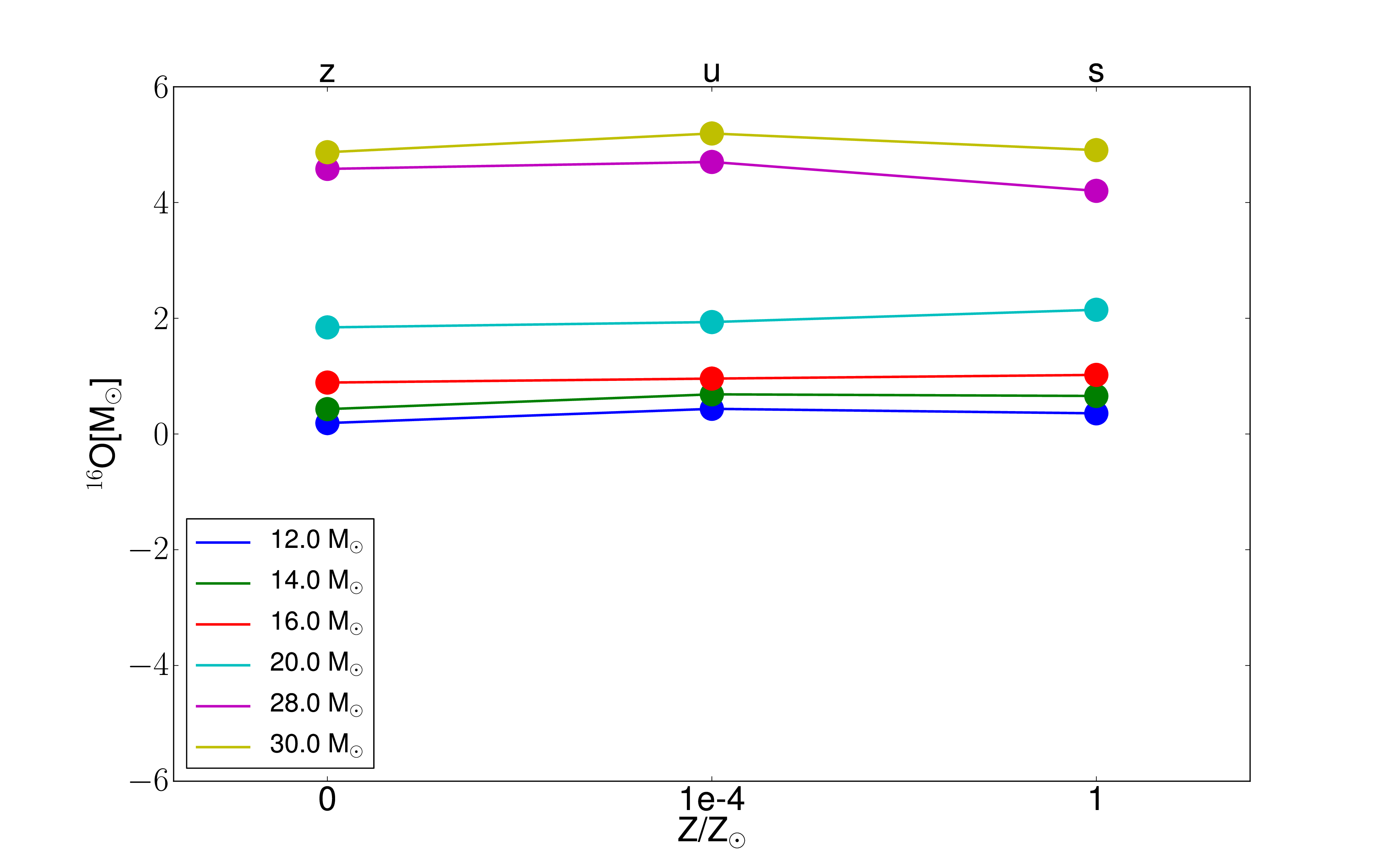}
	\includegraphics[width=0.48\textwidth]{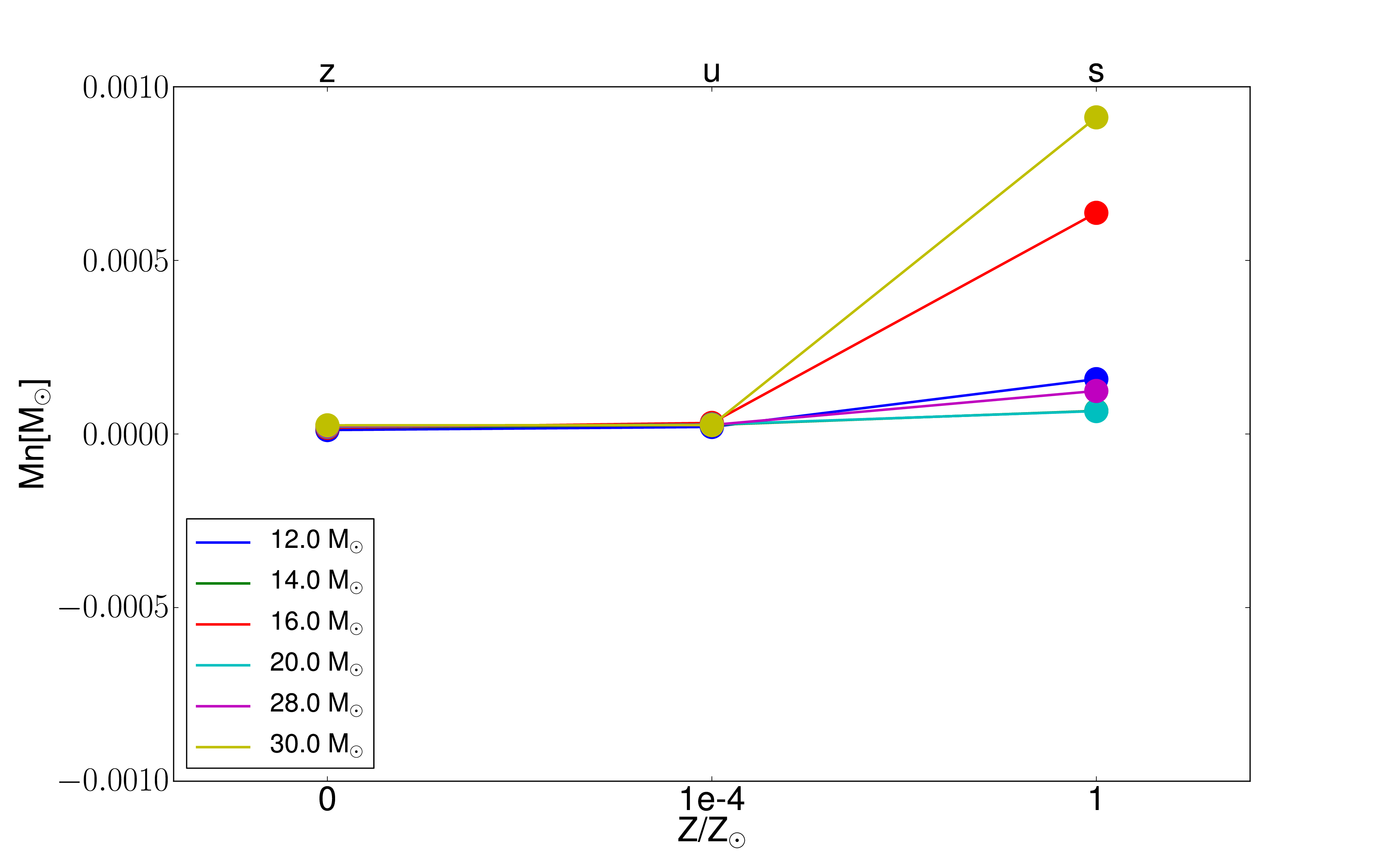} \\
	\includegraphics[width=0.48\textwidth]{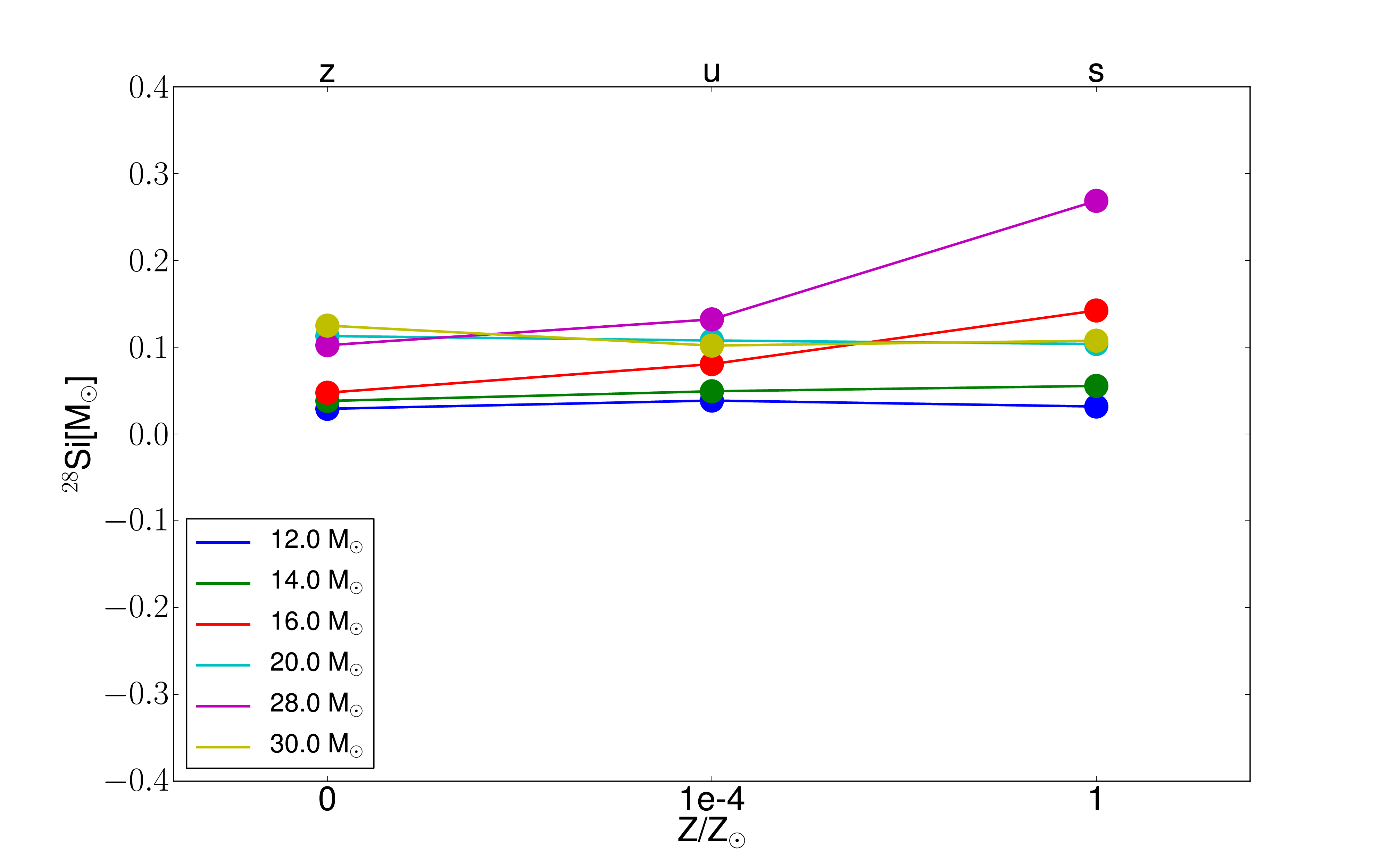}
	\includegraphics[width=0.48\textwidth]{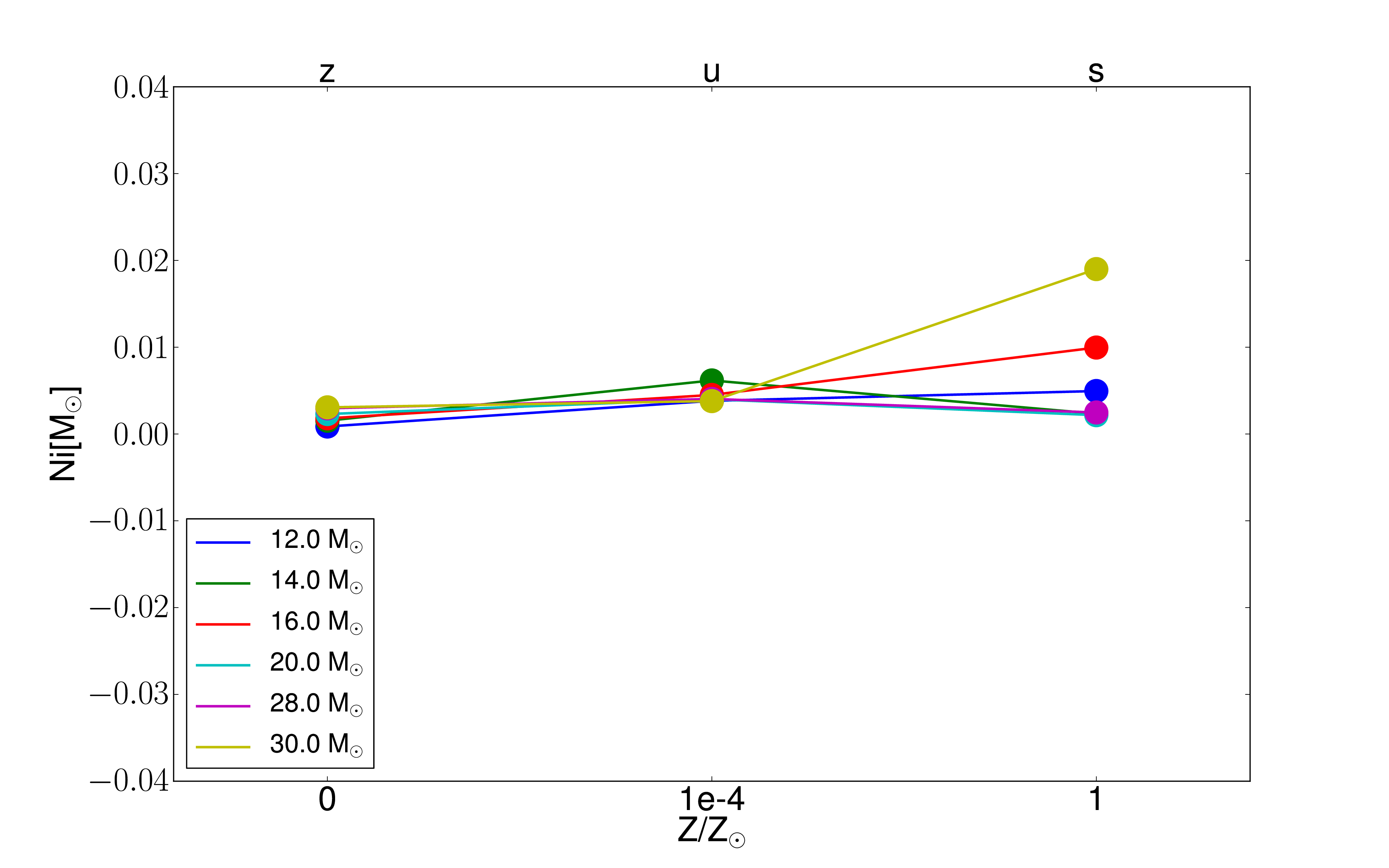} \\
	\includegraphics[width=0.48\textwidth]{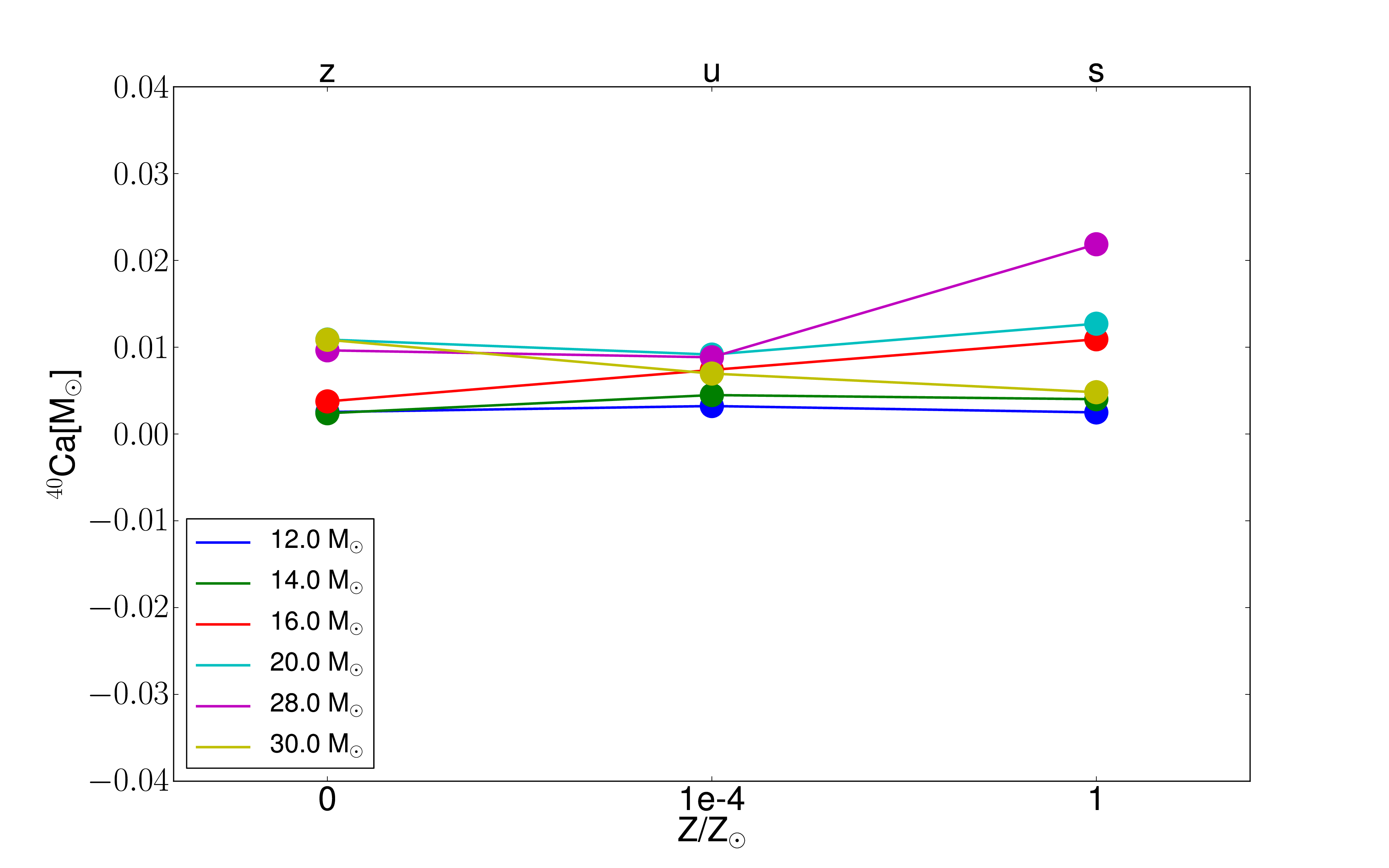}
	\includegraphics[width=0.48\textwidth]{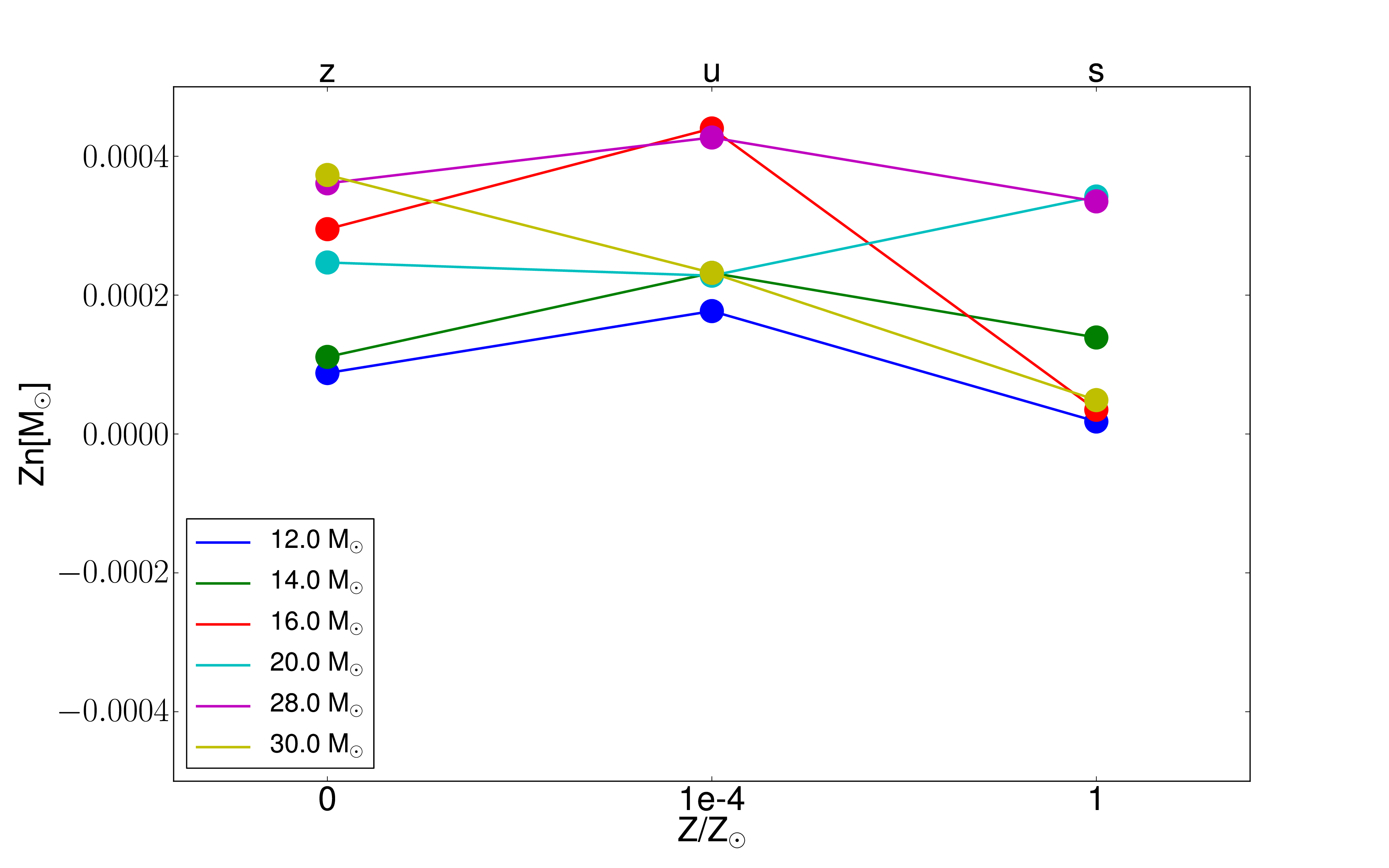} \\	
\caption{Abundances of $^{16}$O, $^{28}$Si, $^{40}$Ca (left column) and elemental Mn, Ni, and Zn (right column) as function of the initial metallicity for models of different ZAMS masses in $M_{\odot}$. 
\label{fig:metallicity}
    }
\end{figure*}

We display in Figure~\ref{fig:metallicity} the abundances of selected stable isotopes ($^{16}$O, $^{28}$Si, and $^{40}$Ca) and of three iron-group elements (Mn, Ni, Zn) for six different ZAMS masses to analyze their dependence on the initial metallicity of the pre-explosion model.
The $^{16}$O is mainly produced in helium and neon burning and is expected to be mostly independent of the initial stellar metallicity. In our models, there is no trend of $^{16}$O abundance with initial metallicity. However, we find higher abundances of $^{16}$O for higher ZAMS mass due to the overall larger stellar mass and larger CO-core. 
The decrease in $^{16}$O yield for the s28.0 model is due to explosive O-burning. Of the models shown, s28.0 is the only model with an extended layer consisting of a mixture of $^{16}$O and $^{28}$Si which allows for $^{16}$O being explosively burned (the other models have a sharp transition from $^{28}$Si to $^{16}$O at the Si-O interface).
For $^{28}$Si and $^{40}$Ca, we find weak trends with initial metallicity: For the higher ZAMS mass models, there are somewhat higher abundances of $^{28}$Si and $^{40}$Ca ejected. In the lower ZAMS mass models, our results do not exhibit any metallicity dependence in those abundances.
The yields of iron group elements, such as Ni, are more sensitive to the details of the explosion, as they are synthesized during explosive burning in the innermost ejecta. The yields of symmetric iron-group nuclei depends mostly on the explosion strength and hence the local peak temperature. For odd-$Z$ elements, the final abundances are also quite sensitive to the local $Y_e$-values, as previously discussed. With very few exceptions, all models of the z-, u-, and s-series only synthesize a very low abundance of Mn. As discussed in Paper~III, the small nuclear network used for modelling the pre-explosion evolution artificially keeps the $Y_e$ closer to 0.5, which is not favorable for the production of Mn. As expected, the Ni yields are mostly independent of the initial metallicity. The higher yields of Ni (and also of other iron-group elements) in some of the solar-metallicity models (e.g.\ s28.0 and s30.0) are due to these models having larger explosion energies than their low/zero metallicity counterparts. 
In our models, zinc is synthesized as $^{64}$Zn in layers with $Y_e>0.5$ and as neutron-rich isotopes of Zn in neutron-rich layers close to the mass cut. The final abundances of Zn are very sensitive to the amount of proton-rich and neutron-rich ejecta, which depends on the explosion strength, the neutrino/anti-neutrino luminosities, the local $Y_e$, and the location of the mass cut. Our models span a range of conditions which is reflected in the range of Zn yields we obtain. We do not find any clear trends with metallicity nor with ZAMS mass.

\section{Comparison with Observations}
\label{sec:observations}

\subsection{The Supernova Landscape} \label{subsec:nomotoplots}

\begin{figure}[]  
	\includegraphics[width=0.48\textwidth]{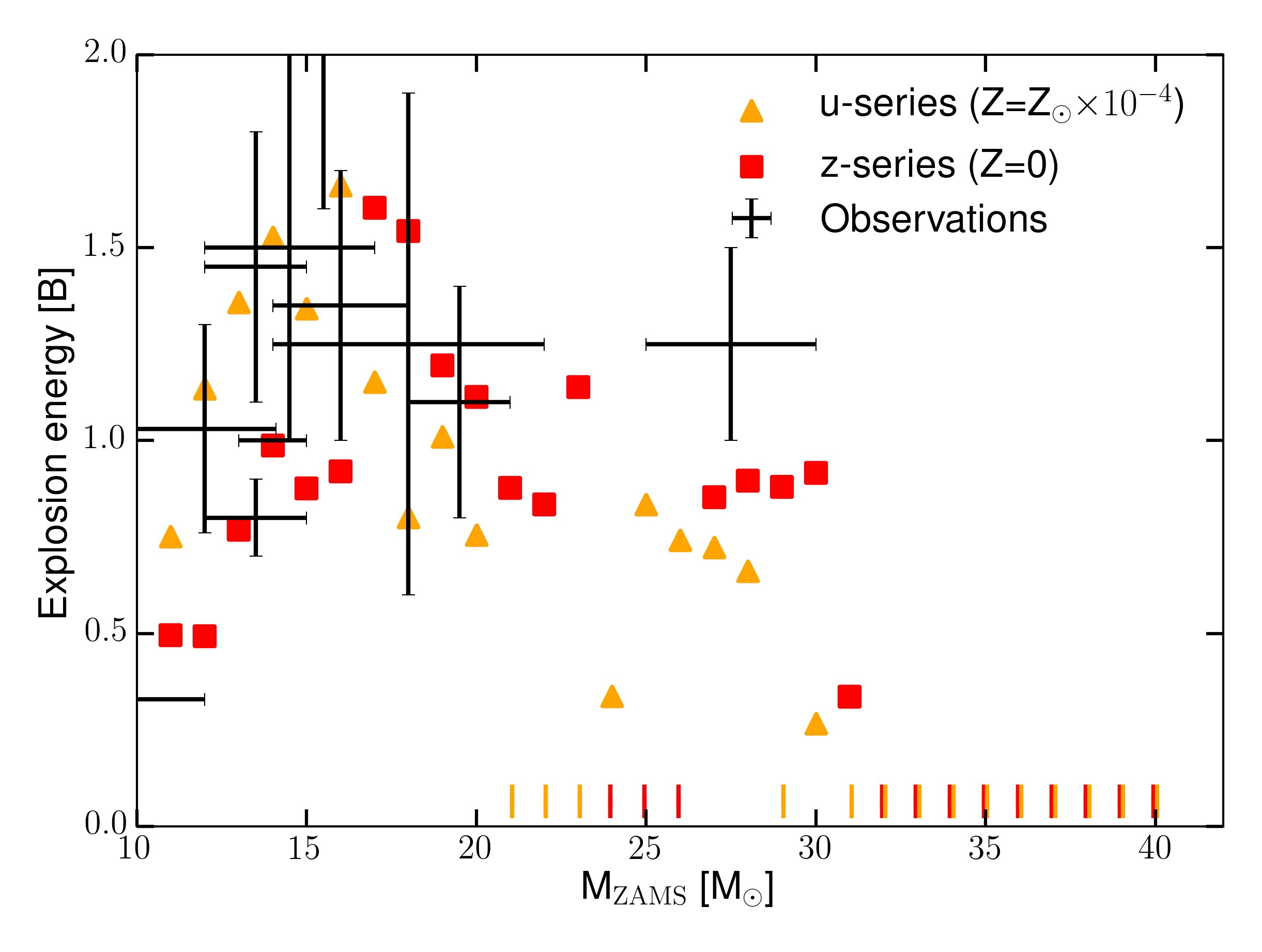}
	\includegraphics[width=0.48\textwidth]{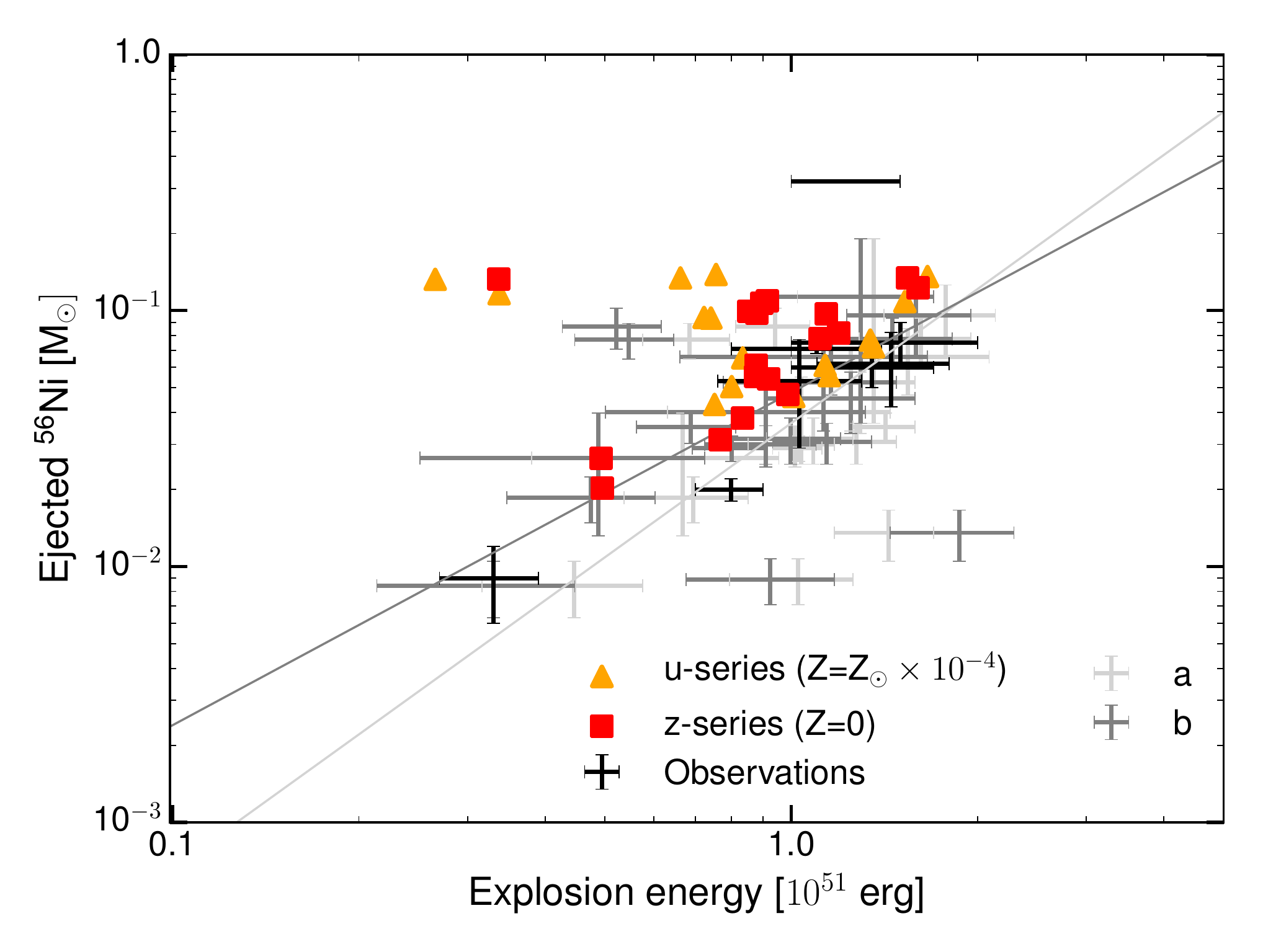}
	\caption{Top: Explosion energies as function of ZAMS mass for observed supernovae (black crosses with error bars taken from Paper~II) and our simulations (red squares for z-series, yellow triangles for u-series) using the standard calibration. The vertical dashes at the bottom of the Figure indicate masses for which a BH was formed. 
	Bottom: Ejected $^{56}$Ni masses as function of explosion energy from our simulations (color symbols) and from observations (black crosses are from Paper~II, grey crosses are from \citet{muller.prieto.ea:2017}; the labels ``a'' and ``b'' denote the two different energy determinations given in \citet{muller.prieto.ea:2017}). The lines are the corresponding fits to the data given in \citet{muller.prieto.ea:2017}.
		\label{fig:nomotoplot}
        }
\end{figure}

First, we compare our simulations to observations of CCSNe. 
In the top panel of Figure~\ref{fig:nomotoplot} we show the explodability for the progenitors in our sample in comparison with observed explosion energies (black crosses). The observational data shown are the same as in Paper~II, complemented with the sample of \cite{muller.prieto.ea:2017}. Note that the observed supernovae are in the local Universe (redshift $z<0.01$). Hence, we do not expect a perfect agreement of the models from low and zero metallicity progenitors with the observational data. Nevertheless, including the observational data in the Figure is useful to identify general trends. The overall trend in explosion energy is very similar to what we found for pre-explosion models at solar metallicity (see Paper~II). The explosion energies for both sets (u and z) increase from lower values for low mass progenitors (``Crab-like SNe'') to explosions with energies $E_{\mathrm{expl}} \approx 0.8$ -- $1.6$~Bethe between  15~$M_{\odot}$ and 21~M$_{\odot}$. We find the strongest explosions around 15~\msun for the u-series and at slightly higher masses around 17~\msun for the z-series. For ZAMS masses above 20~\msun we find predominantly BHs (denoted by short vertical lines along the bottom axis), interspersed with some explosions.  As expected from the higher compactness values, we find more failed explosions and BHs, and overall lower explosion energies, for a given ZAMS mass (especially for ZAMS masses $>25$~$M_{\odot}$) at low and zero metallicity compared to solar metallicity. As already seen in Figure~\ref{fig:prog_compactness}, the compactness curves for the four sets of progenitors are quite similar in shape. The only big difference is a shift of the peak compactness values in the u-series to lower ZAMS masses when compared to the other three series. From this we expect quite similar behaviors from the four series, only with a shift of the peak explosion energies to slightly lower ZAMS masses for the u-series, as seen in Figure~\ref{fig:nomotoplot}.

In the bottom panel of Figure~\ref{fig:nomotoplot} we show the resulting $^{56}$Ni yields against the corresponding explosion energies. When comparing the $^{56}$Ni yields from our simulations with observations, we find that most models agree well with observations (as in the case of solar metallicity models). We find that the amount of $^{56}$Ni increases with ZAMS mass up to $\sim 15$~$M_{\odot}$ and then remains roughly constant, similar to our findings in Paper~III. The almost failing models with $^{56}$Ni yields of $\sim 0.1$~$M_{\odot}$ and with very low explosion energies ($< 0.5$~B) discussed in Section~\ref{subsec:highY-lowE} can be seen above and to the left of the observational data. 
It is worth reminding the reader that the $^{56}$Ni (and the iron group in general) is synthesized from completely dissociated material after shock passage. Hence, the initial metallicity does not directly impact the yield of $^{56}$Ni (and other iron group nuclei). The lower observed explosion energies and $^{56}$Ni ejecta masses originate from progenitors at or below 10~\msun ZAMS mass, which are not included in our samples.

\subsection{Metal-poor Stars} \label{subsec:yields-EMPs}

\begin{figure}[]
	\includegraphics[width=0.48\textwidth]{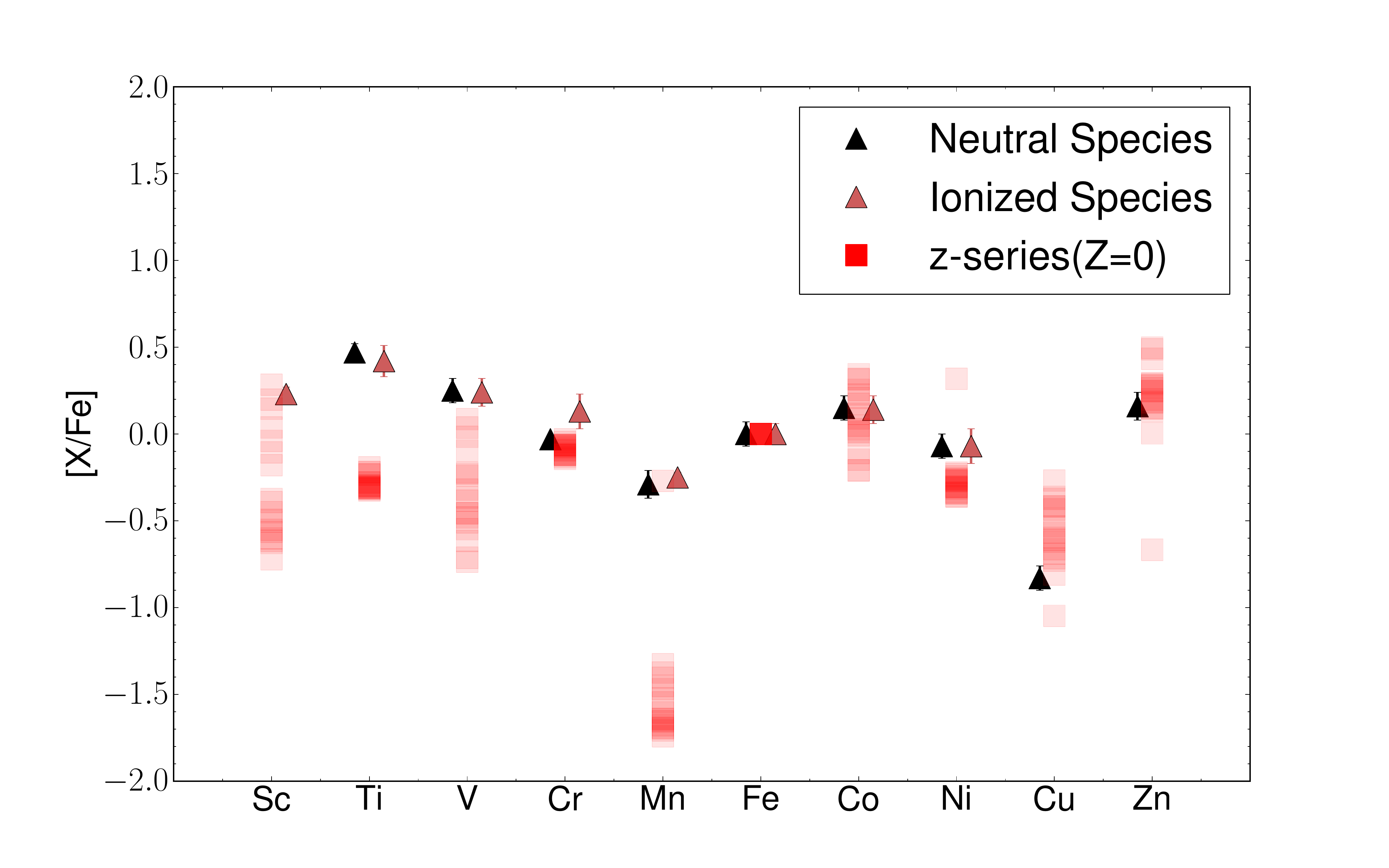} \\
    \includegraphics[width=0.48\textwidth]{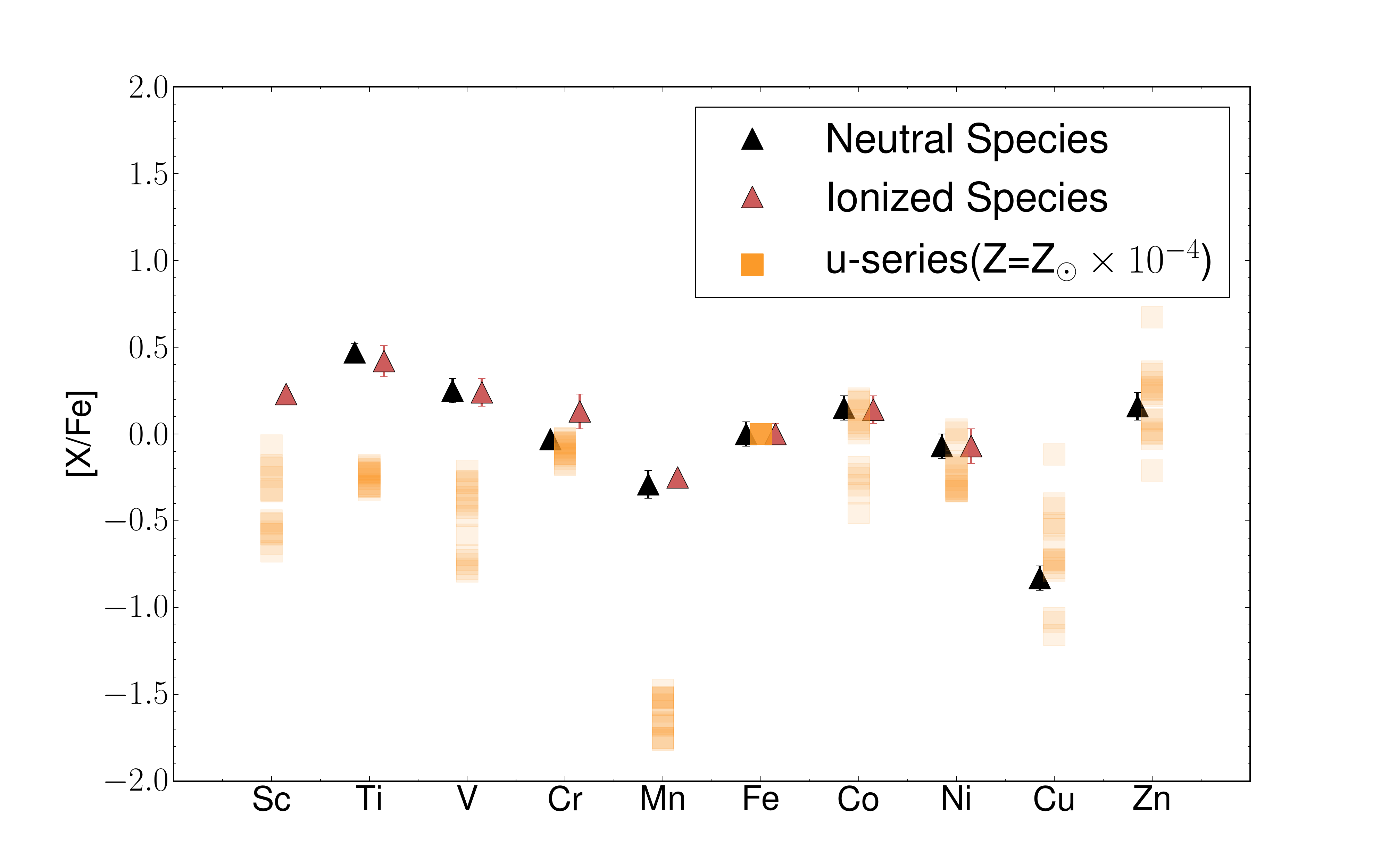}
\caption{Abundances of iron group elements: Observationally-derived abundances for metal-poor star HD~84937 (triangles) together with our results (z-series in the top panel, u-series in the bottom panel). 
		\label{fig:xfe}
    }
\end{figure}

Next, we compare our predicted yields with the observationally derived abundances in metal-poor stars. The atmospheres of these low-mass, long-lived, metal-poor stars carry the signature of one or a few previously exploded CCSNe from massive stars which deposited their yields in the interstellar medium. The iron group elements are of particular interest to test our CCSN nucleosynthesis predictions, as they are made in primary explosive nucleosynthesis processes. In Figure \ref{fig:xfe} we compare our predictions for iron group elements with the abundances of metal-poor stars HD~84937 (u-series in the top panel; z-series in the bottom panel). The abundances of this metal-poor star have recently been determined using improved laboratory data for neutral and singly ionized transitions in iron group elements \citep{sneden16}. Each transparent square represents one of our models. Our results are not weighted with an initial mass function to illustrate how sensitive or robust the results for each element are. The triangles indicate the observational data (neutral and singly ionized species). 
Overall, we find a good agreement between our predictions and the observational data. We do not find significant differences between the u-series and the z-series. 
Scandium and zinc are synthesized at levels comparable to the observed values of [Sc/Fe] and [Zn/Fe], respectively. Both elements are difficult to produce in sufficient amounts in traditional piston and thermal bomb nucleosynthesis calculations which neglect the neutrino interactions and employ a canonical explosion energy of $10^{51}$~erg. Enhanced explosion energies, such as in hypernovae, lead to enhanced production of Sc and Zn, even without the inclusion of neutrino interactions \citep{nomoto06}. A careful treatment of the neutrino interactions robustly leads to enhanced production of Sc and Zn already at canonical explosion energies \citep{cf06a}. Our models (u-series and z-series) co-produce Zn with Fe. We find a smaller spread of [Zn/Fe] values in the u-series and z-series as at solar metallicity (top panel of Figure~9 in Paper~III).
As in the case of the s-series, we find that [Mn/Fe] is significantly lower than the observations in both the u- and the z-series. We attribute this to the relatively small network used in the pre-explosion models of WHW02 at all metallicities, which results in $Y_e$ values of $\sim 0.4995$ in the relevant layers. As discussed above, the production of manganese is quite sensitive to the local $Y_e$ value. Hence, a $Y_e$ value of $\sim 0.4995$ yields small amounts of manganese. The pre-explosion models of the w-series employ a much larger network during the stellar evolution phase, which results in a somewhat lower final $Y_e$ ($\sim 0.4992)$ and hence larger values of [Mn/Fe], see Paper~III for details.

\section{Remnant Properties} \label{sec:remnants}

\begin{figure*}[ht]
	\includegraphics[width=0.48\textwidth]{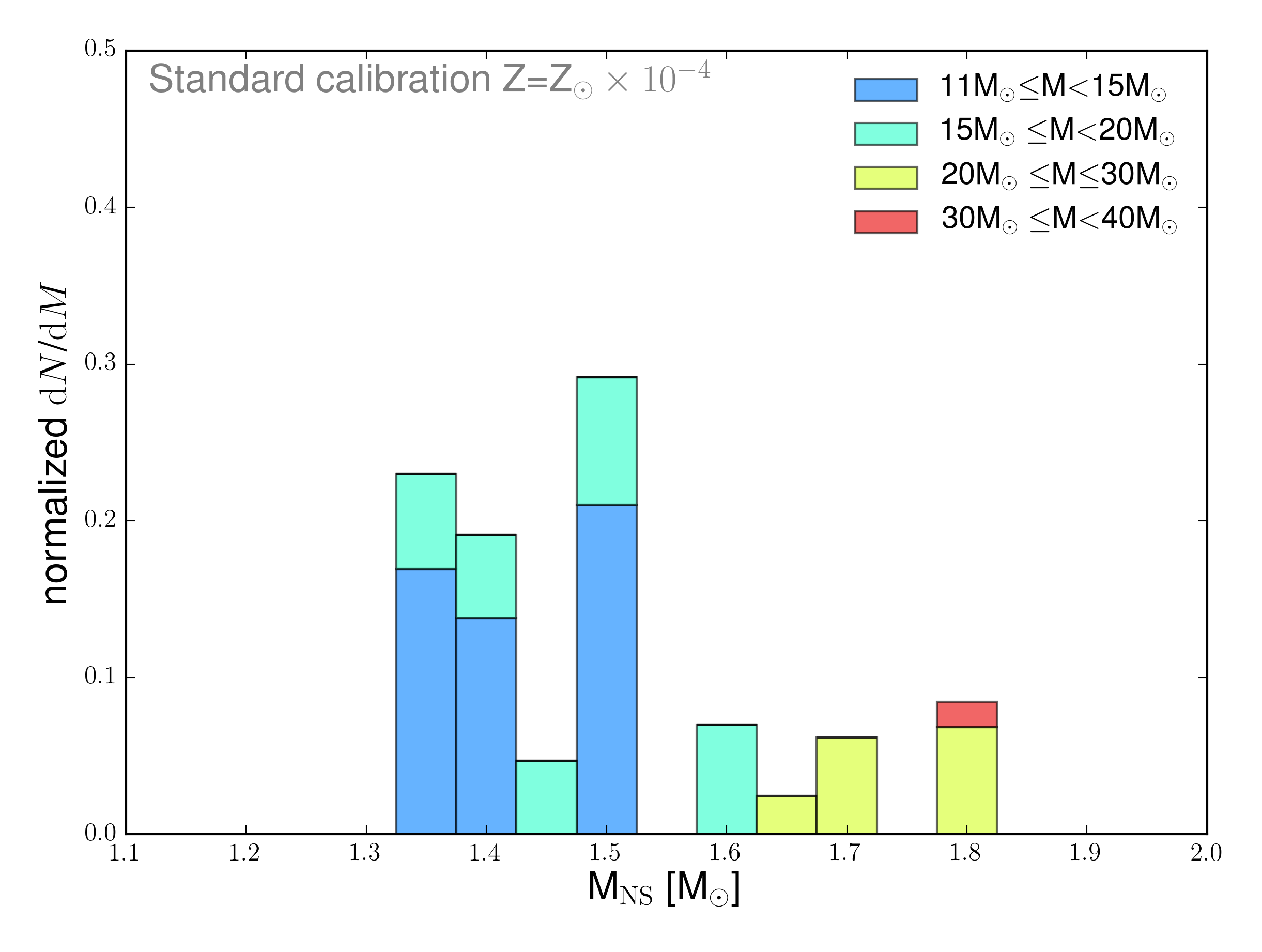}
	\includegraphics[width=0.48\textwidth]{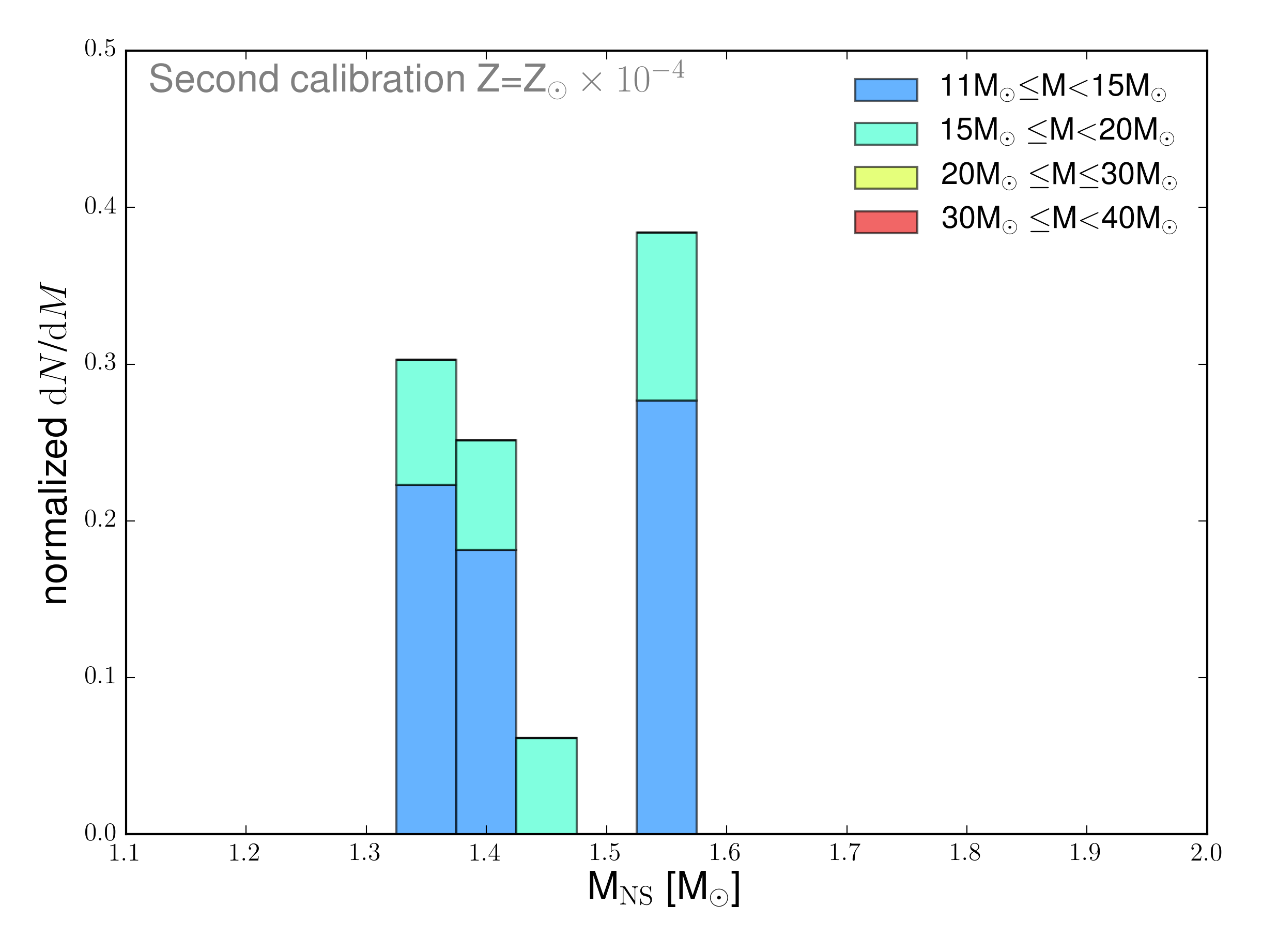} \\
    \includegraphics[width=0.48\textwidth]{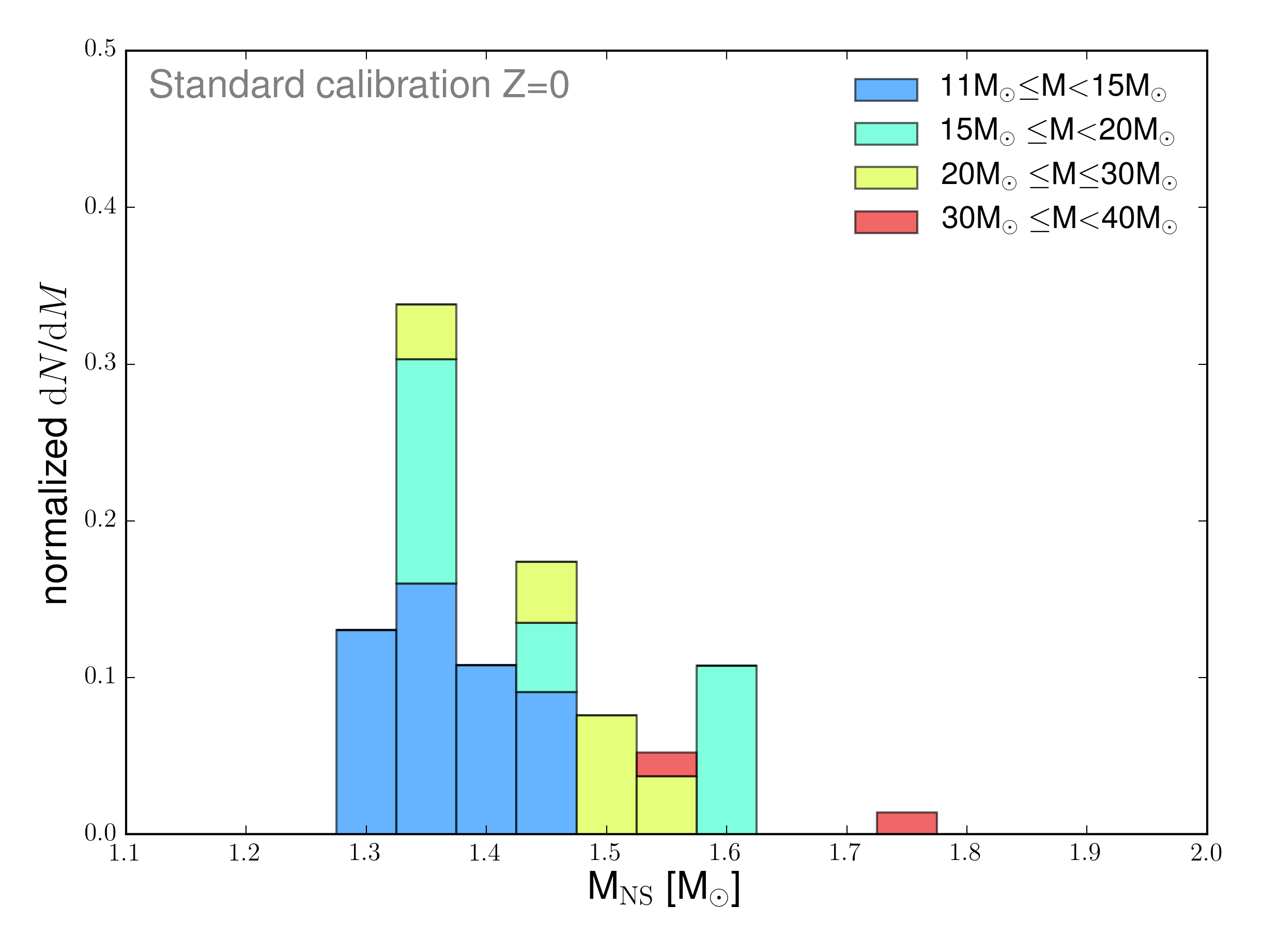}
    \includegraphics[width=0.48\textwidth]{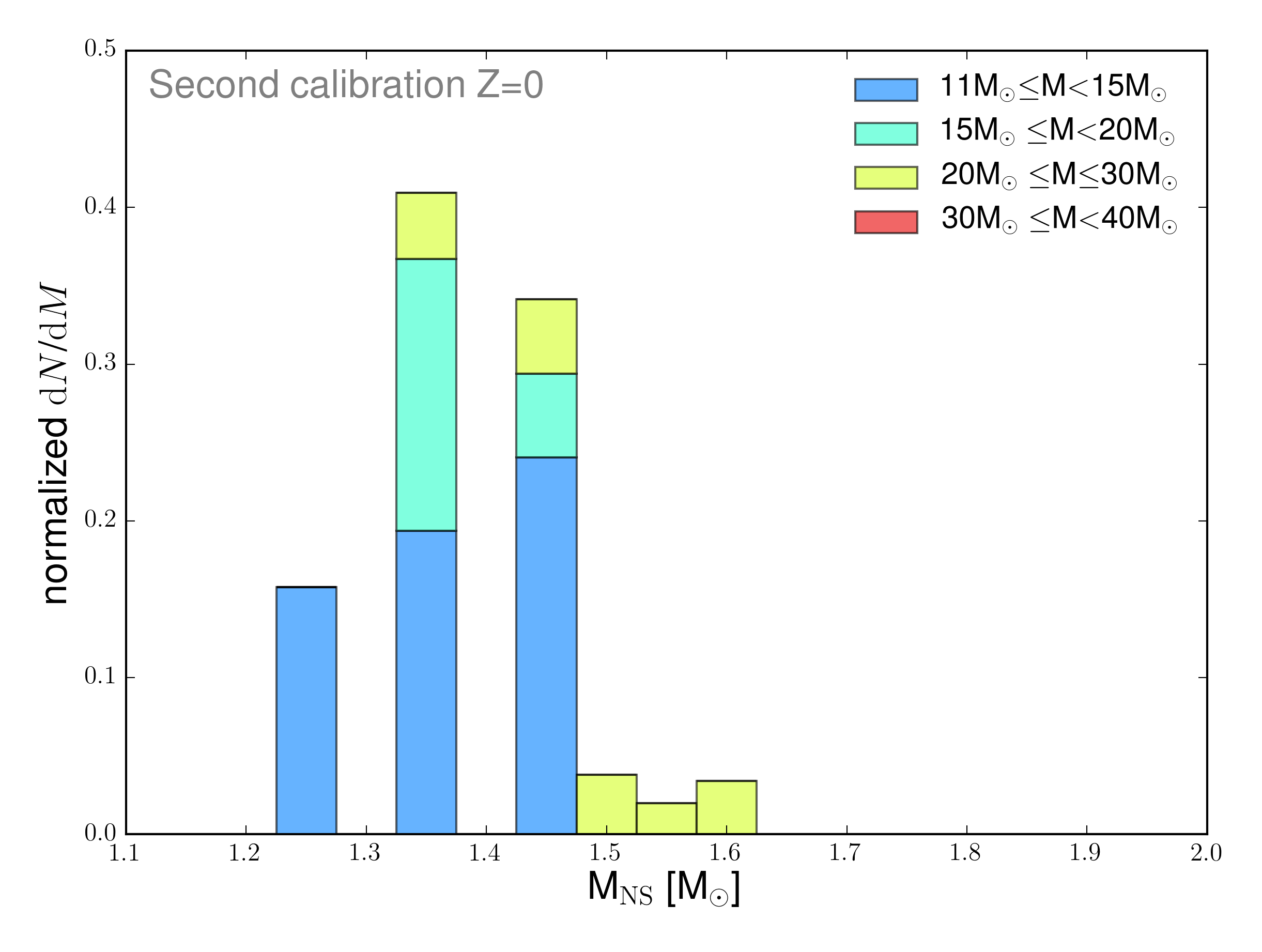}
\caption{Gravitational birth mass distributions of cold NSs for u-series (top panel) and z-series (bottom panel) for the standard calibration (left column) and the second calibration (right column). The colors indicate different ranges of ZAMS masses of the pre-explosion models that contribute to the distribution.
		\label{fig:NS_mass_distr}
    }
\end{figure*}

\begin{figure*}[ht]
	\includegraphics[width=0.48\textwidth]{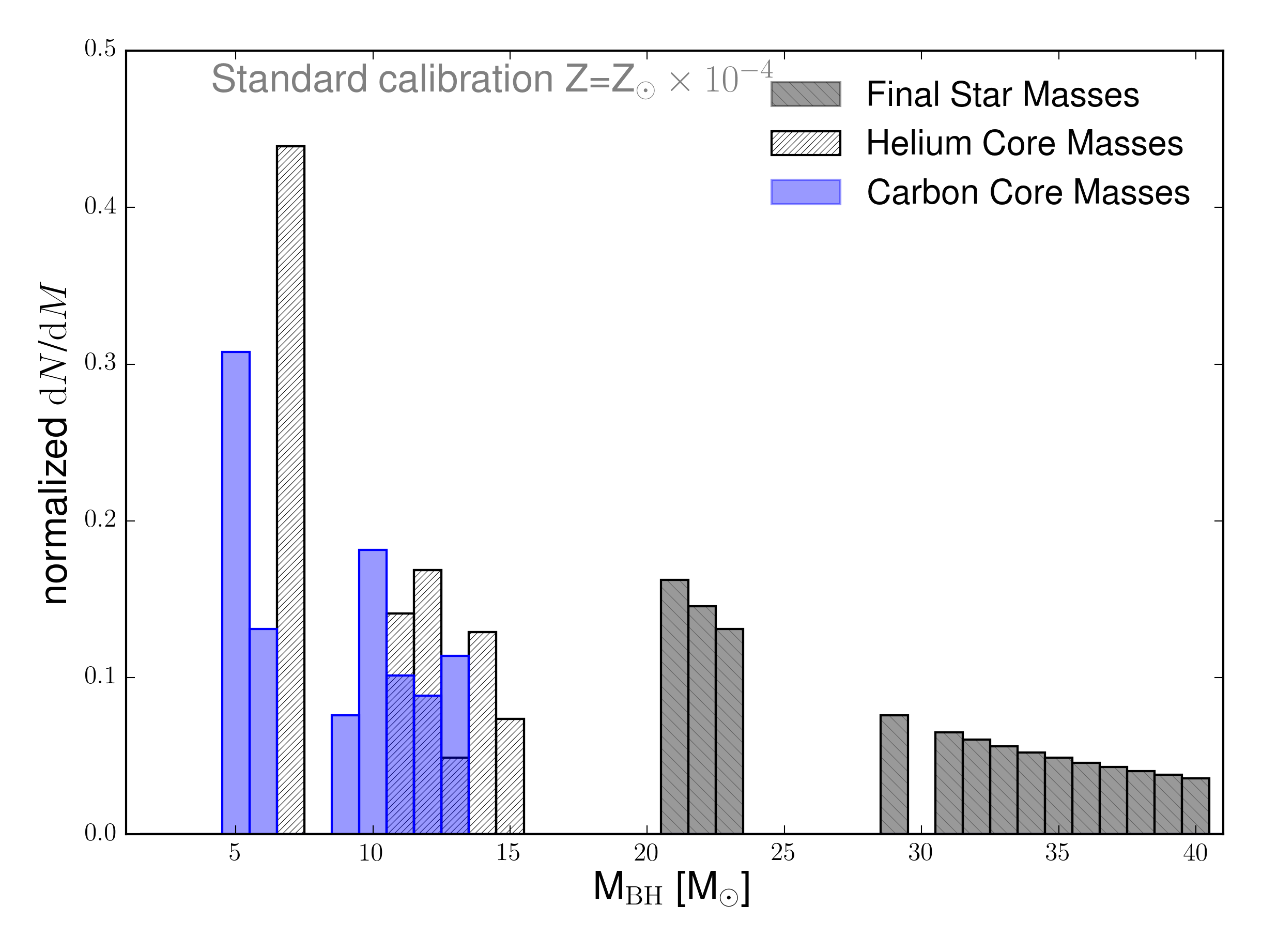}
	\includegraphics[width=0.48\textwidth]{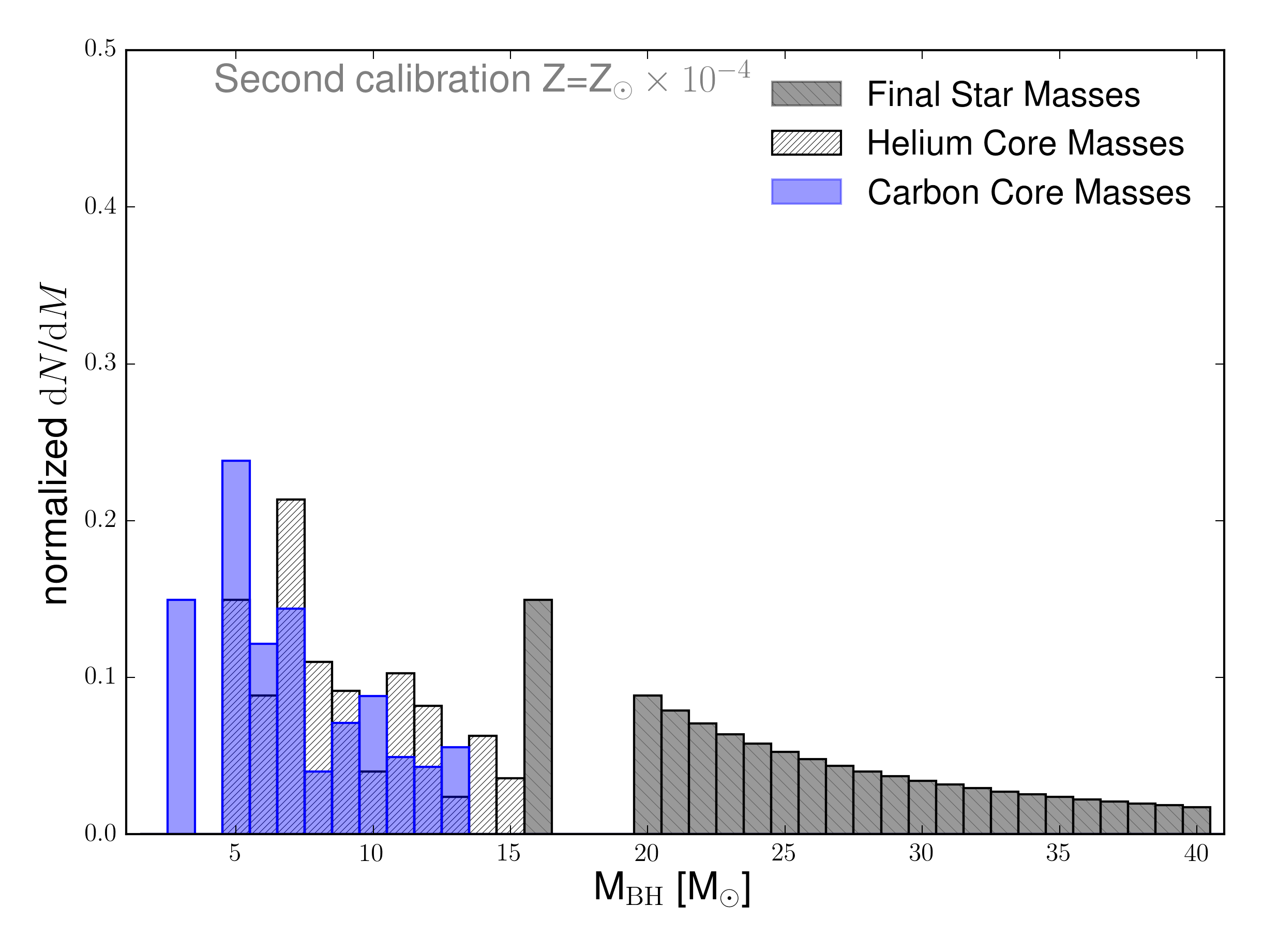} \\
    \includegraphics[width=0.48\textwidth]{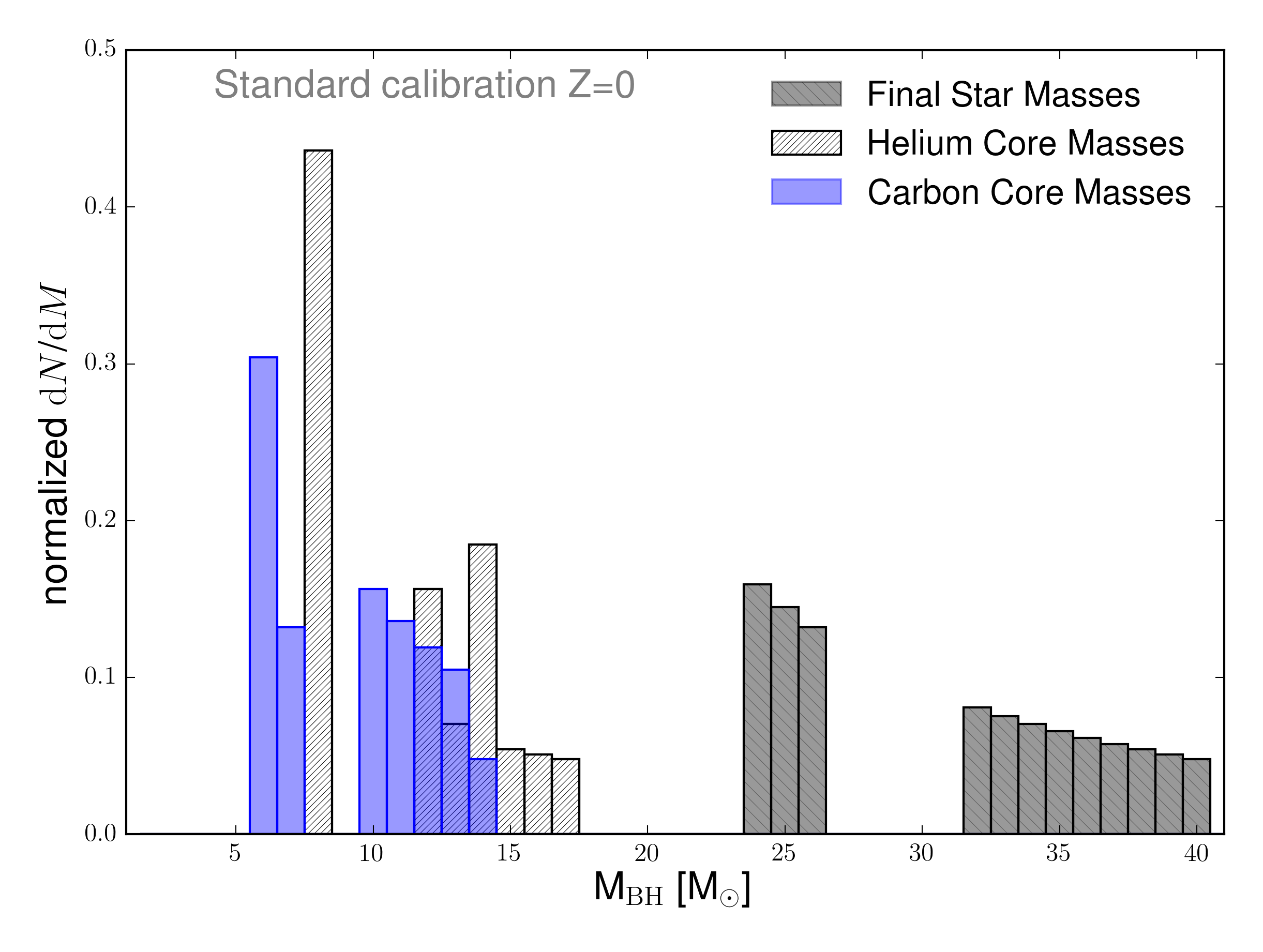}
    \includegraphics[width=0.48\textwidth]{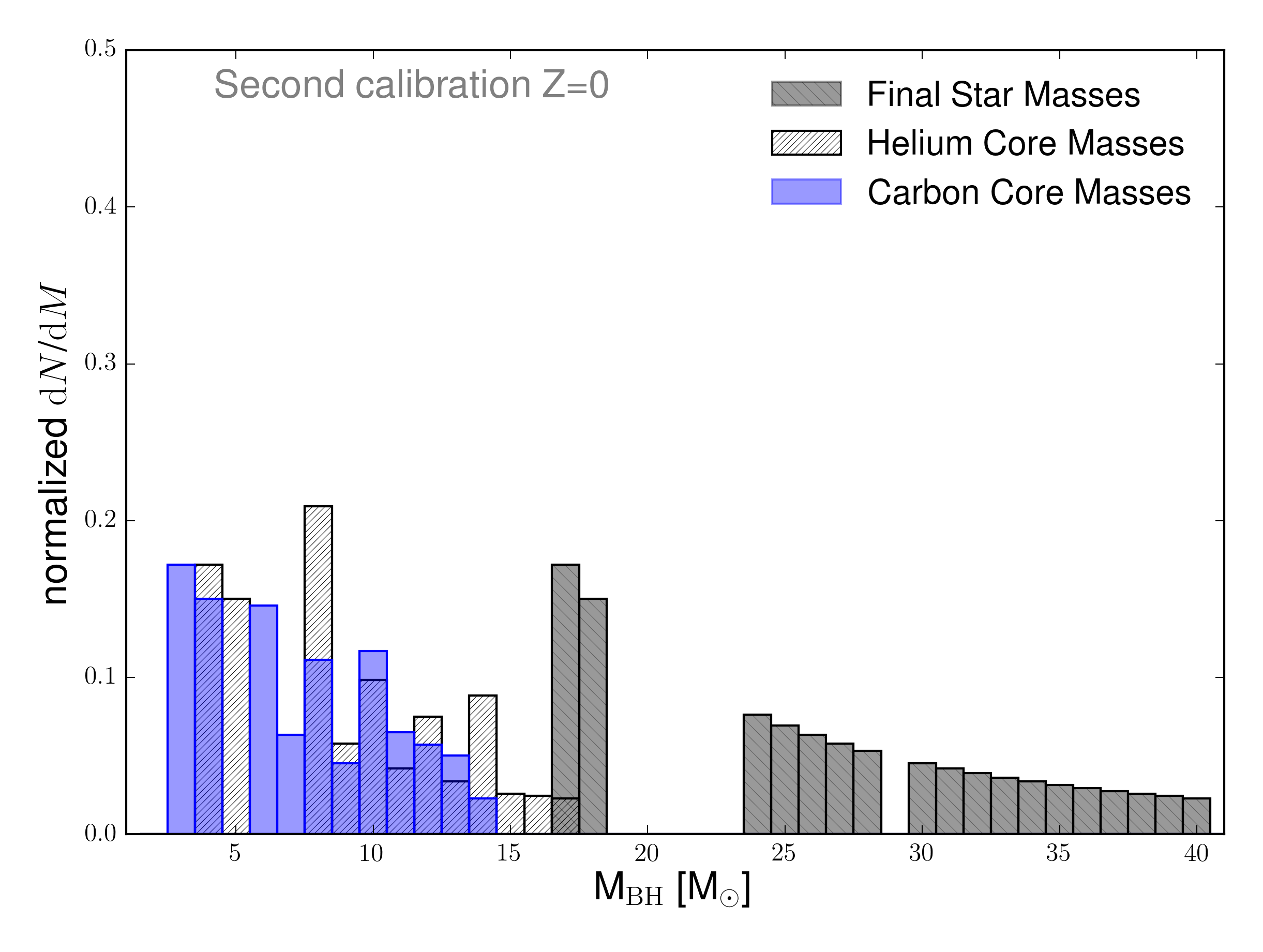}
\caption{Birth mass distributions of BHs for u-series (top panels) and z-series (bottom panels) for the standard calibration (left column) and the second calibration (right column). The different shaded bars indicate three different cases of possible BH mass distributions depending on how much of the initial stellar mass ultimately contributes to the final BH mass.
		\label{fig:bh_massdist}
    }
\end{figure*}

In this Section we present and discuss the mass distribution of the compact remnants formed in our CCSN study. In addition to explosion energies, total ejecta mass and elemental yields, remnant properties represent complementary observables that we can compare our results to. The simulation setup and the procedure to compute the mass distributions are the same as in Paper~II. We follow the full evolution of the PNS and obtain the baryonic mass of the freshly born hot NS. We then compute the corresponding zero-temperature gravitational mass of the NS using the HS(DD2) nuclear equation of state. We obtain the distributions of the birth-masses by weighting the predicted remnant masses as a function of ZAMS mass with the initial mass function (IMF) of massive stars from \citet{salpeter}. This IMF is suitable for the mass range ($M>10$~$M_{\odot}$) of this study.

We show the predicted gravitational birth-mass distribution of cool NSs for the u- and z-series for the standard calibration in Figure~\ref{fig:NS_mass_distr} (left column). The pre-explosion models have masses between 11 and 40~\msun (see Table~\ref{tab:progenitors}). The predicted NS masses are in the range between 1.3 and 1.8~\msun. Different ranges of ZAMS masses of the pre-explosion models that contribute to the distributions are indicated by different colors. In general, the higher ZAMS masses result in higher NS masses. The NS mass range around 1.4~\msun is mainly populated by pre-explosion models in the mass range from 11 to 15~\msun. The NS distribution above 1.4~\msun mainly originates from the ZAMS mass models above 15~\msun. 

The main difference between the u and z-series is how many models contribute to the higher NS masses around $\sim 1.6$~$M_{\odot}$. In the u-series we find some models contributing to the higher NS masses, whereas in the z-series typically these higher ZAMS mass models collapse to BHs and hence do not contribute to the NS distribution. As in our previous study of the solar metallicity samples in Paper~II,  the predicted NS distributions are slightly shifted to higher masses when compared to observed NS distributions \citep{ozel12} . This is due to the lack of low mass pre-explosion models ($M<11$~\msun) in the u- and z-series. Lower mass models are expected to produce lighter NSs. Due to their weighting in the applied IMF, these missing pre-explosion models should considerably contribute to the overall distribution of NS masses and should shift the lower limit of the NS mass distribution. Note that the presented birth mass distributions correspond to the outcome of single-star systems and we do not consider possible effects that are present in binary-star systems (e.g.\ accretion or mass loss). A potential difficulty in comparing theoretically predicted NS masses with observed NS masses is that the precision measurements of NS masses are from binary systems. However, \citet{raithel18} argue that such a comparison is still meaningful since the single-star models can be interpreted to be representative of some close binary scenarios due to the uncertain nature of mass loss.

The predicted NS birth mass distributions for the second calibration are more localized around 1.4~\msun and have an upper mass limit of $M \approx 1.6$~\msun which is lower than for the standard calibration, as can be seen in Figure~\ref{fig:NS_mass_distr}. Pre-explosion models in higher ZAMS mass regions more often collapse to BHs (see below) and do not contribute to the NS birth mass distributions for this calibration.  

If we assume that the almost failing SNe discussed in Section~\ref{subsec:highY-lowE} actually fail to explode and instead form a BH, the NS birth mass distribution we obtain would almost exclude gravitational birth masses above 1.6\msun for non-accreting single-star systems at low metallicity. If we further assume that also the u20 which has a delayed explosion will fail to explode, then the mass range above 1.6~\msun is excluded from non-accreting single-star systems at low metallicity.

Now we turn to the predicted BH birth mass distributions from our simulations. In the PUSH framework, CCSN simulations that run longer than the time on which PUSH is active and ultimately do not explode as well as the simulations that directly lead to the formation of a BH contribute to the predicted BH birth mass distribution. The mass of the BH formed in failed CCSN explosions depends on the stellar mass at collapse, which is affected by the star's mass loss history. In addition to mass loss, other processes such as the loss of the PNS binding energy in a weak shock \citep{lovegrove13} or the stripping of the envelope by a binary companion before collapse may affect the mass ultimately collapsing to a BH. To investigate the impact of these scenarios on the BH mass distribution, we consider three cases that span the range of outcomes. The most massive BHs for a given ZAMS mass are formed when the entire stellar mass at the onset of collapse ultimately ends up in the BH. The smallest BH masses for a given ZAMS are the results of a fully stripped CO-core collapsing to a BH. We also consider an intermediate case where only the hydrogen envelope is stripped, leaving the He-core to collapse to a BH.

In our simulations, we find continuous regions of BH formation above 31~\msun or 32~\msun for the u- and z-series respectively. We also find an isolated region of BH formation around 21--23~\msun (u-series) or 24--26~\msun (z-series). In Figure~\ref{fig:bh_massdist} (left column) we show the predicted BH birth mass distributions for the standard calibration, obtained by weighting the BH masses obtained in our simulations with the Salpeter IMF. Different shaded regions represent the three cases of mass-stripping the star may have experience prior to forming a BH. Depending on the degree of mass stripping, the predicted BH birth mass distributions range from 5 to 14~\msun if only the bare carbon core collapses to 21--40 \msun if the entire final stellar mass collapses to a BH. In contrast to the pre-explosion models at solar metallicity, those at low and zero metallicity do not experience much mass loss during their pre-supernova evolution and as a result their final mass at collapse can be similar to their initial ZAMS mass. This means that BHs as heavy as $\sim 40$~\msun could be formed in our low and zero-metallicity samples (see Figure~\ref{fig:prog_u02_z02_cores}). Note that the z-series does not include pre-explosion models with masses beyond 40~\msun. All u-series models above 40~\msun that are included in our study led to black hole formation.

For the second calibration considerably more BHs are formed (see Figures~\ref{fig:rectangles-engine} and \ref{fig:bh_massdist}). This shifts the resulting BH birth mass distribution to lower BH masses. The BH birth mass distribution resulting from stipped CO-cores have an upper mass limit of $\sim 14$~\msun (similar to the standard calibration) and populate almost all masses between 3~\msun and 14~\msun. The distributions from the case where only the hydrogen envelope is stripped, and hence the He-core mass determines the BH mass, span a range between 4 and 17~\msun. The most massive BHs (of the order of the initial ZAMS masses of the collapsing stars) are obtained in the case where the entire pre-explosion stellar mass forms the BH. In this case, the BH birth mass distribution is not continuous in mass because the ZAMS mass range of models forming BHs is also not continuous.

Next, we compute and present the fraction of stars that ultimately form BHs for all four samples of pre-explosion models and for both calibrations. Following Paper~II, we consider a mass range from 8 to 150~$M_{\odot}$ for the estimate, assume that stars between 8~$M_{\odot}$ and the lowest ZAMS mass in each pre-explosion series successfully explode and leave behind a NS as a remnant, and that the fate of the star with the highest ZAMS mass of each series is continued to 150~$M_{\odot}$. For our estimate we use again the Salpeter IMF. In Table~\ref{tab:bhfrac} we summarize the predictions for all series and both calibrations. Furthermore, we also include the fraction of mass from the CCSN progenitors that ultimately ends up in the BHs. We list the values for the case that the full pre-explosion models collapse to BHs and the values for the case that only the CO-cores contribute to the BH mass.

\begin{table}    
\begin{center}
    \caption{Fraction of core collapses forming BHs and fraction of initial stellar mass of massive stars bound in BHs for the standard calibration (I) and the `second calibration' (II). The range of the mass fraction bound in BHs indicates the three cases also shown in Figure~\ref{fig:bh_massdist}.
    \label{tab:bhfrac} 
    }
\begin{tabular}{ccccl}
\tableline \tableline
Series & Metallicity & Calibration & BH Fraction & Fraction of Mass \\
       &      $Z/Z_{\odot}$       &       &      & bound in BH \\
\tableline
z & 0                   & I & $\sim 18\%$ & $\sim 16-45 \%$ \\ 
u & $10^{-4}$ & I & $\sim 20\%$ & $\sim 18-48 \%$ \\
s & 1 & I & $\sim  5\%$ & $\sim  1-3 \%$  \\ 
w & 1         & I & $\sim  8\%$ & $\sim 5-6 \%$ \\
z & 0                    & II & $\sim 27\%$ & $\sim 18-55\%$ \\ 
u & $10^{-4}$  & II & $\sim 32\%$ & $\sim 22-61\%$ \\
s & 1          & II & $\sim 16\%$ & $\sim 4-7\%$ \\ 
w & 1         & II & $\sim 21\%$ & $\sim 8-14\%$ \\	
\tableline
\end{tabular}
\end{center}
\end{table}

\section{Summary and Discussion} \label{sec:discuss}

In this paper, we have simulated the death of two series of pre-SN models either as successful CCSN explosion or as the collapse to a black hole and, for the successful explosion, we have computed the detailed nucleosynthesis yields. 
The pre-SN models represent massive stars at low initial metallicity ($Z= 10^{-4} Z_{\odot}$) and zero initial metallicity ($Z=0$). We have used the PUSH method, which was first introduced in \citet{push1}, together with the standard calibration obtained in \citet{push2}. The work presented here complements the results obtained for pre-SN models at solar metallicity presented in \citet{push2} and \citet{push3}.

The main findings of our study are:
\begin{enumerate}
    \item As a whole, the resulting explosion energies for the low and zero metallicity pre-SN models are in agreement with observations of CCSNe. The results of the two series are similar, and they are also very similar to our results obtained with the same method for pre-SN models at solar metallicity. 
    As in our earlier works, we find that models with lower compactness tend to explode early, with lower explosion energy, and with a lower remnant mass.  A similar conclusion has recently been found by \citet{2019arXiv190904152B} using 3D models.
    \item The pre-SN model series at low and zero metallicity are more prone to BH formation than those at solar metalllicity. In addition, we find a few almost failing models at low/zero metallicity, which exhibit very low explosion energies together with $\sim 0.1$~$M_{\odot}$ of $^{56}$Ni. These models are located next to regions of BH formation and experience extended periods of mass accretion until the explosion finally sets in late. Unlike all other models, the almost failing supernova do not follow the observed $^{56}$Ni-$E_{\mathrm{expl}}$ relationship.
    \item We have shown and discussed that the compactness allows to infer some interesting explosion properties, however it does not tell the full story, as already found in \citet{push2}. In particular, we found a monotonic correlation between the compactness and the remnant mass of the exploding models. For models with compactness above $\sim 0.3$ we identify two branches of explosion energies, corresponding to models of similar compactness but located to either side of the peak in compactness (see Figure~\ref{fig:prog_compactness}).
    \item We find the same trends with compactness for iron group isotopes and elements for the low/zero metallicity pre-SN series of this work as for the pre-SN series at solar metallicity discussed in \cite{push3}. For symmetric isotopes (e.g.\ $^{56}$Ni and $^{44}$Ti) and iron-group elements (e.g.\ Ni and Cr) dominated by symmetric isotopes the yeilds exhibit linear trends with compactness. For asymmetric isotopes (e.g.\ $^{57}$Ni and $^{58}$Ni) and elements dominated by asymmetric isotopes (e.g.\ Mn) the yields depend more strongly on the local electron fraction, which measures the neutron excess. 
    \item We combined the predicted yields from the low and zero metallicity models with those from the models at solar metallicity and analyzed the metallicity dependence of the yields. For alpha elements which have significant contributions from pre-explosion hydrostatic burning we find almost no metallicity dependence. However, the oxygen yields have a strong dependence on the CO-core mass and hence the ZAMS mass. For iron group elements, the local electron fraction and the location of the mass cut affect the yields more strongly than the initial stellar metallicity. Machine-readable tables of all yields are included in this paper.
    \item We compare our results for iron-group nuclei to observationally derived abundances of a metal-poor star. Overall, we find a good agreement. We find large variations between models for some elements (e.g.\ Sc and Zn). Mn is underproduce in all our models.
    \item The predicted NS masses are broadly consistent with observations, as it was the case for series at solar metallicity in \citet{push2}. For the low/zero metallicity pre-SN samples, we find BH masses up to $40$~$M_{\odot}$, which provides a potential explanation for the BHS observed with LIGO/VIRGO. If we assume that the almost failing SNe do not explode, we exclude NS masses above 1.6~$M_{\odot}$. This is similar to the findings in \citet{raithel18}, where they do not find any NSs masses with $m>1.7$~$M_{\odot}$ from the simulations presented in \citet{sukhbold16}.
    \item We calculated the fraction of BHs as well as the fraction of mass ultimately bound in BHs for all four series of pre-explosion models. The fraction of mass that could ultimately be bound in BHs can be as high as $\sim$45-46~$\%$ for the low-zero metallicity samples when using the standard calibration ($\sim$55-61~$\%$ for the second calibration). For the samples at solar metallicity, the mass fraction in BHs is considerably smaller. 
\end{enumerate}

\acknowledgments  
The work at NC State was supported by United States Department of Energy, Office of Science, Office of  Nuclear Physics (award number DE-FG02-02ER41216). 
The effort at the Universit\"at Basel was supported by the Schweizerischer Nationalfond and by the ERC Advanced Grant ``FISH''. 
This work has benefited from the ``ChETEC'' COST Action (CA16117), supported by COST (European Cooperation in Science and Technology).
KE acknowledges the support from GSI.
AP acknowledges supports from the INFN project "High Performance data Network" funded by the Italian CIPE.
\software{Agile \cite{Liebendoerfer.Agile}, CFNET \cite{cf06a}, Matplotlib \citep{matplotlib}}

\clearpage
\bibliographystyle{yahapj}
\bibliography{references_push}

\clearpage
\pagebreak

\appendix

\section{Tables of explosion properties}
\label{appendix:hydro}

The explosion properties obtained from the standard PUSH calibration for the u- and z-series (see Figure~\ref{fig:properties-zams}) are provided as machine-readable Tables. Only the exploding models are included in the Tables. Parts of the Tables of the u-series and z-series are shown here for guidance regarding their form and content (see Tables~\ref{tab:explproperties} and \ref{tab:explproperties2}). The Tables are published in machine readable format.

\input{tab_appendix_explproperties.tex}

\section{Tables of complete isotopic yields} 
\label{appendix:yields}

Tables~\ref{tab:finab-u} and \ref{tab:finab-z} give the detailed isotopic composition of the post-processed ejecta and the total amount of post-processed ejecta for all exploding models of the u- and z-series\footnote{go.ncsu.edu/astrodata}.

\input{tab_appendix_yields.tex}

\end{document}

%% file: tab_appendix_explproperties.tex
\begin{table*}
	\begin{center}
		\caption{Explosion properties for the u-series
        	\label{tab:explproperties}
        }
		\begin{tabular}{lccccc}
			\tableline \tableline
			$M_{\rm ZAMS}$ & $M_{\rm collapse}$ & $E_{\rm expl}$  & $t_{\rm expl}$ & $M_{\rm ^{56}Ni}$ &  $M_{\rm remn}$ \\ 
			($M_{\odot}$)  & ($M_{\odot}$) & (B) & (s) &  ($M_{\odot}$) & $M_{\odot}$) \\
			\tableline
	11.0 & 11.00 & 0.75 & 0.51 & 4.30E-02 & 1.51 \\
    12.0 & 12.00 & 1.13 & 0.44 & 6.10E-02 & 1.56 \\
			\tableline
		\end{tabular}
	\end{center}
    \tablecomments{Table \ref{tab:explproperties} is published in its entirety in the machine-readable format. A portion is shown here for guidance regarding its form and content.}
\end{table*}

\begin{table*}
	\begin{center}
		\caption{Explosion properties for the z-series
        	\label{tab:explproperties2}
        }
		\begin{tabular}{lccccc}
			\tableline \tableline
			$M_{\rm ZAMS}$ & $M_{\rm collapse}$ & $E_{\rm expl}$  & $t_{\rm expl}$ & $M_{\rm ^{56}Ni}$ &  $M_{\rm remn}$ \\ 
			($M_{\odot}$)  & ($M_{\odot}$) & (B) & (s) &  ($M_{\odot}$) & $M_{\odot}$) \\
			\tableline
	11.0 & 11.00 & 0.50 & 0.49 & 2.02E-02 & 1.50 \\
    12.0 & 12.00 & 0.49 & 0.46 & 2.64E-02 & 1.41 \\
			\tableline
		\end{tabular}
	\end{center}
    \tablecomments{Table \ref{tab:explproperties2} is published in its entirety in the machine-readable format. A portion is shown here for guidance regarding its form and content.}
\end{table*}

%% file: tab_appendix_yields.tex
\begin{table*}
	\begin{center}
		\caption{Isotopic yields of stable isotopes in M~$_{\odot}$ for the u-series
        	\label{tab:finab-u}
        }
		\begin{tabular}{llllllllllll}
			\tableline \tableline
			Isotope & \multicolumn{10}{c}{Integrated processed yields for model} \\
			        & u11.0 & u12.0 & u13.0 & u14.0 & u15.0 & u16.0 & u17.0 & u18.0 & u19.0 & u20.0 & u21.0\\
			(-) & (M$_{\odot}$) & (M$_{\odot}$) & (M$_{\odot}$) & (M$_{\odot}$) & (M$_{\odot}$) & (M$_{\odot}$) & (M$_{\odot}$) & (M$_{\odot}$) & (M$_{\odot}$) & (M$_{\odot}$) & (M$_{\odot}$) \\
			\tableline
		$^{1}$H	& 9.47E-05 & 1.24E-04 & 1.56E-04 & 2.00E-04 & 1.72E-04 & 2.16E-04 & 2.00E-04 & 1.17E-04 & 1.24E-04 & 1.99E-04 & 1.27E-04 \\
		$^{2}$H	& 1.53E-07 & 1.47E-07 & 1.87E-07 & 1.84E-07 & 1.13E-07 & 1.61E-07 & 1.42E-17 & 1.21E-07 & 9.46E-08 & 2.35E-08 & 1.39E-14 \\
		$^{3}$He & 1.85E-09 & 1.78E-09 & 2.27E-09 & 2.24E-09 & 1.36E-09 & 1.93E-09 & 1.85E-12 & 1.46E-09 & 1.14E-09 & 2.81E-10 & 1.13E-14\\
		$^{4}$He & 1.15E-02 & 1.55E-02 & 1.89E-02 & 2.21E-02 & 1.87E-02 & 2.46E-02 & 1.36E-02 & 1.30E-02 & 1.31E-02 & 2.04E-02 & 1.69E-02\\
		$^{6}$Li & 5.42E-15 & 4.33E-19 & 2.41E-19 & 1.39E-19 & 2.18E-19 & 4.41E-19 & 3.45E-19 & 3.09E-19 & 3.31E-19 & 5.51E-19 & 2.93E-16\\
		$^{7}$Li & 3.92E-11 & 5.15E-11 & 6.22E-11 & 9.06E-11 & 4.86E-11 & 5.82E-11 & 3.82E-11 & 2.77E-11 & 2.37E-11 & 1.47E-11 & 1.98E-12\\
		$^{9}$Be & 1.52E-13 & 1.85E-13 & 2.55E-13 & 2.93E-13 & 3.11E-13 & 5.64E-13 & 1.96E-20 & 2.23E-13 & 2.03E-13 & 2.43E-13 & 8.30E-24\\
		$^{10}$B & 3.37E-13 & 3.67E-13 & 4.85E-13 & 5.45E-13 & 6.69E-13 & 1.29E-12 & 9.80E-17 & 4.37E-13 & 4.27E-13 & 5.99E-13 & 1.99E-15\\
		$^{11}$B & 2.13E-11 & 2.48E-11 & 3.04E-11 & 4.27E-11 & 2.44E-11 & 2.97E-11 & 1.22E-11 & 1.11E-11 & 9.48E-12 & 8.93E-12 & 1.38E-12\\
		$^{12}$C & 4.85E-04 & 1.07E-03 & 3.58E-03 & 4.78E-03 & 1.06E-03 & 3.75E-03 & 3.89E-04 & 9.25E-04 & 5.15E-04 & 4.68E-03 & 2.67E-03\\
		$^{13}$C & 4.40E-10 & 6.91E-10 & 1.59E-09 & 1.25E-09 & 8.63E-10 & 9.93E-10 & 6.62E-10 & 4.94E-10 & 4.95E-10 & 7.69E-10 & 5.44E-10\\
		$^{14}$N & 7.65E-09 & 6.85E-09 & 2.25E-08 & 2.87E-08 & 8.02E-09 & 2.45E-08 & 8.38E-09 & 5.97E-09 & 6.11E-09 & 3.55E-08 & 9.95E-09\\
		$^{15}$N & 3.52E-08 & 5.95E-08 & 8.15E-08 & 9.92E-08 & 4.92E-08 & 1.32E-07 & 2.90E-08 & 2.16E-08 & 2.21E-08 & 1.44E-07 & 4.23E-08\\
		$^{16}$O & 1.42E-01 & 2.15E-01 & 2.33E-01 & 3.11E-01 & 3.49E-01 & 4.85E-01 & 2.53E-01 & 2.76E-01 & 3.09E-01 & 5.08E-01 & 6.64E-01\\
		$^{17}$O & 4.02E-09 & 4.50E-09 & 6.13E-09 & 6.23E-09 & 7.14E-09 & 6.51E-09 & 6.75E-09 & 5.36E-09 & 5.37E-09 & 8.21E-09 & 7.34E-09\\
		$^{18}$O & 9.46E-10 & 1.04E-09 & 1.25E-09 & 1.17E-09 & 8.86E-10 & 7.70E-10 & 1.07E-09 & 6.17E-10 & 6.33E-10 & 3.41E-10 & 2.14E-10\\
		$^{19}$F & 1.15E-09 & 2.03E-10 & 5.54E-10 & 1.54E-10 & 1.66E-10 & 5.95E-11 & 2.38E-10 & 3.92E-11 & 4.08E-11 & 3.81E-11 & 1.35E-09\\
		$^{20}$Ne & 2.74E-02 & 3.68E-02 & 3.53E-02 & 4.65E-02 & 3.22E-02 & 6.14E-02 & 2.53E-04 & 2.52E-03 & 1.23E-03 & 6.59E-02 & 2.07E-02\\
		$^{21}$Ne & 1.82E-08 & 1.90E-08 & 1.66E-07 & 2.77E-07 & 9.06E-09 & 2.02E-07 & 5.12E-10 & 1.28E-09 & 6.87E-10 & 3.92E-07 & 1.61E-08\\
		$^{22}$Ne & 9.75E-07 & 9.28E-07 & 2.26E-06 & 2.37E-06 & 2.61E-06 & 2.09E-06 & 1.08E-06 & 7.30E-07 & 7.61E-07 & 1.07E-06 & 2.12E-07\\
			\tableline
		\end{tabular}
	\end{center}
    \tablecomments{The amount of mass above the mass cut which reaches temperatures $>1.75$~GK and hence has been postprocessed with the nuclear reaction network for each model is: 
1.799 M$_{\odot}$ (u11), 
1.989 M$_{\odot}$ (u12),
2.128 M$_{\odot}$ (u13),
2.278 M$_{\odot}$ (u14),
2.287 M$_{\odot}$ (u15),
2.663 M$_{\odot}$ (u16),
2.061 M$_{\odot}$ (u17),
2.093 M$_{\odot}$ (u18),
2.122 M$_{\odot}$ (u19),
2.988 M$_{\odot}$ (u20),
3.195 M$_{\odot}$ (u24),
2.638 M$_{\odot}$ (u25),
3.118 M$_{\odot}$ (u26),
3.171 M$_{\odot}$ (u27),
3.277 M$_{\odot}$ (u28),
3.065 M$_{\odot}$ (u30).
    Table \ref{tab:finab-u} is published in its entirety in the machine-readable format. A portion is shown here for guidance regarding its form and content. }
\end{table*}

\begin{table*}
	\begin{center}
		\caption{Isotopic yields of stable isotopes in M~$_{\odot}$ for the z-series
        	\label{tab:finab-z}
        }
		\begin{tabular}{llllllllllllllllll}
			\tableline \tableline
			Isotope & \multicolumn{10}{c}{Integrated processed yields for model} \\
			        & z11.0 & z12.0 & z13.0 & z14.0 & z15.0 & z16.0 & z17.0 & z18.0 & z19.0 & z20.0 & z24.0\\
			(-) & (M$_{\odot}$) & (M$_{\odot}$) & (M$_{\odot}$) & (M$_{\odot}$) & (M$_{\odot}$) & (M$_{\odot}$) & (M$_{\odot}$) & (M$_{\odot}$) & (M$_{\odot}$) & (M$_{\odot}$) & (M$_{\odot}$) \\
			\tableline
		$^{1}$H	& 5.97E-05 & 9.21E-05 & 1.10E-04 & 1.41E-04 & 1.35E-04 & 1.55E-04 & 2.22E-04 & 2.08E-04 & 1.98E-04 & 2.04E-04 & 1.66E-04\\
		$^{2}$H	& 8.54E-08 & 2.83E-13 & 1.70E-07 & 2.45E-07 & 1.37E-07 & 1.13E-07 & 1.97E-07 & 1.13E-07 & 7.91E-08 & 9.29E-08 & 1.06E-07\\
		$^{3}$He & 1.02E-09 & 5.23E-12 & 2.06E-09 & 3.00E-09 & 1.65E-09 & 1.36E-09 & 2.36E-09 & 1.35E-09 & 9.50E-10 & 1.12E-09 & 1.28E-09\\
		$^{4}$He & 7.90E-03 & 7.94E-03 & 1.11E-02 & 1.41E-02 & 1.38E-02 & 1.30E-02 & 2.47E-02 & 2.42E-02 & 1.61E-02 & 1.62E-02 & 1.37E-02\\
		$^{6}$Li & 4.53E-18 & 6.80E-15 & 9.39E-19 & 5.50E-19 & 2.37E-19 & 5.74E-19 & 3.65E-19 & 4.56E-19 & 2.55E-19 & 2.21E-19 & 2.77E-19\\
		$^{7}$Li & 3.46E-11 & 5.24E-11 & 3.80E-11 & 5.01E-11 & 5.44E-11 & 5.31E-11 & 6.98E-11 & 5.21E-11 & 2.80E-11 & 4.07E-11 & 3.78E-11\\\
		$^{9}$Be & 3.66E-13 & 6.03E-20 & 2.38E-13 & 2.49E-13 & 1.98E-13 & 3.13E-13 & 5.87E-13 & 4.93E-13 & 2.10E-13 & 2.08E-13 & 1.80E-13\\
		$^{10}$B & 9.05E-13 & 3.05E-14 & 4.72E-13 & 4.21E-13 & 3.89E-13 & 6.83E-13 & 1.29E-12 & 1.13E-12 & 4.60E-13 & 4.37E-13 & 3.63E-13\\
		$^{11}$B & 1.48E-11 & 2.23E-11 & 1.92E-11 & 2.52E-11 & 2.44E-11 & 2.00E-11 & 3.57E-11 & 2.65E-11 & 1.13E-11 & 1.64E-11 & 1.46E-11\\
		$^{12}$C & 3.13E-03 & 1.34E-03 & 1.02E-03 & 1.24E-03 & 1.80E-03 & 2.22E-03 & 3.89E-03 & 6.80E-03 & 5.70E-03 & 2.24E-03 & 3.57E-03\\
		$^{13}$C & 6.65E-10 & 6.63E-10 & 8.46E-10 & 1.18E-09 & 2.00E-10 & 9.58E-10 & 1.05E-09 & 8.14E-10 & 1.03E-09 & 6.38E-10 & 9.53E-10\\
		$^{14}$N & 1.66E-08 & 1.09E-08 & 8.11E-09 & 8.42E-09 & 1.07E-08 & 1.42E-08 & 2.04E-08 & 4.34E-08 & 2.65E-08 & 1.22E-08 & 1.67E-08\\
		$^{15}$N & 2.76E-08 & 2.65E-08 & 4.10E-08 & 4.16E-08 & 7.78E-08 & 9.01E-08 & 1.32E-07 & 2.04E-07 & 2.75E-08 & 5.96E-08 & 2.19E-08\\
		$^{16}$O & 5.25E-02 & 5.34E-02 & 1.22E-01 & 1.84E-01 & 1.68E-01 & 1.78E-01 & 3.92E-01 & 4.87E-01 & 3.13E-01 & 3.09E-01 & 1.67E-01\\
		$^{17}$O & 2.99E-09 & 5.71E-09 & 5.54E-09 & 6.10E-09 & 5.35E-09 & 5.47E-09 & 7.14E-09 & 6.94E-09 & 6.58E-09 & 7.22E-09 & 7.10E-09\\
		$^{18}$O & 1.12E-09 & 2.04E-09 & 1.34E-09 & 1.29E-09 & 9.61E-10 & 1.16E-09 & 9.40E-10 & 7.13E-10 & 6.31E-10 & 7.61E-10 & 9.17E-10\\
		$^{19}$F & 3.74E-08 & 5.33E-10 & 5.75E-11 & 2.54E-10 & 4.44E-11 & 2.08E-09 & 9.67E-11 & 5.08E-11 & 4.23E-11 & 5.12E-11 & 4.14E-11\\
		$^{20}$Ne & 1.16E-02 & 1.83E-02 & 2.62E-02 & 2.02E-02 & 4.51E-02 & 4.74E-02 & 6.78E-02 & 1.03E-01 & 7.72E-02 & 8.27E-02 & 4.67E-02\\
		$^{21}$Ne & 1.31E-07 & 1.04E-07 & 1.19E-08 & 1.21E-08 & 5.18E-08 & 8.32E-08 & 1.32E-07 & 5.71E-07 & 5.68E-07 & 7.21E-08 & 2.84E-07\\
		$^{22}$Ne & 1.15E-06 & 1.02E-06 & 4.89E-07 & 9.95E-07 & 7.32E-07 & 1.37E-06 & 2.21E-06 & 2.88E-06 & 2.22E-06 & 1.10E-06 & 1.18E-06\\
			\tableline
		\end{tabular}
	\end{center}
    \tablecomments{The amount of mass above the mass cut which reaches temperatures $>1.75$~GK and hence has been postprocessed with the nuclear reaction network for each model is:
1.627 M$_{\odot}$ (z11),
1.583 M$_{\odot}$ (z12),
1.792 M$_{\odot}$ (z13),
1.951 M$_{\odot}$ (z14),
1.887 M$_{\odot}$ (z15),
1.902 M$_{\odot}$ (z16),
2.547 M$_{\odot}$ (z17),
2.745 M$_{\odot}$ (z18),
2.461 M$_{\odot}$ (z19),
2.332 M$_{\odot}$ (z20),
2.042 M$_{\odot}$ (z21),
2.118 M$_{\odot}$ (z22),
2.565 M$_{\odot}$ (z23),
3.170 M$_{\odot}$ (z27),
2.967 M$_{\odot}$ (z28),
2.631 M$_{\odot}$ (z29),
2.775 M$_{\odot}$ (z30),
3.041 M$_{\odot}$ (z31).
    Table \ref{tab:finab-z} is published in its entirety in the machine-readable format. A portion is shown here for guidance regarding its form and content. }
\end{table*}